\documentclass[journal]{IEEEtran} 
\usepackage{xcolor,soul,framed} 

\colorlet{shadecolor}{yellow}

\usepackage[cmex10]{amsmath}
\usepackage{array}
\usepackage{mdwmath}
\usepackage{enumitem}
\usepackage{amsmath}
\usepackage{amsfonts}
\usepackage{caption}
\usepackage{amsfonts}
\usepackage{algorithmicx}
\usepackage{amssymb}
\usepackage{mdwtab}
\usepackage{eqparbox}
\usepackage{url}
\usepackage{multirow}
\usepackage{algorithm}
\usepackage{graphicx}%
\usepackage{multirow}%
\usepackage{amsmath,amssymb,amsfonts}%
\usepackage{hyperref}
\usepackage{amsthm}%
\usepackage{mathrsfs}%
\usepackage{xcolor}%
\usepackage{textcomp}%
\usepackage{manyfoot}%
\usepackage{booktabs}%
\usepackage{subcaption}
\usepackage{algorithm}%
\usepackage{multirow}
\usepackage{algorithmicx}%
\usepackage{algpseudocode}%
\usepackage{listings}%
\usepackage{makecell}%
\usepackage{caption}
\usepackage{longtable}
\usepackage{tikz}
\usetikzlibrary{shapes.geometric, arrows.meta, positioning, fit, backgrounds, calc}

\tikzset{
  startstop/.style={
    rectangle, rounded corners=4pt, minimum width=3.2cm, minimum height=0.55cm,
    text centered, draw=black, fill=black!10, font=\footnotesize\bfseries
  },
  process/.style={
    rectangle, minimum width=3.2cm, minimum height=0.55cm,
    text centered, draw=black, fill=white, font=\footnotesize,
    text width=3.1cm
  },
  decision/.style={
    diamond, aspect=2.4, minimum width=3.2cm, minimum height=0.6cm,
    text centered, draw=black, fill=white, font=\footnotesize,
    inner sep=1pt, text width=2.4cm
  },
  io/.style={
    trapezium, trapezium left angle=75, trapezium right angle=105,
    minimum width=3.2cm, minimum height=0.55cm,
    text centered, draw=black, fill=black!5, font=\footnotesize,
    text width=2.8cm
  },
  arrow/.style={-{Stealth[length=5pt]}, thick},
  dasharrow/.style={-{Stealth[length=5pt]}, thick, dashed},
  lane/.style={draw=black!30, fill=black!3, rounded corners=3pt
  },
  oracle/.style={
    rectangle, minimum width=2.6cm, minimum height=0.42cm,
    text centered, draw=black, fill=black!5, font=\scriptsize,
    text width=2.5cm
  },
  dasharrow/.style={-{Stealth[length=4pt]}, thick, dashed},
  lane/.style={draw=black!30, fill=black!3, rounded corners=3pt},
  phase/.style={rectangle, rounded corners=2pt, draw=black!35, dashed,
                inner sep=3pt, font=\tiny\itshape},
}

\begin{document}
\bstctlcite{IEEEexample:BSTcontrol}
    \title{\huge A Tutorial on Practical QKD Post-Processing Algorithms
Using a GUI-Based Simulation Framework 
}
  \author{Ushnik Nath, Neha Sharma,
      Ramniwas Meena, and 
      Neel Kanth Kundu,~\IEEEmembership{Member,~IEEE}

      \thanks{This work has been submitted to the IEEE for possible publication. Copyright
may be transferred without notice, after which this version may no longer be
accessible. }
      
     \thanks{This work was supported in part by the National Quantum Mission of India, DST INSPIRE Faculty Fellowship (Reg. No.: IFA22-ENG 344), ANRF Prime Minister Early Career Research Grant (ANRF/ECRG/2024/000324/ENS), and IIT Delhi's New Faculty Seed Grant.}

\thanks{U. Nath, N. Sharma and R. Meena are with the Center for Applied Research in Electronics (CARE), Indian Institute of Technology Delhi, New Delhi 110016, India (e-mails: ushniknath@gmail.com, nehasharmablp461@gmail.com,  ram.21phy@gmail.com.)}

\thanks{N. K. Kundu is with CARE and the Bharti School of Telecommunication Technology and Management, Indian Institute of Technology Delhi, New Delhi-110016, India (e-mail: neelkanth@iitd.ac.in). He is also an honorary fellow at the Department of Electrical and Electronic Engineering, University of Melbourne, Melbourne, VIC, Australia.}

      }
      

\maketitle

\begin{abstract}
Recent experimental demonstrations of Quantum Key Distribution (QKD) have attracted significant attention for their ability to establish information-theoretically secure encryption keys for next-generation communication systems. As practical deployment accelerates, the realization of compact, scalable, and high-throughput discrete-variable (DV) QKD systems has become increasingly important. While advances in photonic integration are enabling low-size, low-weight, and low-power QKD hardware, the security and performance of these systems rely heavily on efficient post-processing techniques to generate identical, secure keys. This tutorial presents a comprehensive algorithm-level review of practical post-processing techniques for DV-QKD, with particular emphasis on information reconciliation methods based on Cascade, Winnow, low-density parity-check (LDPC), and polar codes. We also analyze the theoretical asymptotic and finite-size secret key rates (SKRs) of DV-QKD and benchmark these bounds against the practical SKRs achieved using actual post-processing algorithms. Furthermore, we present a server-client model to implement the entire QKD post-processing pipeline. The performance is evaluated in terms of QBER and SKR across various system parameters, considering the complete post-processing pipeline, including sifting, information reconciliation, privacy amplification, and authentication. This tutorial also presents a graphical user interface (GUI) for the QKD post-processing methods, enabling researchers, industry practitioners, and academic users to visualize and analyze the impact of different post-processing techniques and system parameters, such as QBER, on the SKR performance of DV-QKD systems. 
\end{abstract}


\maketitle

\section{INTRODUCTION}
\IEEEPARstart{S}{ecure} communication forms the foundation of today's digital ecosystem. It enables the secure exchange of information across a wide range of applications, including financial services, private communications, cloud computing, and data transmission. Modern communication systems primarily rely on asymmetric cryptographic algorithms to ensure secure communication \cite{simmons1979symmetric}. One of the most widely deployed and established schemes is the RSA algorithm \cite{rivest1978rsa}. In these algorithms, security relies on assumptions related to computational complexity. In other words, solving mathematical problems like integer factorization or the discrete logarithm problem is considered computationally infeasible \cite{shor1994algorithms}. As a result, cryptographic systems are fundamentally secure because they assume that adversaries possess computational limitations. Advances in algorithmic techniques or the emergence of large-scale quantum computing could render hard problems easier to solve \cite{ma2005practical}. Consequently, there is an urgent need for alternative approaches to enhance security that extend beyond classical computational methods. 

Unlike conventional cryptography, which depends on computational complexity, quantum cryptography utilizes quantum mechanics to enable information-theoretically secure quantum communication. One of the basic principles of quantum mechanics is the no-cloning theorem, which states that an unknown quantum state cannot be copied \cite{wootters1982nocloning}. This security is rooted in the Heisenberg uncertainty principle, which limits the precision of simultaneous measurements of non-commuting observables \cite{heisenberg1927uncertainty}. Consequently, any attempt to intercept or measure quantum information would cause disturbances. Quantum Key Distribution (QKD) has emerged as a promising approach for achieving information-theoretically secure communication \cite{jain2011device}. There are various implementations of QKD protocols that can broadly be classified into discrete-variable (DV-QKD) and continuous-variable (CV-QKD) approaches, depending on the nature of the quantum states used for encoding information. 

In this work, we focus primarily on the BB84 DV-QKD protocol proposed by Bennett and Brassard in 1984, one of the earliest and most influential protocols in QKD \cite{bennett1984bb84}. In the BB84 scheme, the sender (Alice) prepares quantum states in randomly chosen bases and transmits them to the receiver (Bob) over a quantum channel. Although quantum transmission provides a means of detecting the presence of an eavesdropper, the transmission process alone is not sufficient to generate a secure secret key. In a practical quantum communication system, the transmitted quantum states are influenced by channel noise, photon loss, detector imperfections, and other hardware-related limitations \cite{rozenman2026free}. However, the raw keys obtained by Alice and Bob may differ and cannot be used directly as a shared secret key \cite{blum1983exchange}. A sequence of classical post-processing procedures is therefore performed to transform the raw measurement outcomes into a common and secure key. These procedures generally include basis sifting, error-rate estimation, information reconciliation, and privacy amplification. Among these stages, information reconciliation plays a central role because it enables Alice and Bob to identify and correct discrepancies between their respective raw keys while carefully limiting the amount of information revealed over the public communication channel. Several information reconciliation approaches have been investigated for QKD systems, with Cascade and Winnow being among the earlier and widely adopted protocols \cite{brassard1993Cascade,bennett1995privacy}.
More recent approaches employ structured error-correcting codes, such as low-density parity-check (LDPC) codes \cite{Gallager_ldpc} and polar codes \cite{renes2012polar}, to perform information reconciliation.

The choice of reconciliation scheme is not merely an implementation detail; it directly influences the performance of the QKD system. In particular, the amount of information disclosed during reconciliation, the computational and communication requirements, and the protocol's ability to operate reliably at different error rates all affect the achievable secret key rate. These considerations have motivated continued research into post-processing methods for DV-QKD, with particular emphasis on improving reconciliation efficiency while limiting information leakage and maintaining reliable key agreement under realistic channel conditions. The design of efficient post-processing procedures is increasingly important for realizing QKD systems that can operate reliably under practical deployment conditions \cite{Xu2020}. 

Although substantial research has focused on optimizing individual stages of QKD post-processing, several fundamental challenges remain open. In particular, the design of post-processing algorithms that minimize classical computational and communication overhead while preserving rigorous security guarantees remains an important research problem \cite{Luo2024}. Furthermore, the impact of finite-key effects on the achievable secret key rate, relative to the corresponding asymptotic secret key rate under realistic operating conditions, requires careful consideration \cite{Tomamichel2012}. Another critical aspect is establishing the relationship among the number of transmitted quantum signals, the length of the final identical secret key shared by the legitimate users, and the corresponding security parameter \cite{rezki2017secret}. Addressing these challenges is essential for the development and deployment of practically secure and efficient QKD systems.

We address these challenges in this work with the main contributions of this tutorial summarized as follows:
\begin{itemize}
    \item We present a unified, algorithmic-level tutorial on DV-QKD post-processing that covers four widely used error reconciliation protocols: Cascade, Winnow, LDPC, and Polar codes, together with privacy amplification, within a single consistent framework. This work provides step-by-step algorithmic descriptions suitable for direct implementation, addressing a gap that is particularly important for the practical deployment of QKD post-processing.
    \item We present an actor-based server-client model and  review the theoretical asymptotic and finite-size SKR expressions for the BB84 protocol. We benchmark various simulated post-processing methods on this protocol. We compare the practical SKRs achieved using actual post-processing algorithms against the corresponding theoretical finite-size SKR estimates, directly assessing how closely practical implementations approach their theoretical limits.
    \item We implement and comparatively evaluate the four reconciliation protocols under a common set of performance metrics: QBER, reconciliation efficiency, and final SKR, across varying channel attenuation, transmission distance, and mean photon number.
    \item We develop an interactive, web-based GUI that allows researchers, industry professionals, and academic users to visualize and analyze the effect of QKD system parameters such as channel attenuation, transmission distance, and noise on the QBER and SKR  performance metrics across different post-processing methods.
\end{itemize}


The remainder of this paper is organized as follows. Section~\ref{sec:Related_Work} presents reviews of related work. Section~\ref{sec:Modeling_of_DV-QKD_Systems} models the physical components of a DV-QKD system and introduces the associated performance metrics. Section~\ref{sec:BB84_protocol} presents the BB84 protocol along with its QBER and SKR analysis. Section~\ref{sec:Error_Reconciliation_Protocols} details the reconciliation protocols with their algorithmic implementation steps.
In section~\ref{sec:Privacy_Amplification} privacy amplification is discussed. Section~\ref{Simulation_Results} presents the numerical and GUI-based simulation results. Finally, section~\ref{sec:Conclusion} concludes the paper with directions for future work.

\section{Related Work}\label{sec:Related_Work}
QKD has undergone substantial advancement over the past several decades. The growing demand for secure communication has driven continued development in this area. In parallel with experimental advances, simulation frameworks have been developed to support cost-effective evaluation and validation of QKD protocols \cite{gkouliaras2024nuqkd}. Most QKD implementations remain protocol-specific, requiring specialized expertise and configuration for differences in encoding schemes, channel conditions, and security assumptions. Various simulation approaches have therefore been developed, with their limitations summarized in Table~I. Initial studies largely concentrated on mathematical modeling and security analysis, providing theoretical bounds and performance limits for QKD systems \cite{lydersen2010hacking, renner2008security}.

Software-based simulators were later introduced to reproduce QKD protocol behavior under practical channel conditions, including noise effects and reconciliation procedures \cite{elkouss2009efficient}. Platform-oriented simulators expanded this scope by supporting scalable experiments and network-level studies across distributed environments \cite{quantuminspire}. Educational and visualization tools have also been developed to make QKD concepts easier to understand and accessible. Recent studies have increasingly examined post-processing operations, particularly error correction and privacy amplification, to enhance key generation efficiency and reduce resource requirements \cite{hybrid2018, Sharma2025Comparative}. Nevertheless, these developments remain largely fragmented, with most existing approaches concentrating on specific stages of the QKD workflow rather than integrating the complete process within a unified framework \cite{oukaira2025quantum}.

A key limitation in existing work is the absence of integrated frameworks that link QKD key generation with configurable post-processing within an interactive platform \cite{aquina2025critical}. This limitation forms the basis for the proposed GUI-driven framework, which integrates the complete process from sifted key generation to the establishment of an authenticated and identical key. QKD simulation has been widely studied as a means of reducing reliance on specialized quantum hardware and evaluating system performance under controlled conditions. Although extensive work has addressed QKD protocol design and security analysis, relatively limited attention has been given to systematic comparison using practical performance measures such as QBER and SKR. This leaves a gap between theoretical protocol development and its practical evaluation, particularly from an engineering perspective. 

Although QKD simulation frameworks have progressed from mathematical models to software and platform-based implementations, relatively little work has focused on combining analytical evaluation with an intuitive user experience. Software-based simulators often face limitations such as restricted scalability, manual configuration, and limited GUI support, particularly in early-stage implementations. Educational and visualization tools such as CrypTool have addressed usability through interactive demonstrations of QKD processes \cite{cryptool}; however, their protocol coverage and parameter configurability are generally limited. Existing studies have simulated protocols including BB84, B92, DPS-QKD, and coherent-state QKD to examine noise, security, hardware imperfections, and post-processing performance. Platform-based tools such as QuCCs, Quantum-Sim, qkdSim, EnQuad, and NuQKD provide cost-effective environments for QKD simulation and validation before physical deployment \cite{BB84Depolarizing,QuantumSim,EnQuad,NuQKD}. However, these platforms are typically designed for specific protocols and may require substantial configuration. The study presented in \cite{pappalardo2023qkd} further considers post-processing reconciliation, focusing on the optimization of reconciliation and privacy amplification within QKD systems. While it provides useful results on computational efficiency and resource utilization, it lacks an integrated algorithmic framework for QKD post-processing and an interactive simulation environment. Furthermore, these prior works have not compared the practical SKR obtained from the QKD post-processing algorithms against the theoretical asymptotic and finite-size SKR results.

To overcome these limitations, this work presents an integrated framework for QKD post-processing steps with practical implementation. The proposed system couples a user-friendly GUI with a flexible simulation backend, allowing users to execute QKD protocols, apply different post-processing techniques, and perform privacy amplification to obtain an identical final key. Users can initially provide the sifted key obtained from the quantum communication phase of the BB84 protocol, after which the key reconciliation process is performed. Four information reconciliation techniques, namely Cascade, Winnow, LDPC, and polar codes, are evaluated. The simulation framework examines the execution time for information reconciliation, privacy amplification (PA), and the final SKR as output measures. The analysis also considers parameters including block size, distance, mean photon number, measured QBER, and the percentage of key leakage from the sifted key. This enables the generation of an identical key after the QKD post-processing steps and the comparison of the practically achievable SKR with the theoretical asymptotic and finite-size SKR bounds. 


\begin{table*}[htbp!]
\centering
\caption{Overview of BB84 QKD simulation and post-processing frameworks}
\footnotesize
\renewcommand{\arraystretch}{1.05}

\resizebox{\textwidth}{10cm}{
\begin{tabular}{|p{2.4cm}|p{2.2cm}|p{4cm}|p{3cm}|c|}
\hline
\textbf{Category} & \textbf{Tools and Models} & \textbf{Contribution Summary} & \textbf{Limitation} & \textbf{Year} \\
\hline

\multirow{4}{2cm}{\parbox{2cm}{\centering \textbf{Mathematical}\\\textbf{Simulation}}}
& QuTiP & Simulates quantum states, operators, and open quantum-system dynamics & Simplified adversarial models  & 2012 \cite{johansson2012qutip} \\
& Qiskit & Quantum-circuit simulation and implementation& Simulation does not inherently provide an end-to-end DV-QKD post-processing pipeline. & 2025 \cite{javadiabhari2024qiskit} \\
& MATLAB-based Simulation & Modeling QKD protocols under various noise conditions & Does not provide an integrated, user-configurable post-processing workflow & 2010 \cite{Sharma2026matlab} \\

\hline

\multirow{4}{2cm}{\parbox{2cm}{\centering \textbf{Software-Based}\\\textbf{Simulation}}}
& MATLAB Models & BB84 simulation including sifting and noise modeling & Limited scalability & 2009 \cite{buttler1998practical} \\

& Python Tools & Statistical QKD simulation Framewok & Limited focus on QKD post-processing& 2019 \cite{lardier2019quantum} \\
& Qiskit Models & Circuit-level BB84 simulation with noise models & No full pipeline integration & 2021 \cite{grossi2021serverless} \\
& AI Simulators & Adaptive QKD simulation using intelligent optimization & Early-stage development & 2025 \cite{banerjee2025mlqkd} \\
& Hybrid Frameworks & Combined classical-quantum simulation systems & High implementation complexity & 2026 \cite{devani2026quantum} \\
\hline

\multirow{4}{2cm}{\parbox{2cm}{\centering \textbf{Platform-Based}\\\textbf{Simulation}}}
& Network Simulators & Multi-node QKD communication frameworks & Limited post-processing flexibility & 2016 \cite{qkdnetwork2016} \\
& Modular Platforms & Flexible architectures for protocol testing & Partial workflow integration & 2017 \cite{mailloux2017modeling} \\
& Cloud Platforms & Remote-access quantum simulation systems & Restricted parameter control & 2018 \cite{silva2018qiskit} \\
& Decoy-State Models & A universal simulating framework QKD & Protocol-focused only & 2020 \cite{fan2020universal} \\
\hline


\multirow{4}{2cm}{\centering \textbf{Education}\\\textbf{Visualization}\\\textbf{Simulation}}
& CrypTool Project & Interactive visualization of BB84 protocol operations & Not suitable for evaluation & 2013 \cite{cryptool} \\
& Python Demos & Educational QKD implementations & Limited scalability & 2023\cite{sang2023hands} \\
& Web Tools & Browser-based QKD visualization platforms & Lack of rigorous validation & 2026 \cite{nath2026tutorial} \\
\hline

\multirow{4}{2cm}{\parbox{2cm}{\centering \textbf{Post-Processing}\\\textbf{Reconciliation}\\\textbf{Simulation}}}
& Cascade Protocol & Interactive error correction in QKD systems & High communication overhead & 2006 \cite{pedersen2013high} \\
& LDPC-Based Methods & Forward error correction for efficient reconciliation & Complex parameter tuning & 2011 \cite{milicevic2017key} \\
& Privacy Amplification & Hash-based key compression techniques & Depends on entropy estimation & 2013 \cite{privacy2013} \\
& Hybrid Reconciliation & Combined reconciliation and authentication workflows & Limited dynamic configuration & 2018 \cite{sarabu2018framework} \\
& Resource-Efficient QKD & Optimization of post-processing pipelines & No GUI integration & 2024 \cite{wang2026short} \\
& Adaptive Post-Processing & Intelligent reconciliation strategy selection & No standardized frameworks & 2026 \cite{sanz2026extending} \\
\hline
\end{tabular}}
\end{table*}

\section {DV-QKD System Model} \label{sec:Modeling_of_DV-QKD_Systems}
\begin{figure}
    \centering \includegraphics[width=\linewidth]{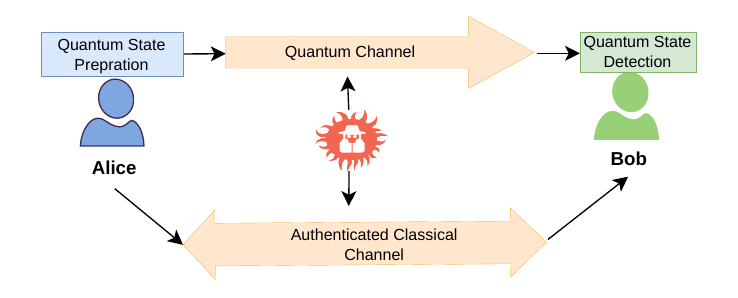}
    \caption{Quantum Key Distribution}\label{sim_opt_eff_classFinv}
\end{figure}

In DV-QKD, secure key establishment is achieved by encoding classical information into quantum states and transmitting them over a quantum channel. The overall process can be modeled through three primary components: the source where the quantum state is prepared, the quantum channel for the transmission, and the detection system where the quantum state is measured. Fig. \ref{sim_opt_eff_classFinv} shows a schematic of a general QKD system. Furthermore, the details of each module of the QKD system are explained in the sub-section below.

\subsection{Source}
The BB84 prepare-and-measure QKD protocol can employ weak coherent pulses (WCPs) as a practical approximation to an ideal single-photon source. The photon number in a WCPs pulse follows a Poisson distribution. For a pulse with mean photon number $\mu$, the probability of containing exactly $n_{\mathrm{ph}}$ photons is given by:

\begin{equation}
P(n_{\mathrm{ph}})=\frac{\mu^{n_{\mathrm{ph}}} e^{-\mu}}{n_{\mathrm{ph}}!} \label{eq:photon_ number}
\end{equation}
where $P(n_{\mathrm{ph}})$ denotes the probability that a pulse contains exactly $n_{\mathrm{ph}}$ photons, and $\mu$ denotes the mean photon number per pulse.

\subsection{Channel}
In the proposed simulation framework, the quantum channel is modeled as an optical-fiber link. The total channel transmission efficiency, $\eta_{\mathrm{ch}}$, depends on the transmission medium and the distance ($d$) between Alice and Bob. For the fiber optic channel, transmission loss is represented using exponential attenuation, with $\alpha_f$ the fiber attenuation coefficient denoted in dB/km. A base channel efficiency, $\eta_0$, is also included to account for practical losses arising from imperfect optical coupling, connector losses, and alignment errors. The effective fibre transmission efficiency is therefore given by:
\begin{equation}
\eta_{\mathrm{ch}}(d) = \eta_0\, 10^{-\alpha_f d/10}.\label{eq:fiber_trans}
\end{equation}

\subsection{Detector}
The photon detector model is defined in terms of the relevant detector parameters \cite{huang2017modeling}. The detector efficiency, $\eta_{\mathrm{det}}$, specifies the probability of registering an incident photon, and $\eta_{\mathrm{sys}}$ denotes the overall system efficiency. The combined channel and detector efficiency between Alice and Bob is expressed as:
\begin{equation}
\eta_{\mathrm{sys}}=\eta_{\mathrm{ch}}\eta_{\mathrm{det}}. \label{eq:channel_trans}
\end{equation}
The dark count rate, $\delta_{\mathrm{dc}}$, characterizes spurious detection events occurring without incident photons and is modeled as a Poisson process. For a detection interval $t_d$, the probability of observing at least one dark count is
\begin{equation}
P_{\mathrm{dark}}=1-e^{-\delta_{\mathrm{dc}}t_d}.
\end{equation}
For $n_{\mathrm{ph}}$ incident photons, the probability of registering at least one photon is
\begin{equation}
P_{\mathrm{detect}}(n_{\mathrm{ph}}) = 1-(1-\eta_{\mathrm{det}})^{n_{\mathrm{ph}}}.
\end{equation}

\subsection{Performance Metrics }\label{sec:Performance_ Metrics}

In DV-QKD, system performance and security are commonly evaluated using the QBER and SKR. QBER measures the mismatch between the transmitted and received key bits and reflects the effects of channel noise, device imperfections, and possible eavesdropping. In contrast, SKR represents the rate of secure key generation after the required post-processing steps. These two parameters provide a practical basis for assessing transmission accuracy, noise tolerance, key generation efficiency, and security under realistic operating conditions.
\subsubsection{QBER}
It measures the fraction of errors between the sifted keys obtained by Alice and Bob, and serves as an indicator of channel quality and potential security threats. It is defined as
\begin{equation}
Q = \frac{n_{\mathrm{err}}}{n_{\mathrm{sift}}}.
\end{equation}
where $n_{\mathrm{err}}$ denotes the number of mismatched bits and $n_{\mathrm{sift}}$ is the sifted key length. An increase in QBER may indicate eavesdropping, since an eavesdropper's interaction with the quantum states introduces additional disturbances.

\subsubsection{SKR}
It represents the amount of secure key that Alice and Bob can obtain per transmitted signal after error correction and protection against information leakage. It can be expressed using the Devetak-Winter bound as \cite{devetak2005distillation}:
\begin{equation}
R_{sk} = I_{AB} - \chi_{AE}.
\end{equation}
where, $I_{AB}$ denotes the classical mutual information shared by Alice and Bob, and $\chi_{AE}$ represents the Holevo information between Alice and Eve. Thus, the achievable SKR is determined by the information shared by the legitimate users and the information potentially available to Eve.

\section{BB84 QKD Protocol}
\label{sec:BB84_protocol}

\subsection{Protocol Description}
\label{subsec:protocol_description}

The BB84 protocol was introduced in 1984 by Bennett and Brassard~\cite{bennett2014quantum}. It allows two legitimate parties, Alice and Bob, to establish a shared secret key whose security is guaranteed by the laws of quantum mechanics rather than computational assumptions. For polarization encoding, Alice prepares each qubit in one of two conjugate bases: the rectilinear $(+)$ basis, $\{\leftrightarrow,\updownarrow\}$, or the diagonal $(\times)$ basis, $\left\{\vcenter{\hbox{\rotatebox{45}{$\leftrightarrow$}}},\vcenter{\hbox{\rotatebox{135}{$\leftrightarrow$}}}\right\}$. The mapping between classical bit values and polarization states, shared publicly between Alice and Bob, is summarized in Table~\ref{tab:bb84_encoding} 

\begin{table}[htp]
\centering
\caption{Polarization encoding scheme for the BB84 protocol.}
\label{tab:bb84_encoding}
\renewcommand{\arraystretch}{1.3}
\begin{tabular}{|c|c|c|}
\hline
\textbf{Basis} & \textbf{Bit value} & \textbf{Polarization state} \\
\hline
\multirow{2}{*}{Rectilinear $(+)$} & 0 & $\leftrightarrow$ (Horizontal) \\
 & 1 & $\updownarrow$ (Vertical) \\
\hline
\multirow{2}{*}{Diagonal $(\times)$} & 0 & $\vcenter{\hbox{\rotatebox{45}{$\leftrightarrow$}}}$ (Diagonal) \\
 & 1 & $\vcenter{\hbox{\rotatebox{135}{$\leftrightarrow$}}}$ (Anti-diagonal) \\
\hline
\end{tabular}
\end{table}

\begin{figure}[htp]
    \centering
    \includegraphics[width=\linewidth]{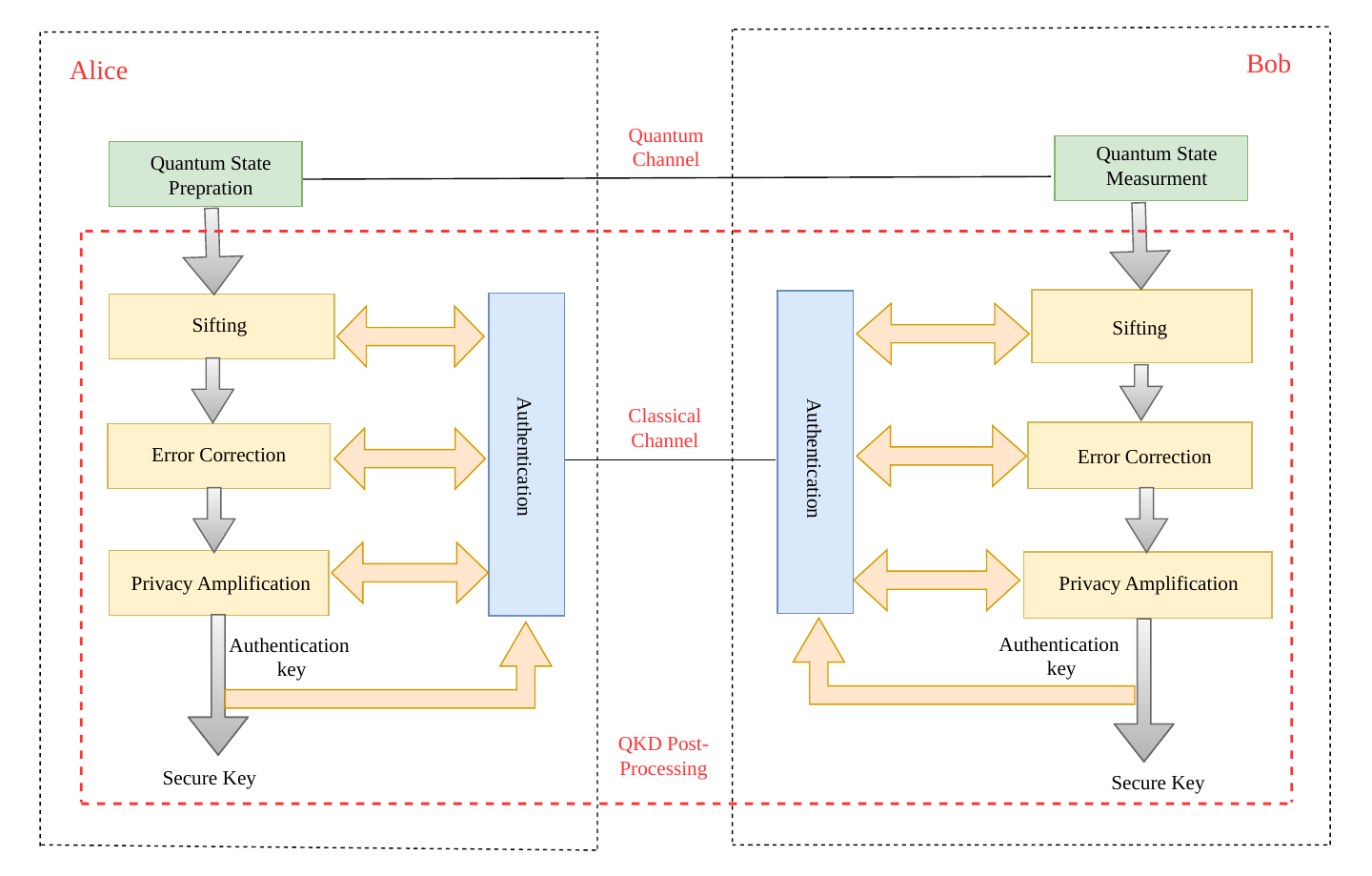}
    \caption{DV-QKD protocol and post-processing pipeline.}
    \label{dv_qkd_ppp}
\end{figure}

Quantum transmission alone is not sufficient to produce a usable secret key because, in practice, quantum channels introduce noise, photon loss, and detector imperfections, and an eavesdropper may retain partial information about the transmitted states. The full protocol, therefore, extends beyond quantum transmission into a classical post-processing stage, which includes error correction and privacy amplification, that converts the raw sifted bits into a shorter, identical, and secure final key. Fig. \ref{dv_qkd_ppp} gives an overview of the complete QKD pipeline, which proceeds through the following steps. \\

\subsubsection{Quantum Communication Phase}
\begin{enumerate}[label=\alph*)]
    \item \textit{Bit and basis generation:} Alice generates $N_{\mathrm{tx}}$ random classical bits. For each bit, she independently and randomly selects one of the two conjugate bases, rectilinear or diagonal.

    \item \textit{State preparation:} Using the encoding rule in Table~\ref{tab:bb84_encoding}, Alice maps each (bit, basis) pair to the corresponding polarisation state and prepares a photon in that state.

    \item \textit{Quantum transmission:} Alice transmits the $N_{\mathrm{tx}}$ prepared qubits to Bob, one at a time, over the quantum channel.

    \item \textit{Basis measurement:} For each received qubit, Bob independently and randomly selects a measurement basis (rectilinear or diagonal) and records the measurement outcome.

\end{enumerate}
    
\subsubsection{Classical Communication Phase}  
\begin{enumerate}[label=\alph*), start=1]
    \item \textit{Sifting:} Over an authenticated classical channel, Bob announces the basis used for each measurement. Alice compares these to her preparation basis and identifies the positions where the two agree.
    When Bob's measurement basis matches Alice's preparation basis, the original bit is recovered with certainty (in the absence of noise or eavesdropping). When the basis differs, the qubit collapses probabilistically into one of the two states of Bob's basis; for example, a horizontally polarized qubit $(\leftrightarrow)$ encoding bit 0 in the rectilinear basis, if measured in the diagonal basis, collapses randomly into either the $\vcenter{\hbox{\rotatebox{45}{$\leftrightarrow$}}}$ or $\vcenter{\hbox{\rotatebox{135}{$\leftrightarrow$}}}$ state.  Let $n_{\mathrm{sift}}$ denote the number of matching positions. The resulting sifted keys are denoted by $\mathbf{x}_A$ and $\mathbf{x}_B$, each of length $n_{\mathrm{sift}}$. Because the two bases are chosen independently and uniformly, approximately half of the $N_{\mathrm{tx}}$ transmitted bits are discarded at this stage, regardless of the underlying encoding scheme.
    
    \item \textit{Parameter Estimation:}
     After sifting, Alice and Bob share sifted keys $\mathbf{x}_A$ and $\mathbf{x}_B$. They randomly select a set $\mathcal{Y}$ of positions and publicly compare the corresponding bit values to estimate the quantum bit error rate (QBER). If the estimated QBER exceeds the security threshold, the protocol is aborted. Otherwise, they discard the disclosed bits and keep the remaining $n_{\mathrm{key}}=n_{\mathrm{sift}}-n_{\mathrm{test}}$ bits for information reconciliation, where $n_{\mathrm{test}}=|\mathcal{Y}|$.

    \item \textit{Error Reconciliation:} 
    Alice and Bob correct the discrepancies remaining in their $n_{\mathrm{key}}$ retained bits while minimizing the information leaked over the authenticated classical channel. Several reconciliation methods are used in practical QKD systems: Cascade and Winnow, which build on the interactive reconciliation framework of Brassard and Salvail~\cite{brassard1993secret}, LDPC-based reconciliation~\cite{elkouss2011information}, and polar-code-based reconciliation~\cite{renes2012polar}. All four error correction protocols and their algorithmic implementation are described in detail in Section ~\ref{sec:Cascade_Error_Reconciliation_Protocol}--\ref{sec:Polar_Code-Based_Reconciliation_Protocol_for_Quantum_Key_Distribution}. In all cases, the objective is to bring Alice's and Bob's key strings into exact agreement before privacy amplification is applied.

    \item \textit{Privacy Amplification:} 
    Even after successful reconciliation, an eavesdropper may hold residual information about the key, whether from channel interception or from the parity bits disclosed during the error reconciliation step. Alice and Bob remove this residual information by compressing the reconciled key $K_{\mathrm{rec}}$ into a shorter final key using a universal hash function~\cite{bennett1995privacy,renner2005universally}:
    \begin{equation}
    K_{\mathrm{final}} = H_{PA}(K_{\mathrm{rec}}), \label{eq:privacy_amp}
    \end{equation}
    where $H_{PA}(\cdot)$ is drawn from a universal$_2$ hash family, and commonly implemented via random Toeplitz matrices, which admit efficient linear-time hashing for large key blocks. The framework evaluated in this work uses a Toeplitz-SHAKE256 construction for this step, detailed in Section~\ref{sec:Privacy_Amplification}. The compression rate is determined by the information leaked during the reconciliation and parameter estimation phase. This ensures that any information retained by an eavesdropper about $K_{\mathrm{final}}$ becomes statistically negligible, yielding the final secret key shared between Alice and Bob.
\end{enumerate}



\subsection{BB84 QBER Analysis}
\label{subsec:qber_analysis}

In our system model three physically distinct effects contribute to QBER. The first is optical misalignment, captured by a baseline error $e_{\mathrm{mis}}$ that persists even under a noiseless channel. Since neither Alice's preparation optics nor Bob's measurement apparatus can be perfectly aligned with the ideal polarization basis. The second is background and detector dark counts, characterized by the probability $p_{\mathrm{bg}}$: with this probability, Bob's detector registers a click uncorrelated with the state Alice actually sent. Because such spurious detections carry no information about the encoded bit, they are equally likely to be read as correct or incorrect, contributing an error probability of $1/2$ whenever they occur. The third is photon loss during propagation through the optical-fibre channel and imperfect detector, governed by the total efficiency $\eta_{\mathrm{sys}}$ of Eq.~\eqref{eq:channel_trans}. These three quantities, $e_{\mathrm{mis}}$, $p_{\mathrm{bg}}$, and $\eta_{\mathrm{sys}}$, combine in a common phenomenological model for the overall gain $G_\mu$, the detection probability for signal intensity $\mu$, and the observed QBER $Q_\mu$ as
\begin{align}
G_\mu &= p_{\mathrm{bg}} + 1 - e^{-\eta_{\mathrm{sys}} \mu}, \label{eq:Gain}\\
G_\mu Q_\mu &= e_{\mathrm{mis}}\left(1 - e^{-\eta_{\mathrm{sys}} \mu}\right) + \frac{1}{2}p_{\mathrm{bg}}, \label{eq:qber_raw}
\end{align}
so that
\begin{equation}
Q_\mu = \frac{e_{\mathrm{mis}}\left(1 - e^{-\eta_{\mathrm{sys}} \mu}\right) + \frac{1}{2}p_{\mathrm{bg}}}{p_{\mathrm{bg}} + 1 - e^{-\eta_{\mathrm{sys}} \mu}}. \label{eq:Qber}
\end{equation}

For the ideal BB84 QKD protocol with one-way classical post-processing, secure key generation is possible only when this QBER remains below the asymptotic threshold of $11\%$.  If the QBER is above this value, the asymptotic SKR becomes zero~\cite{shor2000simple}.

\subsection{BB84 SKR Analysis}


\subsubsection{General SKR Expression}

For BB84 with WCPs in the asymptotic regime, the SKR per transmitted pulse is lower-bounded by \cite{gottesman2002security}
\begin{equation}
R_{\mathrm{sk}} \geq q \left[ G_1 \left(1 - H_2(Q_1)\right) - G_\mu f_{\mathrm{EC}}(Q_\mu) H_2(Q_\mu) \right], 
\label{eq:skr_general}
\end{equation}
where $q = {1}/{2}$ is the basis reconciliation factor, $G_\mu$ and $Q_\mu$ are the gain and QBER from Eqs.~\eqref{eq:Gain}--\eqref{eq:Qber}, $G_1$ and $Q_1$ are the single-photon gain and error rate, $f_{\mathrm{EC}}(Q_\mu)\geq1$ is the error-correction efficiency, and $H_2(\cdot)$ is the binary entropy function.

\subsubsection{Gain and Detection Model}

Since only single-photon pulses are immune to photon-number-splitting attacks, Eq.~\eqref{eq:skr_general} requires the single-photon gain rather than the total gain $G_\mu$:
\begin{equation}
G_1 = Y_1\, \mu\, e^{-\mu}, \label{eq:q1}
\end{equation}
where $Y_1$ is the single-photon yield and $\mu e^{-\mu}$ is the probability of a single-photon emission.

\subsubsection{Simplified SKR Expression}

In an ideal single-photon scenario with symmetric noise, the SKR reduces to
\begin{equation}
R_{\mathrm{sk}} \approx q \eta_{\mathrm{sys}} \left(1 - 2H_2(Q)\right), \label{eq:skr_asy}
\end{equation}
where $Q$ denotes the QBER, $q$ is a basis-sifting factor, and $\eta_{\mathrm{sys}}$ is the overall system efficiency.

\subsection{Finite Size SKR Analysis}
The SKR expression in Eq.~\eqref{eq:skr_asy} assumes an asymptotically large number of exchanged signals. In practice, only a finite block of $N_{\mathrm{sig}}$ signals is available, so the secret key length must instead be bounded using composable security against statistical fluctuations in parameter estimation.
\subsubsection{Finite-Key Length}
Following the smooth min-/max-entropy framework from \cite{Tomamichel2012}, for a block of $n_{\mathrm{key}}$ key-basis and $n_{\mathrm{test}}$ test-basis signals, the extractable secret key length satisfies
\begin{equation}
    L_{\mathrm{sec}}\leq n_{\mathrm{key}} \left[1 - 2H_2(Q+\nu)\right]-\mathrm{leak}_{\mathrm{EC}}-\log_2 \left(\frac{2}{\varepsilon_{\mathrm{corr}}\varepsilon_{\mathrm{sec}}^2}\right) \label{eq:skr_finite}
\end{equation}
where 
\[
\nu= \sqrt{\frac{(n_{\mathrm{key}}+n_{\mathrm{test}})(n_{\mathrm{test}}+1)}{(n_{\mathrm{key}}n_{\mathrm{test}}^2)} \ln{\frac{2}{\varepsilon_{\mathrm{sec}}}}}
\]
is the Serfling-derived correction factor accounting for finite sampling,
\[
\mathrm{leak}_{\mathrm{EC}}= f_{\mathrm{EC}} n_{\mathrm{key}} H_2(Q)
\]
is the information leakage due to the classical error correction with $f_{\mathrm{EC}}\approx1.1$ denoting the reconciliation parameter. Here, $\varepsilon_{\mathrm{corr}}$ is the allowed failure probability of the error-correction step, i.e., the probability that Alice and Bob do not obtain identical keys. $\varepsilon_{\mathrm{sec}}$ is the allowed failure probability of secrecy, i.e., the probability that the final key is not perfectly secret. The third term [ $-\log_2\!\left(\frac{2}{\varepsilon_{\mathrm{corr}}\varepsilon_{\mathrm{sec}}^2}\right)$ ] is the finite-key security penalty. It subtracts bits from the key length to guarantee correctness and secrecy at these levels.

\subsubsection{Finite-Size Key Rate}
The corresponding finite-size SKR per transmitted pulse is given by
\[
R_{\mathrm{finite}}=\frac{L_{\mathrm{sec}}}{N_{\mathrm{sig}}},\quad N_{\mathrm{sig}}=n_{\mathrm{key}}+n_{\mathrm{test}}
\]
which reduces to $R_{\mathrm{sk}}=1-2H_2(Q)$, the bracketed term in Eq.~\eqref{eq:skr_asy}, as $N_{\mathrm{sig}}\rightarrow\infty$, since $\nu\rightarrow 0$ and $\mathrm{leak}_{\mathrm{EC}}\rightarrow H_2(Q)$. At practical block sizes ($N_{\mathrm{sig}}\sim10^4-10^8$), $R_{\mathrm{finite}}$ falls well below this asymptotic value, so finite-key bounds such as Eq. \eqref{eq:skr_finite} are required to certify secure operation of realistic BB84 QKD systems rather than relying on the asymptotic bound in Eq.~\eqref{eq:skr_asy} alone.

\section{Error Reconciliation Protocols} \label{sec:Error_Reconciliation_Protocols}
In the error reconciliation protocol, mainly four protocols are discussed: Cascade, Winnow, LDPC, and polar codes.

\subsection{Cascade Protocol}
\label{sec:Cascade_Error_Reconciliation_Protocol}

The Cascade protocol is an interactive error reconciliation technique used in DV-QKD systems to correct mismatched bits between the sifted keys of Alice and Bob. The protocol operates through multiple reconciliation passes using parity comparisons, recursive binary search, deterministic shuffling, and backtracking operations to efficiently locate and correct transmission errors while minimizing information leakage over the authenticated classical channel \cite{choudhary2023cracking}. The protocol assumes that Alice and Bob possess correlated binary strings generated after the sifting stage. Due to channel imperfections and detector noise, the sifted keys may contain discrepancies quantified by the QBER. Cascade reconciliation aims to eliminate these discrepancies before privacy amplification.

\subsubsection{Protocol Parameters}

The protocol utilizes the following parameters:

\begin{itemize}
    \item $\mathbf{x}_B$ : Bob's initial sifted key sequence.
    \item $Q$ : Estimated QBER.
    \item $N_{\mathrm{pass}}$ : Total number of reconciliation passes.
    \item $\hat{\mathbf{x}}_B$ : Reconciled key estimate.
    \item $\mathcal{O}$ : Classical communication interface used for parity exchange with Alice.
\end{itemize}

\subsubsection{Initialization Phase}
Let the sifted key length be denoted by:
\begin{equation}
k = |\mathbf{x}_B|
\end{equation}
Bob first creates a duplicate copy of the sifted key:
\begin{equation}
\hat{\mathbf{x}}_B \leftarrow \mathbf{x}_B
\end{equation}
An index sequence containing all key positions is then initialized as
\begin{equation}
\mathcal{I} \leftarrow (0,1,\ldots,k-1)
\end{equation}
A storage structure $\mathcal{C}_{\mathrm{pc}}$ is maintained to preserve parity-correct blocks identified during each reconciliation pass, with $\mathcal{C}_{\mathrm{pc}}[p]$ storing the parity-correct blocks for pass $p$.

\subsubsection{Adaptive Block Size Selection}
During the first reconciliation pass, the block size is selected according to the estimated QBER as \cite{mehic2017implementation}

\begin{equation}
\kappa \leftarrow \left\lfloor \frac{0.73}{Q} \right\rfloor
\end{equation}
For subsequent passes, the block size is doubled.
\begin{equation}
\kappa \leftarrow 2\kappa
\end{equation}
The increasing block size improves large-scale error localization efficiency during later iterations.

\subsubsection{Deterministic Shuffling}
Beginning from the second pass onward, the key indices are deterministically shuffled using a predefined seed value.
\begin{equation}
\mathcal{I} \leftarrow \pi_p(\mathcal{I})
\end{equation}
The purpose of deterministic shuffling is to redistribute clustered errors into different blocks across successive passes.

\subsubsection{Block Partitioning}
The shuffled indices are partitioned into contiguous blocks of length $\kappa$:
\begin{equation}
\mathcal{B} \leftarrow (\mathcal{B}_1,\ldots,\mathcal{B}_{N_{\mathrm{blk}}})
\end{equation}
where 
\begin{equation}
\mathcal{B}_\ell = \mathcal{I}_{((\ell-1)\kappa+1):\min(\ell \kappa,k)}, \quad N_{\mathrm{blk}} = \lceil k/\kappa \rceil
\end{equation}

\subsubsection{Parity Comparison}
Let $\mathbf{x}_A=(x_{A,0},\dots,x_{A,k-1})$ denote Alice's sifted key, and let the parity oracle be:
\begin{equation}
\mathcal{O}(S) = \bigoplus_{j\in S} x_{A,j}
\quad\text{for any } S\subseteq\{0,\dots,k-1\}.
\end{equation}
For every block $\mathcal{B}_\ell$, Alice computes the parity value
\begin{equation}
a_\ell = \mathcal{O}(\mathcal{B}_\ell) = \bigoplus_{j\in \mathcal{B}_\ell} x_{A,j},
\end{equation}
where $\oplus$ denotes modulo-2 addition.
Similarly, Bob computes the corresponding local parity:
\begin{equation}
b_\ell = \bigoplus_{j\in \mathcal{B}_\ell}\hat{x}_{B,j}.
\end{equation}

\subsubsection{Error Detection}
The parity values are compared over the authenticated classical channel. Let $\mathcal{E}$ denote the set of block indices with mismatched parity.
\begin{itemize}
    \item If $a_\ell=b_\ell$, the block is considered parity-correct.
    \item If $a_\ell\neq b_\ell$, the block contains an odd number of errors and is marked as erroneous; include the $\ell$-th block in $\mathcal{E}$.
\end{itemize}

\subsubsection{Binary Search Error Localization}
For every $\ell\in\mathcal{E}$, the Cascade protocol calls
$\textsc{BinarySearch}(\mathcal{B}_\ell,p,\hat{\mathbf{x}}_B,\mathcal{O},\mathcal{C}_{\mathrm{pc}})$
to identify the faulty bit. The block is divided into two halves:
\begin{equation}
\mathcal{B}_\ell^{L} \leftarrow \mathcal{B}_\ell[{\rm lo}:mid],\qquad \mathcal{B}_\ell^{R} \leftarrow \mathcal{B}_\ell[mid+1:hi].
\end{equation}
Parity comparison is then performed only on the left half:
\begin{equation}
\alpha = \mathcal{O}(\mathcal{B}_\ell^{L}),\qquad
\beta = \bigoplus_{i\in \mathcal{B}_\ell^{L}}\hat{x}_{B,i}.
\end{equation}
\begin{itemize}
    \item If $\alpha\neq\beta$, the error exists in the left half; set $hi\leftarrow mid$.
    \item Otherwise, the error exists in the right half.
\end{itemize}
The recursive subdivision process continues until a single erroneous bit location is identified.

\subsubsection{Bit Correction}
After locating the erroneous index, the incorrect bit is inverted according to:
\begin{equation}
\hat{x}_{B,j^*} \leftarrow 1-\hat{x}_{B,j^*}.
\end{equation}
This operation corrects the mismatch between Alice and Bob.

\subsubsection{Cascade Backtracking Mechanism}
A correction performed during a later reconciliation pass may alter the parity status of blocks previously considered correct. Therefore, Cascade introduces a backtracking mechanism to revisit affected blocks. Previously corrected blocks containing the newly flipped bit are identified and re-evaluated. If the number of flipped bits inside a block becomes odd, the block is reclassified as erroneous, and binary search is again applied to localize additional hidden errors. This cascading correction process continues iteratively until no further inconsistencies remain.

The Cascade protocol presented in Algorithm \ref{alg:cascade} remains one of the most widely implemented reconciliation techniques in practical DV-QKD systems, due to its high correction accuracy and robustness against transmission noise. By combining parity verification (Algorithm \ref{alg:biconf}), recursive binary search, deterministic shuffling, and iterative backtracking, the protocol efficiently reconciles noisy quantum keys while preserving communication security.

\makeatletter
\newenvironment{breakablealgorithm}
  {\begin{center}
     \refstepcounter{algorithm}%
     \hrule height.8pt depth0pt \kern2pt
     \renewcommand{\caption}[2][\relax]{%
      {\raggedright\textbf{\ALG@name~\thealgorithm} ##2\par}%
      \ifx\relax##1\relax
         \addcontentsline{loa}{algorithm}{\protect\numberline{\thealgorithm}##2}%
      \else
         \addcontentsline{loa}{algorithm}{\protect\numberline{\thealgorithm}##1}%
      \fi
      \kern2pt\hrule 
     }
  }
  {\kern2pt\hrule\relax
  \end{center}}
\makeatother

\begin{breakablealgorithm}
\caption{Cascade Error Correction Protocol}
\label{alg:cascade}
\begin{algorithmic}[1]
\footnotesize

\Require
  $\mathbf{x}_B \in \{0,1\}^k$: Bob's raw key bit-string of length $k$ \\
  $Q \in (0,1)$: Estimated quantum bit error rate (QBER) \\
  $N_{\mathrm{pass}} \in \mathbb{Z}_{>0}$: Total number of reconciliation passes \\
  $\mathcal{O}(S) \triangleq \bigoplus_{i \in S} x_{A,i}$ for any $S \subseteq \{0,\dots,k-1\}$: Alice parity oracle where $\mathbf{x}_A \in \{0,1\}^k$ is Alice's key \\
  $\pi_p: \{0,\dots,k-1\} \to \{0,\dots,k-1\}$: permutation used in pass $p$
 
\Ensure
  $\hat{\mathbf{x}}_B \in \{0,1\}^k$: Bob's corrected key estimate
 
\State $\hat{\mathbf{x}}_B \leftarrow \mathbf{x}_B$
\State $\mathcal{I} \leftarrow (0, 1, \dots, k-1)$ \Comment{ordered list of all indices}
\State $\mathcal{C}_{\mathrm{pc}} \leftarrow \{\}$ \Comment{dictionary of parity-correct blocks}
 
\For{$p \leftarrow 1$ \textbf{to} $N_{\mathrm{pass}}$}
 
    \State $\mathcal{C}_{\mathrm{pc}}[p] \leftarrow [\,]$
 
    \If{$p = 1$}
        \State $\kappa \leftarrow \left\lfloor 0.73\,/\,Q \right\rfloor$
    \Else
        \State $\kappa \leftarrow 2\kappa$
    \EndIf
 
    \If{$p > 1$}
        \State $\mathcal{I} \leftarrow \pi_p(\mathcal{I})$
    \EndIf
 
    \State $N_{\mathrm{blk}} \leftarrow \lceil k / \kappa \rceil$
 
    \State $\mathcal{B} \leftarrow (\mathcal{B}_1, \ldots, \mathcal{B}_{N_{\mathrm{blk}}})$
    \State \hspace{1em} where $\mathcal{B}_\ell = (\mathcal{I}_{(\ell-1)\kappa}, \mathcal{I}_{(\ell-1)\kappa+1}, \dots, \mathcal{I}_{\min(\ell \kappa, k)-1})$ 
    \State \hspace{2em} for $\ell=1,\dots,N_{\mathrm{blk}}$
   
    \State $\mathbf{a} \leftarrow \left( a_\ell \right)_{\ell=1}^{N_{\mathrm{blk}}}$
    \State \hspace{1em} i.e., $a_\ell = \mathcal{O}(\mathcal{B}_\ell) = \bigoplus_{i \in \mathcal{B}_\ell} x_{A,i}$ for $\ell=1,\dots,N_{\mathrm{blk}}$
   
    \State $\mathbf{b} \leftarrow \left( b_\ell \right)_{\ell=1}^{N_{\mathrm{blk}}}$
    \State \hspace{1em} i.e., $b_\ell = \bigoplus_{i \in \mathcal{B}_\ell} \hat{x}_{B,i}$ for $\ell=1,\dots,N_{\mathrm{blk}}$
 
    \State $\mathcal{E} \leftarrow \{\ell \in [N_{\mathrm{blk}}] : a_\ell \neq b_\ell\}$
   
    \State $\mathcal{C}_{\mathrm{pc}}[p] \leftarrow \mathcal{C}_{\mathrm{pc}}[p] \cup \{\mathcal{B}_\ell : \ell \notin \mathcal{E}\}$
 
    \State $\mathcal{J}_{\mathrm{flip}} \leftarrow [\,]$
 
    \ForAll{$\ell \in \mathcal{E}$}
        \State $j^* \leftarrow \textsc{BinarySearch}(\mathcal{B}_\ell,\, p,\, \hat{\mathbf{x}}_B,\, \mathcal{O},\, \mathcal{C}_{\mathrm{pc}})$
        \State $\hat{x}_{B,j^*} \leftarrow 1 - \hat{x}_{B,j^*}$
        \State Append $j^*$ to $\mathcal{J}_{\mathrm{flip}}$
    \EndFor
 
    \If{$p > 1$ \textbf{ and } $|\mathcal{J}_{\mathrm{flip}}| > 0$}
 
        \State $\mathcal{Q}_{\mathrm{bt}} \leftarrow [\,]$
 
        \For{$p' \leftarrow p-1$ \textbf{ downto } $1$}
            \ForAll{$\mathcal{S} \in \mathcal{C}_{\mathrm{pc}}[p']$}
                \If{$\mathcal{S} \cap \mathcal{J}_{\mathrm{flip}} \neq \emptyset$ \textbf{ and } $\mathcal{S} \notin \mathcal{Q}_{\mathrm{bt}}$}
                    \State Append $\mathcal{S}$ to $\mathcal{Q}_{\mathrm{bt}}$
                \EndIf
            \EndFor
        \EndFor
 
        \While{$\mathcal{Q}_{\mathrm{bt}} \neq \emptyset$}
 
            \State $\mathcal{S}^* \leftarrow \arg\min_{\mathcal{S} \in \mathcal{Q}_{\mathrm{bt}}} |\mathcal{S}|$
            \State $\mathcal{Q}_{\mathrm{bt}} \leftarrow \mathcal{Q}_{\mathrm{bt}} \setminus \{\mathcal{S}^*\}$
            \State $p^* \leftarrow$ pass index such that $\mathcal{S}^* \in \mathcal{C}_{\mathrm{pc}}[p^*]$
 
            \If{$|\mathcal{S}^* \cap \mathcal{J}_{\mathrm{flip}}|$ is odd}
 
                \State $\mathcal{C}_{\mathrm{pc}}[p^*] \leftarrow \mathcal{C}_{\mathrm{pc}}[p^*] \setminus \{\mathcal{S}^*\}$
 
                \State $j^* \leftarrow \textsc{BinarySearch}(\mathcal{S}^*,\, p^*,\, \hat{\mathbf{x}}_B,\, \mathcal{O},\, \mathcal{C}_{\mathrm{pc}})$
                \State $\hat{x}_{B,j^*} \leftarrow 1 - \hat{x}_{B,j^*}$
                \State Append $j^*$ to $\mathcal{J}_{\mathrm{flip}}$
 
                \For{$p' \leftarrow p$ \textbf{ downto } $1$}
                    \If{$p' \neq p^*$}
                        \ForAll{$\mathcal{S} \in \mathcal{C}_{\mathrm{pc}}[p']$}
                            \If{$j^* \in \mathcal{S}$ \textbf{ and } $\mathcal{S} \notin \mathcal{Q}_{\mathrm{bt}}$}
                                \State Append $\mathcal{S}$ to $\mathcal{Q}_{\mathrm{bt}}$
                            \EndIf
                        \EndFor
                    \EndIf
                \EndFor
 
            \EndIf
 
        \EndWhile
 
    \EndIf
 
\EndFor
 
\State \Return $\hat{\mathbf{x}}_B$

\end{algorithmic}
\end{breakablealgorithm}

\begin{breakablealgorithm}
\caption{Binary Search Subroutine (BICONF)}
\label{alg:biconf}
\begin{algorithmic}[1]
\footnotesize

\Require
  $\mathcal{B} = (j_0, j_1, \dots, j_{r-1})$: Ordered index block with $|\mathcal{B}| = r$, known to contain an odd number of errors \\
  $p$: Pass number under which $\mathcal{B}$ is being resolved \\
  $\hat{\mathbf{x}}_B \in \{0,1\}^k$: Bob's current key estimate \\
  $\mathcal{O}(S) \triangleq \bigoplus_{i \in S} x_{A,i}$ for any $S \subseteq \{0,\dots,k-1\}$: Alice parity oracle where $\mathbf{x}_A \in \{0,1\}^k$ is Alice's key \\
  $\mathcal{C}_{\mathrm{pc}}$: Dictionary of parity-correct blocks (mutated in place)
 
\Ensure
  $j^* \in \{0,\dots,k-1\}$: Index of a single erroneous bit
 
\State ${\rm lo} \leftarrow 0$,\quad ${\rm hi} \leftarrow r - 1$
 
\While{$lo < hi$}
 
    \State $mid \leftarrow \lfloor (lo + hi) / 2 \rfloor$
 
    \State $\mathcal{B}^L \leftarrow (j_{lo}, j_{lo+1}, \dots, j_{mid-1})$
    \State $\mathcal{B}^R \leftarrow (j_{mid+1}, j_{mid+2}, \dots, j_{hi-1})$
 
    \State $\alpha \leftarrow \mathcal{O}(\mathcal{B}^L) = \bigoplus_{i \in \mathcal{B}^L} x_{A,i}$
 
    \State $\beta \leftarrow \bigoplus_{i \in \mathcal{B}^L} \hat{x}_{B,i}$
 
    \If{$\alpha \neq \beta$}
        
        \State $hi \leftarrow mid$
        \State Append $\mathcal{B}^R$ to $\mathcal{C}_{\mathrm{pc}}[p]$
        
    \Else
        
        \State $lo \leftarrow mid + 1$
        \State Append $\mathcal{B}^L$ to $\mathcal{C}_{\mathrm{pc}}[p]$
       
    \EndIf
 
\EndWhile
 
\State \Return $j_{lo}$

\end{algorithmic}
\end{breakablealgorithm}

\subsection{Winnow Protocol}
\label{sec:Winnow_Error_Reconciliation_Protocol}
The Winnow protocol is an efficient error reconciliation technique used in QKD systems to correct mismatches between the sifted keys of Alice and Bob while minimizing information leakage over the authenticated classical communication channel. Unlike fully interactive reconciliation methods, Winnow combines parity checking with Hamming syndrome decoding to reduce communication overhead and improve reconciliation efficiency.

In DV-QKD systems, the sifted keys generated after the quantum transmission stage may contain discrepancies due to channel noise, detector imperfections, and environmental disturbances. Winnow reconciles these discrepancies through block partitioning, parity exchange, syndrome calculation, and iterative error correction, resulting in identical secret keys.

\subsubsection{Protocol Parameters}

The Winnow protocol utilizes the following parameters:

\begin{itemize}
    \item $\mathbf{x}_B \in \{0,1\}^k$ : Bob's initial sifted key sequence of length $k$.
    \item $Q \in (0,1)$ : Estimated quantum bit error rate (QBER).
    \item $N_{\mathrm{pass}} \in \mathbb{Z}_{>0}$ : Total number of reconciliation passes.
    \item $\hat{\mathbf{x}}_B \in \{0,1\}^k$ : Reconciled key estimate.
    \item $\mathcal{O}(S) = \bigoplus_{i \in S} x_{A,i}$ : Classical parity oracle for Alice's key $\mathbf{x}_A \in \{0,1\}^k$, where $S \subseteq \{0,\dots,k-1\}$.
\end{itemize}

\subsubsection{Initialization Phase}

Let the key length be denoted by:
\begin{equation}
k = |\mathbf{x}_B|
\end{equation}

Bob initializes the integer representation of his key as:
\begin{equation}
\sigma \leftarrow \sum_{i=0}^{k-1} x_{B,i} \cdot 2^{k-1-i}
\end{equation}

The index permutation is initialized as:
\begin{equation}
\pi \leftarrow (0,1,\dots,k-1)
\end{equation}

\subsubsection{Adaptive Block Formation}
During the first reconciliation pass, the block size is selected according to the estimated QBER:
\begin{equation}
\kappa \leftarrow 2^{\lfloor \log_2(\lfloor 0.5/Q \rfloor) \rfloor},
\qquad
\kappa \leftarrow \max(4,\kappa)
\end{equation}
This selection reduces the probability of multiple errors occurring within the same block. For subsequent reconciliation passes, the block size is doubled:
\begin{equation}
\kappa \leftarrow 2\kappa
\end{equation}

\subsubsection{Deterministic Shuffling}

For pass $p>1$, the key indices are deterministically shuffled using a predefined seed value:
\begin{equation}
\pi \leftarrow \mathrm{Shuffle}(\pi,p)
\end{equation}
This redistribution spreads clustered errors across different blocks during successive reconciliation passes.

\subsubsection{Block Partitioning}

The shuffled key indices are partitioned into contiguous blocks of size $\kappa$:
\begin{equation}
N_{\mathrm{blk}} \leftarrow \lceil k/\kappa \rceil,
\qquad
\mathcal{B}_\ell \leftarrow \pi[(\ell-1)\kappa : \min(\ell\kappa,k)]
\end{equation}
for $\ell = 1,\dots,N_{\mathrm{blk}}$.
Each block is independently processed during reconciliation.

\subsubsection{Parity Exchange}

For every block $\mathcal{B}_\ell$, Alice computes the parity value:
\begin{equation}
a_\ell \leftarrow \mathcal{O}(\mathcal{B}_\ell)
= \bigoplus_{i\in\mathcal{B}_\ell} x_{A,i}
\end{equation}
Similarly, Bob calculates the local parity:
\begin{equation}
b_\ell \leftarrow \bigoplus_{i\in\mathcal{B}_\ell} x_{B,i}
\end{equation}
where $\oplus$ denotes modulo-2 addition. If $a_\ell \neq b_\ell$, the block is marked as erroneous and added to the list of bad blocks:
\begin{equation}
\mathcal{E} \leftarrow \{\ell \in [N_{\mathrm{blk}}] : a_\ell \neq b_\ell\}.
\end{equation}

\subsubsection{Hamming Syndrome Construction}

For every erroneous block, the protocol computes Hamming syndrome subsets to localize the erroneous bit. Let the block length be:
\begin{equation}
\kappa_\ell \leftarrow |\mathcal{B}_\ell|
\end{equation}
The number of syndrome bits required is:
\begin{equation}
n_{\mathrm{syn}} \leftarrow \left\lceil \log_2(\kappa_\ell+1) \right\rceil
\end{equation}
For each syndrome bit position $j=0,\dots,n_{\mathrm{syn}}-1$, the subset is selected as:
\begin{equation}
\mathcal{T}_{\ell,j}
\leftarrow
\left\{
\mathcal{B}_\ell[\nu] :
\nu \in [\kappa_\ell],\;
\left\lfloor(\nu+1)/2^j\right\rfloor \text{ is odd}
\right\}.
\end{equation}
Parity values for these subsets are exchanged between Alice and Bob:
\begin{equation}
\alpha_{\ell,j} \leftarrow \mathcal{O}(\mathcal{T}_{\ell,j}),
\qquad
\beta_{\ell,j} \leftarrow \bigoplus_{i \in \mathcal{T}_{\ell,j}} x_{B,i}.
\end{equation}

\subsubsection{Syndrome Difference Reconstruction}

The syndrome difference value is reconstructed by comparing Alice's and Bob's syndrome parities. The syndrome difference is reconstructed as:
\begin{equation}
s_\ell
\leftarrow
\sum_{j=0}^{n_{\mathrm{syn}}-1}
(\alpha_{\ell,j} \oplus \beta_{\ell,j})\cdot 2^j.
\end{equation}
The final syndrome difference $s_\ell$ identifies the error position within the block.

\subsubsection{Error Correction}

The erroneous bit position is determined as:
\begin{equation}
i_{\mathrm{err}} \leftarrow s_\ell - 1
\end{equation}
When the calculated position falls within the valid block boundaries, i.e. $i_{\mathrm{err}} < \kappa_\ell$, the corresponding bit in Bob's integer key representation is flipped as follows
\begin{equation}
\sigma \leftarrow \sigma \oplus 2^{k-1-\mathcal{B}_\ell[i_{\mathrm{err}}]}.
\end{equation}
This bit-flip operation removes the identified discrepancy between Alice's and Bob's keys.

Winnow performs error reconciliation in DV-QKD by using parity checks together with Hamming syndrome decoding presented in Algorithm \ref{alg:winnow}. Compared with highly interactive reconciliation methods, it can reduce communication overhead while retaining reliable error correction. The use of adaptive block sizes, deterministic shuffling, and syndrome-based error localization makes Winnow suitable for secure key reconciliation in practical quantum communication systems.

\begin{breakablealgorithm}
\caption{Winnow Error Correction Protocol}
\label{alg:winnow}
\begin{algorithmic}[1]
\footnotesize

\Require $\mathbf{x}_B \in \{0,1\}^k$,\; $Q \in (0,1)$,\; oracle $\mathcal{O}(S)=\bigoplus_{i\in S} x_{A,i}$ for Alice key $\mathbf{x}_A\in\{0,1\}^k$,\; passes $N_{\mathrm{pass}} \in \mathbb{Z}^+$
\Ensure $\hat{\mathbf{x}}_B \in \{0,1\}^k$

\State $\sigma \leftarrow \displaystyle\sum_{i=0}^{k-1} x_{B,i} \cdot 2^{k-1-i}$ \quad (integer representation of $\mathbf{x}_B$)

\State $\pi \leftarrow (0, 1, \dots, k-1)$ \quad (identity index permutation)

\State $\kappa \leftarrow 2^{\lfloor \log_2(\lfloor 0.5/Q \rfloor) \rfloor}$,\quad $\kappa \leftarrow \max(4,\, \kappa)$

\For{$p = 1, \dots, N_{\mathrm{pass}}$}

    \If{$p > 1$}
        \State $\pi \leftarrow \mathrm{Shuffle}(\pi,\, p)$ 
        \State $\kappa \leftarrow 2\kappa$
    \EndIf

    \State $N_{\mathrm{blk}} \leftarrow \lceil k / \kappa \rceil$

    \For{$\ell = 1, \dots, N_{\mathrm{blk}}$}
        \State $\mathcal{B}_\ell \leftarrow \pi[(\ell-1)\kappa \;:\; \min(\ell\kappa,\, k)]$
    \EndFor

    \For{$\ell = 1, \dots, N_{\mathrm{blk}}$}
        \State $a_\ell \leftarrow \mathcal{O}(\mathcal{B}_\ell) = \bigoplus_{i \in \mathcal{B}_\ell} x_{A,i}$;\quad $b_\ell \leftarrow \bigoplus_{i \in \mathcal{B}_\ell} x_{B,i}$
    \EndFor

    \State $\mathcal{E} \leftarrow \{\ell \in [N_{\mathrm{blk}}] : a_\ell \neq b_\ell\}$ \quad (discordant blocks)

    \For{$\ell \in \mathcal{E}$}

        \State $\kappa_\ell \leftarrow |\mathcal{B}_\ell|$;\quad $n_{\mathrm{syn}} \leftarrow \lceil \log_2(\kappa_\ell+1) \rceil$

        \For{$j = 0, \dots, n_{\mathrm{syn}}-1$}
            \State $\mathcal{T}_{\ell,j} \leftarrow \left\{\mathcal{B}_\ell[\,\nu\,] : \nu \in [\kappa_\ell],\; \left\lfloor(\nu+1) / 2^j\right\rfloor \text{ is odd}\right\}$
            \State $\alpha_{\ell,j} \leftarrow \mathcal{O}(\mathcal{T}_{\ell,j})$;\quad $\beta_{\ell,j} \leftarrow \bigoplus_{i \in \mathcal{T}_{\ell,j}} x_{B,i}$
        \EndFor

        \State $s_\ell \leftarrow \displaystyle\sum_{j=0}^{n_{\mathrm{syn}}-1} (\alpha_{\ell,j} \oplus \beta_{\ell,j})\cdot 2^j$ \quad (syndrome difference)

        \If{$s_\ell > 0$}
            \State $i_{\mathrm{err}} \leftarrow s_\ell - 1$ \quad (error position within block)
            \If{$i_{\mathrm{err}} < \kappa_\ell$}
                \State $\sigma \leftarrow \sigma \oplus 2^{k - 1 - \mathcal{B}_\ell[i_{\mathrm{err}}]}$ \quad (flip bit $\mathcal{B}_\ell[i_{\mathrm{err}}]$ in $\sigma$)
            \EndIf
        \EndIf

    \EndFor

\EndFor

\State $\hat{\mathbf{x}}_B \leftarrow$ binary representation of $\sigma$ with leading zeros to length $k$

\State \Return $\hat{\mathbf{x}}_B$

\end{algorithmic}
\end{breakablealgorithm}

\subsection{LDPC Codes}
\label{sec:5G_NR_LDPC-Based_Reconciliation_Protocol}

Low-Density Parity-Check (LDPC) codes \cite{Gallager_ldpc} are widely recognized as an efficient and the most efficient forward error correction technique for QKD reconciliation due to their near-capacity performance, reduced information leakage, and computationally efficient decoding. In practical DV-QKD systems, LDPC-based syndrome reconciliation enables reliable correction of mismatched sifted keys while minimizing the amount of information exchanged over the authenticated classical channel \cite{elkouss2011information}. In this work, we present a rate-adaptive syndrome reconciliation framework based on 5G New Radio (NR) quasi-cyclic LDPC (QC-LDPC) codes \cite{5GNR}. The framework uses the quasi-cyclic base graph structures, namely Base Graph 1 and Base Graph 2, specified in the 3GPP TS 38.212 standard, along with the protograph lifting procedure. These structures are implemented using the NVIDIA Sionna library \cite{sionna}. To prevent zero-degree column starvation, the lifted 3GPP check equations are folded modulo the segment length and ordered via greedy column-coverage balancing combined with progressive edge-growth (PEG) conditioning \cite{Hu_PEG_2001}. Decoding is performed using damped normalized min-sum syndrome belief propagation \cite{Chen_Fossorier_2002,Pretti_2005} augmented by full-rank reliability-ordered GF(2) ordered statistics decoding of order zero (OSD-0) \cite{Fossorier_Lin_1995} and iterative variable-node pinning, verified by a universal hash \cite{CARTER1979143,Krawczyk-Hugo}. The steps below are organized as follows. First, we define the key parameters and the initialization process. Next, we detail the construction of the rate-adaptive parity-check matrix. Finally, we describe the complete decoding pipeline, including the belief propagation, post-processing, and verification steps.

\subsubsection{Protocol Parameters}
The reconciliation procedure operates on the following quantities:
\begin{itemize}
    \item $\mathbf{x}_B\in\{0,1\}^k$: Bob's noisy sifted key of length $k$.
    \item $\mathbf{x}_A\in\{0,1\}^k$: Alice's sifted key.
    \item $Q\in(0,1)$: estimated quantum bit error rate (QBER).
    \item $\mathcal{O}(S)=\bigoplus_{i\in S}x_{A,i}$: parity oracle for any subset $S\subseteq\{0,\dots,k-1\}$.
    \item $t_{\mathrm{hash}}\in\mathbb{Z}^+$: universal-hash output length.
    \item $\gamma>1$: coarse parity-budget sweep factor.
    \item $\gamma_{\mathrm{fine}}\in(1,\gamma)$: fine parity-budget sweep factor.
    \item $\tau_{\mathrm{syn}}\in\mathbb{Z}^+$: syndrome mismatch threshold for fine rate-adaptive sweeping.
    \item $P_{\mathrm{dec}}\in\mathbb{Z}^+$: maximum number of decoding passes per segment.
    \item $I_{\mathrm{BP}}\in\mathbb{Z}^+$: maximum number of belief-propagation (BP) iterations per decoding pass.
    \item $\kappa_{\mathrm{seg}}\in\mathbb{Z}^+$: maximum segment length.
    \item $\Lambda_{\mathrm{sat}}>0$: log-likelihood-ratio (LLR) saturation value
    \item $\eta_{\mathrm{SW}}>1, \beta_{\mathrm{disp}}>0$: Slepian--Wolf entropy scaling factor and finite-length dispersion scaling factor
    \item $\alpha_{\mathrm{BP}}, \beta_{\mathrm{BP}}\in(0,1)$: normalized min-sum check-node scaling factor and message damping factor
    \item $\omega_{\mathrm{EMA}}\in(0,1)$: exponential moving average (EMA) weight for posterior LLR smoothing
    \item $\eta_{\mathrm{OSD}}>0$: violated-check penalty weight in OSD-0 column ordering
    \item $P_{\mathrm{pin}}\in\mathbb{Z}^+$: maximum number of blind variable-node pinning rounds
    \item $\rho_{\mathrm{pin}}\in(0,1)$: fraction of variable nodes revealed per pinning round
    \item $\eta_{\mathrm{pin}}>0$: violated-check penalty weight in variable-node pinning selection
    \item $s_0\in\mathbb{N}$: pseudo-random generator seed
    \item $\hat{\mathbf{x}}_B\in\{0,1\}^k$: reconciled key estimate
    \item $\Lambda_{\mathrm{leak}}\in\mathbb{N}$: total number of bits revealed through the oracle
    \item $f_{\mathrm{EC}}\in\mathbb{R}^+$: reconciliation efficiency
    \item $\mathcal{F}\subseteq\{1,\dots,M_{\mathrm{seg}}\}$: set of failed segments
\end{itemize}
The binary Shannon entropy and the binary symmetric channel dispersion \cite{Polyanskiy-Poor-Verdu} are defined as 
\begin{equation}
H_{2}(Q)=-Q\log_2Q-(1-Q)\log_2(1-Q),
\end{equation}
\begin{equation}
V_Q=Q(1-Q)\log_2^2\!\left(\frac{1-Q}{Q}\right),
\end{equation}
and the intrinsic channel log-likelihood ratio (LLR) is given by \begin{equation}
L_{\mathrm{ch}}=\ln\!\left(\frac{1-Q}{Q}\right).
\end{equation}

\subsubsection{Initialization and Segment Partitioning}
Let $k=|\mathbf{x}_B|$ A pseudo-random permutation $\pi$ on $\{0,\dots,k-1\}$ is generated from the seed $s_0$ The shuffled key is
\begin{equation}
y_i=x_{B,\pi(i)},\qquad i=0,\dots,k-1.
\end{equation}
The shuffled key is partitioned into $M_{\mathrm{seg}}=k/\kappa_{\mathrm{seg}}$ segments. For $c=1,\dots,M_{\mathrm{seg}}$, the length of segment $c$ is
\begin{equation}
k_c=\left\lfloor\frac{k}{M_{\mathrm{seg}}}\right\rfloor+
\mathbf{1}\!\left(c\le k\bmod M_{\mathrm{seg}}\right),
\end{equation}
where $\mathbf{1}(\cdot)$ denotes the indicator function. The segment offsets are defined recursively by
\begin{equation}
o_1=0,\qquad o_{c+1}=o_c+k_c,
\end{equation}
so that the $c$-th segment is given by
\begin{equation}
\mathbf{y}_c=(y_{o_c},\dots,y_{o_{c+1}-1})\in\{0,1\}^{k_c}.
\end{equation}
The permutation $\pi$ serves two primary purposes. First, it disperses burst errors that may arise from correlated fluctuations in the quantum channel across multiple segments, ensuring that the LDPC decoder encounters approximately independent bit errors within each segment. Second, it randomizes the variable-node positions relative to the check-node structure across segments. Partitioning the sifted key into segments of bounded length $\kappa_{\mathrm{seg}}$ keeps the computational cost of belief-propagation message passing and Gauss-Jordan elimination \cite{press2007numerical} tractable while covering all $k$ bits

\subsubsection{Parity-Check Matrix Construction and Coverage Balancing}
For each segment $c$ of length $k_c$, the underlying 3GPP TS 38.212 mother code length $n_c$ is selected according to the standard base-graph threshold \cite{3gpp38212} :
\begin{equation}
n_c=
\begin{cases}
5k_c, & \text{if } k_c\le 3840 \quad\text{(3GPP Base Graph 2, rate } 1/5\text{)},\\[4pt]
3k_c, & \text{if } k_c> 3840 \quad\text{(3GPP Base Graph 1, rate } 1/3\text{)}.
\end{cases}
\end{equation}
This specification is passed to the 5G NR LDPC encoder $E_c=\mathrm{LDPC5GEncoder}(k_c,n_c)$ from the NVIDIA Sionna library \cite{sionna}, from which the lifted quasi-cyclic parity-check matrix $\mathbf{H}_{\mathrm{raw}}\in\{0,1\}^{m_{\mathrm{raw}}\times n_{\mathrm{raw}}}$ is extracted Because rate-adaptive syndrome reconciliation operates directly on the $k_c$ systematic sifted-key bits rather than transmitting $n_c$ coded symbols, the lifted circulant shifts of $\mathbf{H}_{\mathrm{raw}}$ are mapped onto the domain $\{0,\dots,k_c-1\}$ by folding $B_{\mathrm{fold}}$ contiguous blocks of width $k_c$ over $\mathbb{F}_2$ :
\begin{equation}
\mathbf{H}_{\mathrm{fold}}
=
\bigoplus_{b=0}^{B_{\mathrm{fold}}-1}
\mathbf{H}_{\mathrm{raw}}[:,\,b k_c:(b+1)k_c-1]
\in\{0,1\}^{m_{\mathrm{raw}}\times k_c}.
\end{equation}
From the rows of $\mathbf{H}_{\mathrm{fold}}$, a candidate pool of check rows with Hamming weights between $w_{\min}$ and $w_{\max}$ is formed :
\begin{equation}
\mathcal{R}_{\mathrm{pool}}
=
\left\{
\mathbf{h}\in\mathrm{Rows}(\mathbf{H}_{\mathrm{fold}}):
w_{\min}\le \sum_{i=0}^{k_c-1}h_i\le w_{\max}
\right\}.
\end{equation}
If one were to slice rows sequentially from a punctured 5G NR base graph at small syndrome sizes $m_c\ll k_c$, several variable nodes would remain unconnected (degree zero) in the active subgraph, causing belief propagation to stall at low QBER. To guarantee uniform variable-node coverage across all rate-adaptive operating points, a square parity-check matrix $\mathbf{H}_c\in\{0,1\}^{k_c\times k_c}$ is assembled row by row ($r=0,\dots,k_c-1$) while tracking the cumulative column-degree vector $\mathbf{d}_{\mathrm{col}}\in\mathbb{N}^{k_c}$. Whenever $\mathcal{R}_{\mathrm{pool}}$ is non-empty and all columns satisfy $d_{\mathrm{col},i}\ge w_{\min}$, the next check row $\mathbf{h}^*$ is selected greedily to maximize inverse-degree column coverage while adhering to a target row-degree schedule $d_{\mathrm{tgt}}(r)$ :
\small
\begin{equation}
\mathbf{h}^*
=
\arg\max_{\mathbf{h}\in\mathcal{R}_{\mathrm{pool}}}
\left(
\sum_{i=0}^{k_c-1}\frac{h_i}{1+d_{\mathrm{col},i}}
-\eta_{\mathrm{cov}}\left|\sum_{i=0}^{k_c-1}h_i-d_{\mathrm{tgt}}(r)\right|
\right),
\end{equation}
\normalsize
where $\eta_{\mathrm{cov}}>0$ penalizes deviations from $d_{\mathrm{tgt}}(r)$. If low-degree columns require immediate balancing ($d_{\mathrm{col},i}<w_{\min}$) or $\mathcal{R}_{\mathrm{pool}}$ is exhausted, a row of weight $d_{\mathrm{tgt}}(r)$ is synthesized using a progressive edge-growth (PEG) rule \cite{Hu_PEG_2001} : columns are scanned in ascending order of $d_{\mathrm{col},i}$ (with ties broken pseudorandomly via $s_0$) and added to the active support $\mathcal{C}_{\mathrm{sel}}$ provided they do not create length-$4$ cycles with the preceding three rows :
\begin{equation}
H_c[r',i]+\sum_{\ell\in\mathcal{C}_{\mathrm{sel}}}H_c[r',\ell]<2,
\qquad
\forall r'\in\{r-3,r-2,r-1\}.
\end{equation}
The selected row $\mathbf{h}^*$ is placed into $\mathbf{H}_c[r,:]$, and the column degrees are updated via $\mathbf{d}_{\mathrm{col}}\leftarrow\mathbf{d}_{\mathrm{col}}+\mathbf{h}^*$

\subsubsection{Finite-Length Parity Budget and Oracle Syndrome Generation}
For each segment $c$, the maximum parity budget is $m_c^{\max}=k_c$ The initial number of syndrome bits requested from Alice is determined from a finite-length Slepian--Wolf normal approximation \cite{slepian-wolf-1973,Polyanskiy-Poor-Verdu,Tan_2014} :
\begin{equation}
H_{\mathrm{fl}}
=
\eta_{\mathrm{SW}}k_c H_{2}(Q)
+
\beta_{\mathrm{disp}}\sqrt{k_c V_Q},
\end{equation}
where $\eta_{\mathrm{SW}}>1$ is the entropy overhead factor and $\beta_{\mathrm{disp}}>0$ scales the finite-blocklength dispersion penalty Setting the initial budget to $m_c=H_{\mathrm{fl}}$ ensures that the protocol begins close to the theoretical Slepian--Wolf limit without under-provisioning syndrome information at low QBER

For each check row $j=0,\dots,k_c-1$, let the row support be \begin{equation}
\mathcal{T}_j=
\left\{
i\in\{0,\dots,k_c-1\}:
H_c[j,i]=1
\right\}.
\end{equation}
In any decoding pass $p$, let $|\mathbf{v}|$ denote the number of syndrome bits already collected for segment $c$. Bob requests $m'=m_c-|\mathbf{v}|$ new parity bits by querying the oracle on the permuted index sets 
\begin{equation}
\begin{aligned}
v_j &= \mathcal{O}(S_j), \\
S_j &= \pi\!\left(\{o_c+i:i\in \mathcal{T}_j\}\right), \quad j=|\mathbf{v}|,\dots,m_c-1.
\end{aligned}
\end{equation}
Each oracle query increments the total leakage accumulator by one bit :
\begin{equation}
\Lambda_{\mathrm{leak}}\leftarrow\Lambda_{\mathrm{leak}}+1.
\end{equation}
The accumulated syndrome vector is $\mathbf{v}=(v_0,\dots,v_{m_c-1})\in\{0,1\}^{m_c}$, and the active parity-check submatrix for the current pass is $\mathbf{H}_{\mathrm{act}}=\mathbf{H}_c[0:m_c-1,\,0:k_c-1]\in\{0,1\}^{m_c\times k_c}$.

\subsubsection{Damped Normalized Min-Sum Syndrome Belief Propagation}
Given Bob's noisy segment $\mathbf{y}_c$, the active submatrix $\mathbf{H}_{\mathrm{act}}$, and Alice's syndrome $\mathbf{v}$, decoding is first attempted via damped normalized min-sum belief propagation \cite{Chen_Fossorier_2002,Pretti_2005} on the corresponding Tanner graph \cite{Kschischang2001,Richardson_Urbanke_2008} The intrinsic channel LLR vector $\boldsymbol{\lambda}\in\mathbb{R}^{k_c}$ is initialized according to 
\begin{equation}
\lambda_i=(1-2y_c[i])L_{\mathrm{ch}},
\qquad
i=0,\dots,k_c-1,
\end{equation}
where a positive LLR corresponds to bit $0$ and a negative LLR corresponds to bit $1$ Let $\mathcal{N}(i)$ denote the set of active check nodes adjacent to variable node $i$, and let $\mathcal{N}(a)$ denote the set of variable nodes adjacent to check node $a$ in $\mathbf{H}_{\mathrm{act}}$. At initialization ($i_{\mathrm{BP}}=0$), the check-to-variable messages are set to $M_{a\to i}^{(0)}=0$ and the variable-to-check messages are set to $M_{i\to a}^{(0)}=\lambda_i$ for all edges $(a,i)$ with $H_{\mathrm{act}}[a,i]=1$

For each iteration $i_{\mathrm{BP}}=1,\dots,I_{\mathrm{BP}}$, the raw check-to-variable message $\tilde{M}_{a\to i}$ incorporates Alice's target syndrome bit $v_a\in\{0,1\}$ through the sign factor $(1-2v_a)$ and scales the minimum incoming magnitude by the normalization constant $\alpha_{\mathrm{BP}}\in(0,1)$ \cite{Chen_Fossorier_2002}:

\begin{equation}
\begin{aligned}
\tilde{M}_{a\to i} 
&= \alpha_{\mathrm{BP}}(1-2v_a) 
\left(
\prod_{i'\in\mathcal{N}(a)\setminus\{i\}}
\mathrm{sgn}\!\left(M_{i'\to a}^{(i_{\mathrm{BP}}-1)}\right)
\right) \\
&\quad \min_{i'\in\mathcal{N}(a)\setminus\{i\}}
\left|M_{i'\to a}^{(i_{\mathrm{BP}}-1)}\right|
\end{aligned}
\end{equation}
To suppress oscillations caused by short cycles in the active submatrix, message damping \cite{Pretti_2005} with coefficient $\beta_{\mathrm{BP}}\in(0,1)$ is applied across consecutive iterations :
\begin{equation}
M_{a\to i}^{(i_{\mathrm{BP}})}
=
\beta_{\mathrm{BP}}M_{a\to i}^{(i_{\mathrm{BP}}-1)}
+
(1-\beta_{\mathrm{BP}})\tilde{M}_{a\to i}.
\end{equation}
The posterior LLR at variable node $i$ is then accumulated as 
\begin{equation}
\lambda_i^{\mathrm{post}}
=
\lambda_i+\sum_{a\in\mathcal{N}(i)}M_{a\to i}^{(i_{\mathrm{BP}})}.
\end{equation}
During the second half of the decoding schedule ($i_{\mathrm{BP}}>I_{\mathrm{BP}}/2$), an exponential moving average of the posterior LLRs is maintained with smoothing factor $\omega_{\mathrm{EMA}}\in(0,1)$:
\begin{equation}
\bar{\lambda}_i
=
\begin{cases}
\omega_{\mathrm{EMA}}\bar{\lambda}_i+(1-\omega_{\mathrm{EMA}})\lambda_i^{\mathrm{post}}, & \text{if } i_{\mathrm{BP}}>I_{\mathrm{BP}}/2,\\[4pt]
\lambda_i^{\mathrm{post}}, & \text{otherwise},
\end{cases}
\end{equation}
which provides a stabilized soft-reliability profile $|\bar{\boldsymbol{\lambda}}|$ for subsequent post-processing. A tentative hard decision is formed at each iteration via \begin{equation}
\hat{y}_c[i]=\mathbf{1}\!\left(\lambda_i^{\mathrm{post}}<0\right),
\qquad
i=0,\dots,k_c-1,
\end{equation}
and its syndrome mismatch count is evaluated as the Hamming weight of the residual syndrome :
\begin{equation}
\mathrm{syn}
=
\mathrm{wt}_{\mathrm{H}}\!\left(
(\mathbf{H}_{\mathrm{act}}\hat{\mathbf{y}}_c\bmod 2)\oplus\mathbf{v}
\right).
\end{equation}
The decoder tracks the best-so-far estimate $\hat{\mathbf{y}}_c^*$, its reliability vector $\boldsymbol{\rho}_c^*=|\bar{\boldsymbol{\lambda}}|$, and the minimum mismatch $\mathrm{syn}^*$, terminating early if $\mathrm{syn}^*=0$ Otherwise, the extrinsic variable-to-check messages are updated according to \begin{equation}
M_{i\to a}^{(i_{\mathrm{BP}})}
=
\lambda_i^{\mathrm{post}}-M_{a\to i}^{(i_{\mathrm{BP}})},
\end{equation}
with message magnitudes bounded by the saturation limit $\Lambda_{\mathrm{sat}}$.

\subsubsection{OSD-0 Post-Processing}
When belief propagation terminates with a non-zero syndrome mismatch ($\mathrm{syn}>0$) due to trapping sets or linear dependencies among the active rows, the tentative decision $\hat{\mathbf{y}}_c$ and its soft reliability vector $\boldsymbol{\rho}_c$ are passed to an order-zero ordered statistics decoder (OSD-0) \cite{Fossorier_Lin_1995,Panteleev_Kalachev_2021}. Let the residual syndrome vector be 
\begin{equation}
\mathbf{s}_{\mathrm{res}}
=
(\mathbf{H}_{\mathrm{act}}\hat{\mathbf{y}}_c\bmod 2)\oplus\mathbf{v},
\end{equation}
and let the set of violated check nodes be 
\begin{equation}
\mathcal{V}_{\mathrm{chk}}
=
\left\{
a\in\{0,\dots,m_c-1\}:
s_{\mathrm{res},a}=1
\right\}.
\end{equation}
For each variable node $i\in\{0,\dots,k_c-1\}$, the number of violated check equations in which bit $i$ participates is 
\begin{equation}
\delta_i=\sum_{a\in\mathcal{V}_{\mathrm{chk}}}H_{\mathrm{act}}[a,i].
\end{equation}
Each variable node is assigned a composite reliability score that penalizes participation in violated checks while rewarding high posterior LLR magnitude:
\begin{equation}
\psi_i=-\eta_{\mathrm{OSD}}\delta_i+\rho_{c,i},
\qquad
i=0,\dots,k_c-1,
\end{equation}
where $\eta_{\mathrm{OSD}}>0$ is the OSD penalty weight. Sorting the variable nodes in ascending order of $\psi_i$ defines a column permutation $\pi_{\mathrm{OSD}}=\mathrm{Argsort}(\boldsymbol{\psi})$ that places the least reliable, most syndrome-violating bits first. The linear system
\begin{equation}
\mathbf{H}_{\mathrm{act}}[:,\pi_{\mathrm{OSD}}]\,\mathbf{e}_{\mathrm{OSD}}
=
\mathbf{s}_{\mathrm{res}}\pmod 2
\end{equation}
is solved over $\mathbb{F}_2$ via Gauss--Jordan elimination by selecting independent pivot columns sequentially from left to right. If a consistent error pattern $\mathbf{e}_{\mathrm{OSD}}\in\{0,1\}^{k_c}$ exists, the refined segment is obtained by flipping the selected pivot bits,
\begin{equation}
\tilde{y}_c[\pi_{\mathrm{OSD}}(i)]
=
\hat{y}_c[\pi_{\mathrm{OSD}}(i)]\oplus e_{\mathrm{OSD},i},
\qquad
i=0,\dots,k_c-1,
\end{equation}
which resolves the residual syndrome ($\mathrm{syn}_{\mathrm{new}}=0$) in a single algebraic step without additional oracle queries

\subsubsection{Universal-Hash Verification}
Whenever the combined BP and OSD-0 decoder achieves $\mathrm{syn}=0$, the candidate segment $\hat{\mathbf{y}}_c$ is verified using a Carter--Wegman Toeplitz universal hash \cite{CARTER1979143,Krawczyk-Hugo} A pseudo-random bit sequence
\begin{equation}
\mathbf{s}\in\{0,1\}^{k_c+t_{\mathrm{hash}}-1}
\end{equation}
is drawn from the synchronized pseudo-random generator. For each hash index $j=0,\dots,t_{\mathrm{hash}}-1$, define the Toeplitz row window and its support:
\begin{align}
\mathbf{r}_j & =(s_j,s_{j+1},\dots,s_{j+k_c-1}),
\nonumber \\
W_j & =\left\{i\in\{0,\dots,k_c-1\}:\mathbf{r}_j[i]=1\right\}.
\end{align}
Alice's hash tag $\mathbf{h}^{\mathrm{A}}=(h_0^{\mathrm{A}},\dots,h_{t_{\mathrm{hash}}-1}^{\mathrm{A}})$ is obtained via the parity oracle on the permuted supports:
\begin{equation}
h_j^{\mathrm{A}}=\mathcal{O}(V_j),
\qquad
V_j=\pi\!\left(\{o_c+i:i\in W_j\}\right),
\end{equation}
while Bob computes his local hash tag $\mathbf{h}^{\mathrm{B}}=(h_0^{\mathrm{B}},\dots,h_{t_{\mathrm{hash}}-1}^{\mathrm{B}})$ as 
\begin{equation}
h_j^{\mathrm{B}}=\bigoplus_{i\in W_j}\hat{y}_c[i].
\end{equation}
This verification step discloses $t_{\mathrm{hash}}$ bits through the oracle:
\begin{equation}
\Lambda_{\mathrm{leak}}\leftarrow\Lambda_{\mathrm{leak}}+t_{\mathrm{hash}}.
\end{equation}
The segment is accepted if and only if $\mathbf{h}^{\mathrm{A}}=\mathbf{h}^{\mathrm{B}}$ Universal-hash verification is essential because a zero syndrome only guarantees that $\hat{\mathbf{y}}_c$ lies in the same coset as Alice's segment; if the decoder converges to an undetected codeword error, the Toeplitz hash detects the discrepancy with collision probability at most $2^{-t_{\mathrm{hash}}}$ \cite{CARTER1979143,Krawczyk-Hugo}.

\subsubsection{Rate-Adaptive Parity Sweep, Blind Pinning, and Key Reassembly}
If the syndrome remains non-zero or the universal-hash check fails, the parity budget $m_c$ is incremented for the next pass $p+1$ To avoid excessive parity leakage when the decoder is already close to convergence, a two-tier multiplicative sweep schedule is employed based on the minimum observed syndrome mismatch $\mathrm{syn}^*$ :
\begin{equation}
m_c\leftarrow
\begin{cases}
\gamma_{\mathrm{fine}}m_c, & \text{if } \mathrm{syn}^*\le\tau_{\mathrm{syn}},\\[4pt]
\gamma m_c, & \text{if } \mathrm{syn}^*>\tau_{\mathrm{syn}}.
\end{cases}
\end{equation}
Because the incremental syndrome queries reuse all previously revealed parities, only the $m'$ newly activated rows of $\mathbf{H}_c$ are queried from the oracle.

If a segment does not converge after $P_{\mathrm{dec}}$ sweep passes, the protocol invokes up to $P_{\mathrm{pin}}$ rounds of blind variable-node pinning to break persistent trapping sets. In each pinning round $p_{\mathrm{pin}}\in\{1,\dots,P_{\mathrm{pin}}\}$, the residual syndrome $\mathbf{s}_{\mathrm{res}}=(\mathbf{H}_{\mathrm{act}}\hat{\mathbf{y}}_c^*\bmod 2)\oplus\mathbf{v}$ and the violated-check participation counts $\delta_i$ are evaluated around the best estimate $\hat{\mathbf{y}}_c^*$. Let $\mathcal{P}_{\mathrm{pin}}$ denote the set of variable nodes already pinned by weight-$1$ check rows in $\mathbf{H}_{\mathrm{act}}$, and let $\mathcal{U}_{\mathrm{pin}}=\{0,\dots,k_c-1\}\setminus\mathcal{P}_{\mathrm{pin}}$ denote the unpinned candidates. Each unpinned node $i\in\mathcal{U}_{\mathrm{pin}}$ is scored via, 
\begin{equation}
\psi_i=-\eta_{\mathrm{pin}}\delta_i+\rho_{c,i}^*,
\end{equation}
and the $n_{\mathrm{pin}}=\rho_{\mathrm{pin}}k_c$ indices $\mathcal{I}_{\mathrm{pin}}=(i_1,\dots,i_{n_{\mathrm{pin}}})\subseteq\mathcal{U}_{\mathrm{pin}}$ with the smallest scores $\psi_i$ are queried directly from the oracle,
\begin{equation}
v_{\mathrm{pin},\ell}=\mathcal{O}\!\left(\{\pi(o_c+i_\ell)\}\right),
\qquad
\ell=1,\dots,n_{\mathrm{pin}},
\end{equation}
incrementing $\Lambda_{\mathrm{leak}}$ by $n_{\mathrm{pin}}$ bits. The revealed bits overwrite the corresponding entries of $\mathbf{y}_c^{\mathrm{pin}}$ and are appended to the linear system as unit-weight check rows $\mathbf{U}_{\mathrm{pin}}=[\mathbf{e}_{i_1},\dots,\mathbf{e}_{i_{n_{\mathrm{pin}}}}]^\top\in\{0,1\}^{n_{\mathrm{pin}}\times k_c}$ :
\begin{equation}
\mathbf{H}_{\mathrm{act}}\leftarrow
\begin{bmatrix}
\mathbf{H}_{\mathrm{act}}\\
\mathbf{U}_{\mathrm{pin}}
\end{bmatrix},
\qquad
\mathbf{v}\leftarrow
\begin{bmatrix}
\mathbf{v}\\
\mathbf{v}_{\mathrm{pin}}
\end{bmatrix},
\end{equation}
followed by another pass of syndrome BP, OSD-0 refinement, and hash verification. If $\mathrm{ok}=0$ after all $P_{\mathrm{pin}}$ rounds, the segment index is recorded in the failed set:
\begin{equation}
\mathcal{F}\leftarrow\mathcal{F}\cup\{c\}.
\end{equation}

Once all $M_{\mathrm{seg}}$ segments have been processed, the reconciled shuffled key is reassembled via \begin{equation}
\hat{y}_{o_c+i}=\hat{y}_c[i],
\qquad
c=1,\dots,M_{\mathrm{seg}},\quad
i=0,\dots,k_c-1,
\end{equation}
and the inverse permutation restores the original sifted-key ordering :
\begin{equation}
\hat{x}_{B,\pi(i)}=\hat{y}_i,
\qquad
i=0,\dots,k-1.
\end{equation}
Any unresolved segments in $\mathcal{F}$ are synchronized and discarded by both parties prior to privacy amplification. The complete rate-adaptive LDPC protocol is summarized in Algorithm~\ref{alg:ldpc_main} and the 5G NR parity-check matrix construction is detailed in Algorithm~\ref{alg:ldpc_matrix}. Furthermore, the damped normalized min-sum syndrome BP and GF(2) OSD-0 decoders are presented in Algorithms~\ref{alg:ldpc_bp} and~\ref{alg:ldpc_osd}, respectively, and the universal-hash verification subroutine is given in Algorithm~\ref{alg:hash}.

\begin{breakablealgorithm}
\caption{LDPC Protocol}
\label{alg:ldpc_main}
\begin{algorithmic}[1]
\footnotesize

\Require
  $\mathbf{x}_B \in \{0,1\}^k$,
  $Q \in (0,1)$,
  oracle $\mathcal{O}(S) = \bigoplus_{i \in S} x_{A,i}$ with Alice key $\mathbf{x}_A \in \{0,1\}^k$,
  Hash length $t_{\mathrm{hash}} \in \mathbb{Z}^+$,
  Sweep factor $\gamma > 1$,
  Fine sweep factor $\gamma_{\mathrm{fine}} \in (1, \gamma)$,
  Mismatch threshold $\tau_{\mathrm{syn}} \in \mathbb{Z}^+$,
  Pass cap $P_{\mathrm{dec}} \in \mathbb{Z}^+$,
  BP iteration cap $I_{\mathrm{BP}} \in \mathbb{Z}^+$,
  Maximum segment size $\kappa_{\mathrm{seg}} \in \mathbb{Z}^+$,
  LLR saturation $\Lambda_{\mathrm{sat}} > 0$,
  Slepian--Wolf scaling factors $\eta_{\mathrm{SW}} > 1, \beta_{\mathrm{disp}} > 0$,
  BP normalization and damping $\alpha_{\mathrm{BP}}, \beta_{\mathrm{BP}} \in (0,1)$,
  EMA weight $\omega_{\mathrm{EMA}} \in (0,1)$,
  OSD-0 penalty $\eta_{\mathrm{OSD}} > 0$,
  Pinning rounds $P_{\mathrm{pin}} \in \mathbb{Z}^+$,
  Pinning fraction $\rho_{\mathrm{pin}} \in (0,1)$,
  Pinning penalty $\eta_{\mathrm{pin}} > 0$,
  PRNG seed $s_0 \in \mathbb{N}$.

\Ensure
  $\hat{\mathbf{x}}_B \in \{0,1\}^k$ (reconciled estimate),
  $\Lambda_{\mathrm{leak}} \in \mathbb{N}$ (oracle leakage, in bits),
  $f_{\mathrm{EC}} \in \mathbb{R}^+$ (reconciliation efficiency),
  $\mathcal{F} \subseteq \{1,\dots,M_{\mathrm{seg}}\}$ (failed segments).

\State $V_Q \leftarrow Q(1-Q) \log_2^2((1-Q)/Q)$
\State $L_{\mathrm{ch}} \leftarrow \ln((1-Q)/Q)$

\State $\pi \leftarrow \mathrm{Shuffle}(0,1,\dots,k-1; s_0)$
\State $y_i \leftarrow x_{B,\pi(i)}$ for $i=0,\dots,k-1$

\State $M_{\mathrm{seg}} \leftarrow k/\kappa_{\mathrm{seg}}$
\State $k_c \leftarrow \lfloor k/M_{\mathrm{seg}} \rfloor + [c \le k \bmod M_{\mathrm{seg}}], \quad c=1,\dots,M_{\mathrm{seg}}$
\State $o_1 \leftarrow 0, \quad o_{c+1} \leftarrow o_c + k_c$

\State $\Lambda_{\mathrm{leak}} \leftarrow 0, \quad \mathcal{F} \leftarrow \emptyset$

\For{$c=1,\dots,M_{\mathrm{seg}}$}
  \State $\mathbf{y}_c \leftarrow (y_{o_c},\dots,y_{o_{c+1}-1}) \in \{0,1\}^{k_c}$
  \State $\mathbf{H}_c \leftarrow \textsc{ConstructMatrix}(k_c, s_0)$
         \Comment{via NVIDIA Sionna \texttt{LDPC5GEncoder}}
  \State $\mathcal{T}_j \leftarrow \{i \in \{0,\dots,k_c-1\} : H_c[j,i] = 1\}$ for $j=0,\dots,k_c-1$
  \State $m_c^{\max} \leftarrow k_c$
  \State $H_{\mathrm{fl}} \leftarrow \eta_{\mathrm{SW}} k_c H_2(Q) + \beta_{\mathrm{disp}} (k_c V_Q)^{1/2}$
  \State $m_c \leftarrow H_{\mathrm{fl}}$
  \State $p \leftarrow 1, \quad \mathbf{v} \leftarrow \emptyset, \quad \mathrm{ok} \leftarrow 0$
  \State $\hat{\mathbf{y}}_c^* \leftarrow \mathbf{y}_c, \quad \boldsymbol{\rho}_c^* \leftarrow \mathbf{1}_{k_c}, \quad \mathrm{syn}^* \leftarrow k_c$

  \While{$p \le P_{\mathrm{dec}}$ \textbf{and} $\mathrm{ok} = 0$}
    \State $m' \leftarrow m_c - |\mathbf{v}|$
    \If{$m' \le 0$}
      \State \textbf{break}
    \EndIf

    \For{$j = |\mathbf{v}|,\dots,m_c-1$}
      \State $S_j \leftarrow \pi(\{o_c + i : i \in \mathcal{T}_j\})$
      \State $v_j \leftarrow \mathcal{O}(S_j)$
      \State $\Lambda_{\mathrm{leak}} \leftarrow \Lambda_{\mathrm{leak}} + 1$
    \EndFor
    \State $\mathbf{v} \leftarrow (v_0,\dots,v_{m_c-1}) \in \{0,1\}^{m_c}$
    \State $\mathbf{H}_{\mathrm{act}} \leftarrow \mathbf{H}_c[0:m_c-1, \, 0:k_c-1] \in \{0,1\}^{m_c \times k_c}$

    \State $(\hat{\mathbf{y}}_c, \boldsymbol{\rho}_c, \mathrm{syn}) \leftarrow \textsc{SyndromeBP}(\mathbf{y}_c, \mathbf{H}_{\mathrm{act}}, \mathbf{v}, L_{\mathrm{ch}}, \Lambda_{\mathrm{sat}}, I_{\mathrm{BP}}, \alpha_{\mathrm{BP}}, \beta_{\mathrm{BP}}, \omega_{\mathrm{EMA}})$

    \If{$\mathrm{syn} > 0$}
      \State $(\hat{\mathbf{y}}_c, \mathrm{syn}) \leftarrow \textsc{OSD0Solve}(\hat{\mathbf{y}}_c, \mathbf{y}_c, \boldsymbol{\rho}_c, \mathbf{H}_{\mathrm{act}}, \mathbf{v}, Q, \eta_{\mathrm{OSD}})$
    \EndIf

    \If{$\mathrm{syn} \le \mathrm{syn}^*$}
      \State $\mathrm{syn}^* \leftarrow \mathrm{syn}, \quad \hat{\mathbf{y}}_c^* \leftarrow \hat{\mathbf{y}}_c, \quad \boldsymbol{\rho}_c^* \leftarrow \boldsymbol{\rho}_c$
    \EndIf

    \If{$\mathrm{syn} = 0$}
      \State $(\mathrm{ok}, \Lambda_{\mathrm{leak}}) \leftarrow \textsc{HashVerify}(\hat{\mathbf{y}}_c, o_c, t_{\mathrm{hash}}, \pi, \mathcal{O}, \Lambda_{\mathrm{leak}})$
    \EndIf

    \If{$\mathrm{ok} = 0$}
      \If{$m_c \ge m_c^{\max}$}
        \State \textbf{break}
      \EndIf
      \If{$\mathrm{syn}^* \le \tau_{\mathrm{syn}}$}
        \State $m_c \leftarrow \gamma_{\mathrm{fine}} m_c$
      \Else
        \State $m_c \leftarrow \gamma m_c$
      \EndIf
      \State $p \leftarrow p + 1$
    \EndIf
  \EndWhile

  \If{$\mathrm{ok} = 0$}
    \State $\mathbf{y}_c^{\mathrm{pin}} \leftarrow \mathbf{y}_c$
    \For{$p_{\mathrm{pin}} = 1,\dots,P_{\mathrm{pin}}$}
      \State $\mathbf{s}_{\mathrm{res}} \leftarrow (\mathbf{H}_{\mathrm{act}} \hat{\mathbf{y}}_c^* \bmod 2) \oplus \mathbf{v}$
      \State $\mathcal{V}_{\mathrm{chk}} \leftarrow \{a : s_{\mathrm{res},a} = 1\}$
      \State $\delta_i \leftarrow \sum_{a \in \mathcal{V}_{\mathrm{chk}}} H_{\mathrm{act}}[a,i]$ for $i=0,\dots,k_c-1$
      \State $\mathcal{P}_{\mathrm{pin}} \leftarrow \{i \in \{0,\dots,k_c-1\} : \exists \, a \text{ s.t. } \sum_{\ell=0}^{k_c-1} H_{\mathrm{act}}[a,\ell] = 1 \text{ and } H_{\mathrm{act}}[a,i] = 1\}$
      \State $\mathcal{U}_{\mathrm{pin}} \leftarrow \{0,\dots,k_c-1\} \setminus \mathcal{P}_{\mathrm{pin}}$
      \If{$\mathcal{U}_{\mathrm{pin}} = \emptyset$}
        \State \textbf{break}
      \EndIf
      \State $\psi_i \leftarrow -\eta_{\mathrm{pin}} \delta_i + \rho_{c,i}^*$ for $i \in \mathcal{U}_{\mathrm{pin}}$
      \State $n_{\mathrm{pin}} \leftarrow \rho_{\mathrm{pin}} k_c$
      \State $\mathcal{I}_{\mathrm{pin}} \leftarrow (i_1,\dots,i_{n_{\mathrm{pin}}}) \subseteq \mathcal{U}_{\mathrm{pin}}$ sorted by ascending $\psi_i$
      \For{$\ell = 1,\dots,n_{\mathrm{pin}}$}
        \State $v_{\mathrm{pin},\ell} \leftarrow \mathcal{O}(\{\pi(o_c + i_\ell)\})$
        \State $\Lambda_{\mathrm{leak}} \leftarrow \Lambda_{\mathrm{leak}} + 1$
        \State $y_{c,i_\ell}^{\mathrm{pin}} \leftarrow v_{\mathrm{pin},\ell}$
      \EndFor
      \State $\mathbf{U}_{\mathrm{pin}} \leftarrow [\mathbf{e}_{i_1}, \dots, \mathbf{e}_{i_{n_{\mathrm{pin}}}}]^\top \in \{0,1\}^{n_{\mathrm{pin}} \times k_c}$
      \State $\mathbf{H}_{\mathrm{act}} \leftarrow \begin{bmatrix} \mathbf{H}_{\mathrm{act}} \\ \mathbf{U}_{\mathrm{pin}} \end{bmatrix}, \quad \mathbf{v} \leftarrow \begin{bmatrix} \mathbf{v} \\ \mathbf{v}_{\mathrm{pin}} \end{bmatrix}$
      \State $(\hat{\mathbf{y}}_c^*, \boldsymbol{\rho}_c^*, \mathrm{syn}^*) \leftarrow \textsc{SyndromeBP}(\mathbf{y}_c^{\mathrm{pin}}, \mathbf{H}_{\mathrm{act}}, \mathbf{v}, L_{\mathrm{ch}}, \Lambda_{\mathrm{sat}}, I_{\mathrm{BP}}, \alpha_{\mathrm{BP}}, \beta_{\mathrm{BP}}, \omega_{\mathrm{EMA}})$
      \If{$\mathrm{syn}^* > 0$}
        \State $(\hat{\mathbf{y}}_c^*, \mathrm{syn}^*) \leftarrow \textsc{OSD0Solve}(\hat{\mathbf{y}}_c^*, \mathbf{y}_c^{\mathrm{pin}}, \boldsymbol{\rho}_c^*, \mathbf{H}_{\mathrm{act}}, \mathbf{v}, Q, \eta_{\mathrm{OSD}})$
      \EndIf
      \If{$\mathrm{syn}^* = 0$}
        \State $(\mathrm{ok}, \Lambda_{\mathrm{leak}}) \leftarrow \textsc{HashVerify}(\hat{\mathbf{y}}_c^*, o_c, t_{\mathrm{hash}}, \pi, \mathcal{O}, \Lambda_{\mathrm{leak}})$
        \If{$\mathrm{ok} = 1$}
          \State \textbf{break}
        \EndIf
      \EndIf
    \EndFor
    \State $\hat{\mathbf{y}}_c \leftarrow \hat{\mathbf{y}}_c^*$
    \If{$\mathrm{ok} = 0$}
      \State $\mathcal{F} \leftarrow \mathcal{F} \cup \{c\}$
    \EndIf
  \EndIf
\EndFor

\State $\hat{y}_{o_c + i} \leftarrow \hat{y}_c[i]$ for $c=1,\dots,M_{\mathrm{seg}}, \, i \in \{0,\dots,k_c-1\}$
\State $\hat{x}_{B,\pi(i)} \leftarrow \hat{y}_i$ for $i=0,\dots,k-1$

\State $f_{\mathrm{EC}} \leftarrow \Lambda_{\mathrm{leak}} / (k H_2(Q))$
\State \Return $(\hat{\mathbf{x}}_B, \Lambda_{\mathrm{leak}}, f_{\mathrm{EC}}, \mathcal{F})$

\end{algorithmic}
\end{breakablealgorithm}
\begin{breakablealgorithm}
\caption{Parity-Check Matrix Construction (ConstructMatrix)}
\label{alg:ldpc_matrix}
\begin{algorithmic}[1]
\footnotesize

\Require
  segment length $k_c \in \mathbb{Z}^+$,
  PRNG seed $s_0 \in \mathbb{N}$,
  folding factor $B_{\mathrm{fold}} \in \mathbb{Z}^+$,
  degree weight bounds $w_{\min}, w_{\max} \in \mathbb{Z}^+$,
  coverage penalty $\eta_{\mathrm{cov}} > 0$,
  target row-degree schedule $\{d_{\mathrm{tgt}}(r)\}_{r=0}^{k_c-1}$.

\Ensure
  degree-balanced parity-check matrix $\mathbf{H}_c \in \{0,1\}^{k_c \times k_c}$.

\State $n_c \leftarrow \begin{cases} 5 k_c & \text{if } k_c \le 3840 \text{ (3GPP Base Graph 2)} \\ 3 k_c & \text{if } k_c > 3840 \text{ (3GPP Base Graph 1)} \end{cases}$
\State $E_c \leftarrow \mathrm{LDPC5GEncoder}(k_c, n_c)$
       \Comment{NVIDIA Sionna library}
\State $\mathbf{H}_{\mathrm{raw}} \in \{0,1\}^{m_{\mathrm{raw}} \times n_{\mathrm{raw}}} \leftarrow \mathrm{ParityCheckMatrix}(E_c)$
\State $\mathbf{H}_{\mathrm{fold}} \leftarrow \bigoplus_{b=0}^{B_{\mathrm{fold}}-1} \mathbf{H}_{\mathrm{raw}}[:, \, b k_c : (b+1) k_c - 1] \in \{0,1\}^{m_{\mathrm{raw}} \times k_c}$
\State $\mathcal{R}_{\mathrm{pool}} \leftarrow \left\{ \mathbf{h} \in \mathrm{Rows}(\mathbf{H}_{\mathrm{fold}}) : w_{\min} \le \sum_{i=0}^{k_c-1} h_i \le w_{\max} \right\}$

\State $\mathbf{H}_c \leftarrow \mathbf{0}_{k_c \times k_c}, \quad \mathbf{d}_{\mathrm{col}} \leftarrow \mathbf{0}_{k_c}$

\For{$r = 0,\dots,k_c-1$}
  \If{$\mathcal{R}_{\mathrm{pool}} \neq \emptyset$ \textbf{and} $\neg(\exists i \text{ s.t. } d_{\mathrm{col},i} < w_{\min})$}
  
    \begin{align*}
    \mathbf{h}^* \leftarrow \arg\max_{\mathbf{h}\in\mathcal{R}_{\mathrm{pool}}}
    \Bigg( & \sum_{i=0}^{k_c-1} \frac{h_i}{1+d_{\mathrm{col},i}} \\
    & - \eta_{\mathrm{cov}} \left| \sum_{i=0}^{k_c-1} h_i - d_{\mathrm{tgt}}(r) \right| \Bigg)
    \end{align*}

    \State $\mathcal{R}_{\mathrm{pool}} \leftarrow \mathcal{R}_{\mathrm{pool}} \setminus \{\mathbf{h}^*\}$
  \Else
    \State $\mathcal{C}_{\mathrm{sel}} \leftarrow \emptyset$
    \State Sort column indices $\{0,\dots,k_c-1\}$ by ascending $d_{\mathrm{col},i}$ (ties broken via $s_0$)
    \For{each sorted column index $i$}
      \If{$\forall r' \in \{r-3,\dots,r-1\}, \; H_c[r', i] + \sum_{\ell \in \mathcal{C}_{\mathrm{sel}}} H_c[r', \ell] < 2$}
        \State $\mathcal{C}_{\mathrm{sel}} \leftarrow \mathcal{C}_{\mathrm{sel}} \cup \{i\}$
      \EndIf
      \If{$|\mathcal{C}_{\mathrm{sel}}| = d_{\mathrm{tgt}}(r)$}
        \State \textbf{break}
      \EndIf
    \EndFor
    \State $h_i^* \leftarrow [i \in \mathcal{C}_{\mathrm{sel}}]$ for $i=0,\dots,k_c-1$
  \EndIf
  \State $\mathbf{H}_c[r, :] \leftarrow \mathbf{h}^*$
  \State $\mathbf{d}_{\mathrm{col}} \leftarrow \mathbf{d}_{\mathrm{col}} + \mathbf{h}^*$
\EndFor

\State \Return $\mathbf{H}_c$

\end{algorithmic}
\end{breakablealgorithm}
\begin{breakablealgorithm}
\caption{Damped Normalized Min-Sum Syndrome BP Subroutine (SyndromeBP)}
\label{alg:ldpc_bp}
\begin{algorithmic}[1]
\footnotesize

\Require
  noisy segment $\mathbf{y}_c \in \{0,1\}^{k_c}$,
  active matrix $\mathbf{H}_{\mathrm{act}} \in \{0,1\}^{m_c \times k_c}$,
  target syndrome $\mathbf{v} \in \{0,1\}^{m_c}$,
  channel LLR $L_{\mathrm{ch}}$,
  saturation $\Lambda_{\mathrm{sat}}$,
  iteration cap $I_{\mathrm{BP}}$,
  scaling $\alpha_{\mathrm{BP}}$,
  damping $\beta_{\mathrm{BP}}$,
  EMA factor $\omega_{\mathrm{EMA}}$.

\Ensure
  decoded segment $\hat{\mathbf{y}}_c \in \{0,1\}^{k_c}$,
  reliability vector $\boldsymbol{\rho}_c \in (\mathbb{R}^+)^{k_c}$,
  syndrome mismatch $\mathrm{syn} \in \mathbb{N}$.

\State $\lambda_i \leftarrow (1 - 2 y_c[i]) L_{\mathrm{ch}}$ for $i=0,\dots,k_c-1$
\State $\bar{\boldsymbol{\lambda}} \leftarrow \boldsymbol{\lambda}, \quad \hat{\mathbf{y}}_c^* \leftarrow \mathbf{y}_c, \quad \boldsymbol{\rho}_c^* \leftarrow |\boldsymbol{\lambda}|$
\State $\mathrm{syn}^* \leftarrow \mathrm{wt}_{\mathrm{H}}((\mathbf{H}_{\mathrm{act}} \mathbf{y}_c \bmod 2) \oplus \mathbf{v})$
\If{$\mathrm{syn}^* = 0$}
  \State \Return $(\hat{\mathbf{y}}_c^*, \boldsymbol{\rho}_c^*, 0)$
\EndIf

\State $M_{a \to i}^{(0)} \leftarrow 0, \quad M_{i \to a}^{(0)} \leftarrow \lambda_i$ for all $(a,i)$ with $H_{\mathrm{act}}[a,i] = 1$

\For{$i_{\mathrm{BP}} = 1,\dots,I_{\mathrm{BP}}$}
  \For{$a = 0,\dots,m_c-1$ \textbf{and} $i \in \mathcal{N}(a)$}
    \State $\tilde{M}_{a \to i} \leftarrow \alpha_{\mathrm{BP}} (1 - 2 v_a) \left( \prod_{i' \in \mathcal{N}(a) \setminus \{i\}} \mathrm{sgn}\!\left(M_{i' \to a}^{(i_{\mathrm{BP}}-1)}\right) \right) \min_{i' \in \mathcal{N}(a) \setminus \{i\}} \left|M_{i' \to a}^{(i_{\mathrm{BP}}-1)}\right|$
    \State $M_{a \to i}^{(i_{\mathrm{BP}})} \leftarrow \mathrm{clamp}(M_{a \to i}^{(i_{\mathrm{BP}})}, -\Lambda_{\mathrm{sat}}, \Lambda_{\mathrm{sat}})$
  \EndFor

  \For{$i = 0,\dots,k_c-1$}
    \State $\lambda_i^{\mathrm{post}} \leftarrow \lambda_i + \sum_{a \in \mathcal{N}(i)} M_{a \to i}^{(i_{\mathrm{BP}})}$
    \If{$i_{\mathrm{BP}} > I_{\mathrm{BP}}/2$}
      \State $\bar{\lambda}_i \leftarrow \omega_{\mathrm{EMA}} \bar{\lambda}_i + (1 - \omega_{\mathrm{EMA}}) \lambda_i^{\mathrm{post}}$
    \Else
      \State $\bar{\lambda}_i \leftarrow \lambda_i^{\mathrm{post}}$
    \EndIf
    \State $\hat{y}_c[i] \leftarrow [\lambda_i^{\mathrm{post}} < 0]$
  \EndFor

  \State $\mathrm{syn} \leftarrow \mathrm{wt}_{\mathrm{H}}((\mathbf{H}_{\mathrm{act}} \hat{\mathbf{y}}_c \bmod 2) \oplus \mathbf{v})$
  \If{$\mathrm{syn} < \mathrm{syn}^*$}
    \State $\mathrm{syn}^* \leftarrow \mathrm{syn}, \quad \hat{\mathbf{y}}_c^* \leftarrow \hat{\mathbf{y}}_c, \quad \boldsymbol{\rho}_c^* \leftarrow |\bar{\boldsymbol{\lambda}}|$
    \If{$\mathrm{syn}^* = 0$}
      \State \Return $(\hat{\mathbf{y}}_c^*, \boldsymbol{\rho}_c^*, 0)$
    \EndIf
  \EndIf

  \For{$i = 0,\dots,k_c-1$ \textbf{and} $a \in \mathcal{N}(i)$}
    \State $M_{i \to a}^{(i_{\mathrm{BP}})} \leftarrow \mathrm{clamp}(M_{i \to a}^{(i_{\mathrm{BP}})}, -\Lambda_{\mathrm{sat}}, \Lambda_{\mathrm{sat}})$
  \EndFor
\EndFor
\State \Return $(\hat{\mathbf{y}}_c^*, |\bar{\boldsymbol{\lambda}}|, \mathrm{syn}^*)$

\end{algorithmic}
\end{breakablealgorithm}
\begin{breakablealgorithm}
\caption{OSD-0 Subroutine (\textsc{OSD0Solve})}
\label{alg:ldpc_osd}
\begin{algorithmic}[1]
\footnotesize

\Require
  tentative segment $\hat{\mathbf{y}}_c \in \{0,1\}^{k_c}$,
  noisy segment $\mathbf{y}_c \in \{0,1\}^{k_c}$,
  reliability vector $\boldsymbol{\rho}_c \in (\mathbb{R}^+)^{k_c}$,
  active matrix $\mathbf{H}_{\mathrm{act}} \in \{0,1\}^{m_c \times k_c}$,
  target syndrome $\mathbf{v} \in \{0,1\}^{m_c}$,
  QBER $Q$,
  OSD penalty $\eta_{\mathrm{OSD}}$.

\Ensure
  refined segment $\hat{\mathbf{y}}_c \in \{0,1\}^{k_c}$,
  updated syndrome mismatch $\mathrm{syn} \in \mathbb{N}$.

\State $\mathbf{s}_{\mathrm{res}} \leftarrow (\mathbf{H}_{\mathrm{act}} \hat{\mathbf{y}}_c \bmod 2) \oplus \mathbf{v}$
\State $\mathrm{syn} \leftarrow \mathrm{wt}_{\mathrm{H}}(\mathbf{s}_{\mathrm{res}})$
\State $\mathcal{V}_{\mathrm{chk}} \leftarrow \{a \in \{0,\dots,m_c-1\} : s_{\mathrm{res},a} = 1\}$
\State $\delta_i \leftarrow \sum_{a \in \mathcal{V}_{\mathrm{chk}}} H_{\mathrm{act}}[a,i]$ for $i=0,\dots,k_c-1$
\State $\psi_i \leftarrow -\eta_{\mathrm{OSD}} \delta_i + \rho_{c,i}$ for $i=0,\dots,k_c-1$
\State $\pi_{\mathrm{OSD}} \leftarrow \mathrm{Argsort}(\boldsymbol{\psi})$ \Comment{Ascending order of $\psi_i$}
\State Solve $\mathbf{H}_{\mathrm{act}}[:, \pi_{\mathrm{OSD}}] \mathbf{e}_{\mathrm{OSD}} = \mathbf{s}_{\mathrm{res}} \pmod 2$ via Gauss-Jordan elimination over $\mathbb{F}_2$
\If{consistent solution $\mathbf{e}_{\mathrm{OSD}} \in \{0,1\}^{k_c}$ exists}
  \State $\tilde{y}_c[\pi_{\mathrm{OSD}}(i)] \leftarrow \hat{y}_c[\pi_{\mathrm{OSD}}(i)] \oplus e_{\mathrm{OSD},i}$ for $i=0,\dots,k_c-1$
  \State $\mathrm{syn}_{\mathrm{new}} \leftarrow \mathrm{wt}_{\mathrm{H}}((\mathbf{H}_{\mathrm{act}} \tilde{\mathbf{y}}_c \bmod 2) \oplus \mathbf{v})$
  \If{$\mathrm{syn}_{\mathrm{new}} = 0$}
    \State \Return $(\tilde{\mathbf{y}}_c, 0)$
  \EndIf
\EndIf
\State \Return $(\hat{\mathbf{y}}_c, \mathrm{syn})$

\end{algorithmic}
\end{breakablealgorithm}
\begin{breakablealgorithm}
\caption{Hash Verification Subroutine (\textsc{HashVerify})}
\label{alg:hash}
\begin{algorithmic}[1]
\footnotesize

\Require
  $\hat{\mathbf{y}}_c \in \{0,1\}^{k_c}$ (decoded segment),
  $o_c \in \mathbb{N}$ (segment offset),
  $t_{\mathrm{hash}} \in \mathbb{Z}^+$ (hash output length),
  permutation $\pi$,
  oracle $\mathcal{O}$,
  leakage accumulator $\Lambda_{\mathrm{leak}}$.

\Ensure
  verification flag in $\{0,1\}$,
  updated $\Lambda_{\mathrm{leak}}$.

\State $\mathbf{s} \leftarrow \mathrm{PRNG}() \in \{0,1\}^{k_c + t_{\mathrm{hash}} - 1}$
\State $\mathbf{r}_j \leftarrow (s_j, s_{j+1}, \dots, s_{j+k_c-1})$ for $j=0,\dots,t_{\mathrm{hash}}-1$
\State $W_j \leftarrow \{i \in \{0,\dots,k_c-1\} : \mathbf{r}_j[i] = 1\}$ for $j=0,\dots,t_{\mathrm{hash}}-1$
\State $V_j \leftarrow \pi(\{o_c + i : i \in W_j\})$ for $j=0,\dots,t_{\mathrm{hash}}-1$
\State $h^{\mathrm{A}}_j \leftarrow \mathcal{O}(V_j)$ for $j=0,\dots,t_{\mathrm{hash}}-1$
\State $\Lambda_{\mathrm{leak}} \leftarrow \Lambda_{\mathrm{leak}} + t_{\mathrm{hash}}$
\State $h^{\mathrm{B}}_j \leftarrow \bigoplus_{i \in W_j} \hat{y}_c[i]$ for $j=0,\dots,t_{\mathrm{hash}}-1$
\State \Return $([\mathbf{h}^{\mathrm{A}} = \mathbf{h}^{\mathrm{B}}], \Lambda_{\mathrm{leak}})$

\end{algorithmic}
\end{breakablealgorithm}

\subsection{Polar Codes}
\label{sec:Polar_Code-Based_Reconciliation_Protocol_for_Quantum_Key_Distribution}

Polar codes have emerged as an efficient channel coding technique capable of achieving Shannon channel capacity with low-complexity decoding. In QKD systems, polar-code-based reconciliation provides an effective mechanism for correcting discrepancies between the sifted keys of Alice and Bob while minimizing information leakage over the authenticated classical communication channel.

In this work, a Successive Cancellation (SC) decoding-based polar reconciliation framework is presented for DV-QKD systems \cite{jouguet2012high}. The protocol utilizes adaptive reliability estimation, recursive likelihood propagation, oracle-assisted parity queries, and polar transform operations to reconcile Bob's noisy key with Alice's original key. The method dynamically balances decoding confidence and parity disclosure using the FER budget mechanism. The protocol operates on binary key sequences and incorporates recursive F-transform and G-transform operations to estimate the most reliable decoded bit sequence while minimising the number of parity disclosures.

\subsubsection{Protocol Parameters}

The reconciliation framework utilizes the following parameters:
\begin{itemize}
    \item $\mathbf{x}_B \in \{0,1\}^k$ : Bob's noisy sifted key of length $k$.
    \item $Q \in (0,1)$ : Estimated QBER.
    \item $\mathcal{O}(S)=\bigoplus_{i\in S} x_{A,i}$ : Parity oracle where Alice's key is $\mathbf{x}_A\in\{0,1\}^k$.
    \item $\varepsilon_{\mathrm{FER}} \in (0,1)$ : Target frame error rate.
    \item $\theta_c \in (0,1)$ : Confidence threshold parameter.
    \item $n_p$ : Polar transform order, $n_p=\log_2 N_p=\lceil \log_2 k\rceil$.
    \item $N_p$ : Polar code block length, $N_p=2^{n_p}\ge k$; $N_p$ is the smallest power of two not less than $k$, and the algorithm zero-pads the length-$k$ key to length $N_p$.
\end{itemize}

\subsubsection{Key Padding and Initialization}

Let the original sifted key length be:
\begin{equation}
k = |\mathbf{x}_B|
\end{equation}
The nearest power-of-two block length is selected as
\begin{equation}
n_p \leftarrow \lceil \log_2 k \rceil,\qquad N_p \leftarrow 2^{n_p}
\end{equation}
Next, the sifted key is padded with zeros when required:
\begin{equation}
\mathbf{x}_B \leftarrow (\mathbf{x}_B, \underbrace{0,\dots,0}_{N_p-k})
\end{equation}
A reliability matrix and bit-decision matrix are initialized as
\begin{equation}
\mathbf{\Gamma} \in \mathbb{R}^{(n_p+1)\times N_p}
\end{equation}

\begin{equation}
\mathbf{D} \in \{-1,0,1\}^{(n_p+1)\times N_p}
\end{equation}
where $\mathbf{\Gamma}$ stores Log-Likelihood Ratio (LLR)-style reliability values, and $\mathbf{D}$ stores intermediate decoded bit estimates.

\subsubsection{Reliability Initialization}
The initial channel reliability value is calculated according to the observed QBER as
\begin{equation}
\rho \leftarrow 1 - 2Q
\end{equation}
For each received bit:
\begin{equation}
\Gamma_{n_p,i} \leftarrow
\begin{cases}
1.0, & i \ge k \\
\rho\,(1 - 2x_{B,i}), & i < k
\end{cases}
\end{equation}
where $\Gamma_{n_p,i}$ denotes the initial LLR-style reliability value for the $i$-th coordinate at the top stage $n_p$ of the polar decoding tree, before any $f$ or $g$-transform is applied. Positive values indicate confidence toward the bit value $0$, while negative values indicate confidence toward the bit value $1$.

\subsubsection{Polar Transform Dependency Construction}
The protocol computes the dependency structure of the polar generator matrix as
\begin{equation}
\mathbf{G}_{N_p} = \mathbf{F}^{\otimes n_p}
\end{equation}
where
\begin{equation}
\mathbf{F}=
\begin{pmatrix}
1 & 0 \\
1 & 1
\end{pmatrix}
\end{equation}
and $\otimes$ denotes the Kronecker product. The dependency sets determine which encoded symbols contribute to each decoded bit during successive cancellation decoding. The construction uses the half-size
\begin{equation}
h_{\mathrm{dep}} \leftarrow 2^{s},
\qquad s=0,\dots,n_p-1,
\end{equation}
at each stage $s$ of the dependency-set construction.

\subsubsection{Successive Cancellation Decoding}
The decoding process recursively traverses the polar decoding tree. Let \(m_{\mathrm{pol}}\) be the offset index of the current node. Let \(\Gamma_{s,i}\) denote the reliability value at tree level \(s\) and index \(i\), and let \(D_{s,i}\) denote the decoded bit at level \(s\) and index \(i\). Let \(\Delta_{\mathrm{FER}}\) be the remaining frame-error budget, initialised to the target frame error rate \(\varepsilon_{\mathrm{FER}}\). Let \(\mathcal{J}_{m_{\mathrm{pol}}}\) be the dependency set for node offset \(m_{\mathrm{pol}}\). At each leaf node, the reliability value is converted into a probability estimate as
\begin{equation}
p_0 \leftarrow \frac{1}{2}(1 + \Gamma_{0,m_{\mathrm{pol}}}).
\end{equation}
An adaptive confidence threshold is calculated as
\begin{equation}
\xi \leftarrow \min\left(\theta_c,\frac{\Delta_{\mathrm{FER}}}{N_p-m_{\mathrm{pol}}}\right),
\end{equation}
where \(\theta_c\) is the confidence threshold and \(N_p\) is the codeword length. The decoded bit is determined using
\begin{equation}
D_{0,m_{\mathrm{pol}}} \leftarrow
\begin{cases}
0, & p_0 > 1-\xi, \\
1, & p_0 < \xi, \\
\mathcal{O}(\mathcal{Q}_{m_{\mathrm{pol}}}), & \text{otherwise},
\end{cases}
\end{equation}
where
\begin{equation}
\mathcal{Q}_{m_{\mathrm{pol}}} \leftarrow \{i \in \mathcal{J}_{m_{\mathrm{pol}}} : i < k\},
\end{equation}
and \(k\) is the key length. In the first case, update \(\Delta_{\mathrm{FER}} \leftarrow \Delta_{\mathrm{FER}} - (1-p_0)\).  
In the second case, update \(\Delta_{\mathrm{FER}} \leftarrow \Delta_{\mathrm{FER}} - p_0\).
When the decoder confidence is insufficient, the protocol queries Alice for the parity of the corresponding subset of key bits.

\subsubsection{Polar Transform Operations}
During successive cancellation decoding, the polar decoding tree uses two recursive reliability updates: the \(f\)-transform for the left child and the \(g\)-transform for the right child. Let \(s\) be the current tree level, \(m_{\mathrm{pol}}\) the offset index, and
\begin{equation}
h_{\mathrm{tr}} \leftarrow 2^{s-1}.
\end{equation}
The reliability values are denoted by \(\Gamma_{s,i}\), and the decoded bits by \(D_{s,i}\).

\paragraph{F-Transform Operation}
The left-child reliability update is computed using
\begin{equation}
\Gamma_{s-1,m_{\mathrm{pol}}+i} \leftarrow
\Gamma_{s,m_{\mathrm{pol}}+i} \cdot \Gamma_{s,m_{\mathrm{pol}}+h_{\mathrm{tr}}+i},
\quad i=0,\dots,h_{\mathrm{tr}}-1.
\end{equation}
This operation models the XOR combination channel in the polar decoding tree.

\paragraph{G-Transform Operation}
After decoding the left child, the right-child reliability is updated using
\begin{equation}
\Gamma_{s-1,m_{\mathrm{pol}}+h_{\mathrm{tr}}+i} \leftarrow
\begin{cases}
\dfrac{r_u^{\mathrm{pol}} + r_v^{\mathrm{pol}}}{1 + r_u^{\mathrm{pol}} r_v^{\mathrm{pol}}}, & b^{\mathrm{pol}}=0 \\[6pt]
\dfrac{r_v^{\mathrm{pol}} - r_u^{\mathrm{pol}}}{1 - r_u^{\mathrm{pol}} r_v^{\mathrm{pol}}}, & b^{\mathrm{pol}}=1
\end{cases}
\end{equation}
where
\begin{equation}
r_u^{\mathrm{pol}} \leftarrow \Gamma_{s,m_{\mathrm{pol}}+i},\;
r_v^{\mathrm{pol}} \leftarrow \Gamma_{s,m_{\mathrm{pol}}+h_{\mathrm{tr}}+i},\;
b^{\mathrm{pol}} \leftarrow D_{s-1,m_{\mathrm{pol}}+i}.
\end{equation}
The \(g\)-transform incorporates previously decoded information into the right subtree estimation.

\paragraph{Decision Combination}
The decoded child decisions are combined according to
\begin{equation}
D_{s,m_{\mathrm{pol}}+i} \leftarrow D_{s-1,m_{\mathrm{pol}}+i} \oplus D_{s-1,m_{\mathrm{pol}}+h_{\mathrm{tr}}+i},
\end{equation}
\begin{equation}
D_{s,m_{\mathrm{pol}}+h_{\mathrm{tr}}+i} \leftarrow D_{s-1,m_{\mathrm{pol}}+h_{\mathrm{tr}}+i},
\end{equation}
where \(\oplus\) denotes modulo-2 addition.

\subsubsection{Polar Re-Encoding}

After completion of successive cancellation decoding, the estimated bit vector is re-encoded using the polar transform:
\begin{equation}
\hat{\mathbf{c}}_B \leftarrow \hat{\mathbf{u}}_B\,\mathbf{G}_{N_p} \bmod 2
\end{equation}
The corrected codeword is then truncated to the original sifted key length:
\begin{equation}
\hat{\mathbf{x}}_B \leftarrow \hat{\mathbf{c}}_{B,0:k}
\end{equation}

\subsubsection{Parity Disclosure Efficiency}

The number of oracle parity requests is at most $N_p$:
\begin{equation}
\{\text{oracle parity queries}\} \leq N_p
\end{equation}
Each parity request reveals one XOR parity relation. The protocol minimizes parity disclosures by utilizing confidence-based adaptive decoding.

The proposed polar-code-based reconciliation framework provides an efficient and capacity-approaching post-processing solution for DV-QKD systems. The protocol combines successive cancellation decoding, adaptive reliability estimation, recursive polar transforms, and oracle-assisted parity queries to perform error reconciliation reliably while limiting information leakage. The use of adaptive frame error rate management further enables efficient finite-size secure key generation in practical quantum communication systems.

\begin{breakablealgorithm}
\caption{Polar Code Reconciliation for QKD}
\label{alg:polar_main}
\begin{algorithmic}[1]
\footnotesize

\Require $\mathbf{x}_B \in \{0,1\}^k$,\; $Q \in (0,1)$,\; oracle $\mathcal{O}(S)=\bigoplus_{i\in S} x_{A,i}$ for Alice key $\mathbf{x}_A\in\{0,1\}^k$,\; target FER $\varepsilon_{\mathrm{FER}} \in (0,1)$,\; threshold $\theta_c \in (0,1)$
\Ensure $\hat{\mathbf{x}}_B \in \{0,1\}^k$

\State $n_p \leftarrow \lceil \log_2 k \rceil$,\quad $N_p \leftarrow 2^{n_p}$

\State $\mathbf{x}_B \leftarrow (\mathbf{x}_B,\, \underbrace{0,\dots,0}_{N_p-k})$ \quad (zero-pad to length $N_p$)

\vspace{2pt}

\State $\rho \leftarrow 1 - 2Q$

\For{$i = 0, \dots, N_p-1$}
    \State $\Gamma_{n_p,i} \leftarrow \begin{cases} 1.0 & i \geq k \\ \rho\,(1 - 2x_{B,i}) & i < k \end{cases}$
\EndFor

\vspace{2pt}

\State $\mathcal{J}_{m_{\mathrm{pol}}} \leftarrow \{m_{\mathrm{pol}}\}$ for $m_{\mathrm{pol}} = 0,\dots,N_p-1$

\For{$s = 0, \dots, n_p-1$}
    \State $h_{\mathrm{dep}} \leftarrow 2^s$
    \For{$m_{\mathrm{pol}} = 0,\,2h_{\mathrm{dep}},\,4h_{\mathrm{dep}},\dots$}
        \For{$t = 0, \dots, h_{\mathrm{dep}} - 1$}
            \State $\mathcal{J}_{m_{\mathrm{pol}}+t} \leftarrow \mathcal{J}_{m_{\mathrm{pol}}+t} \cup \mathcal{J}_{m_{\mathrm{pol}}+t+h_{\mathrm{dep}}}$
        \EndFor
    \EndFor
\EndFor

\vspace{2pt}

\State $\Delta_{\mathrm{FER}} \leftarrow \varepsilon_{\mathrm{FER}}$ \Comment{remaining frame-error budget}
\State $D_{s,i} \leftarrow -1$ for all $s,i$ \quad (undecided)

\Function{Node}{$m_{\mathrm{pol}},\, s$}

    \If{$s = 0$} \Comment{Leaf decision}
        \State $p_0 \leftarrow \tfrac{1}{2}(1 + \Gamma_{0,m_{\mathrm{pol}}})$
        \State $\xi \leftarrow \min\!\left(\theta_c,\;\; \Delta_{\mathrm{FER}}\,/\,(N_p - m_{\mathrm{pol}})\right)$
        \If{$p_0 > 1 - \xi$}
            \State $D_{0,m_{\mathrm{pol}}} \leftarrow 0$;\quad $\Delta_{\mathrm{FER}} \leftarrow \Delta_{\mathrm{FER}} - (1 - p_0)$
        \ElsIf{$p_0 < \xi$}
            \State $D_{0,m_{\mathrm{pol}}} \leftarrow 1$;\quad $\Delta_{\mathrm{FER}} \leftarrow \Delta_{\mathrm{FER}} - p_0$
        \Else \Comment{Ambiguous: query oracle}
            \State $\mathcal{Q}_{m_{\mathrm{pol}}} \leftarrow \{i \in \mathcal{J}_{m_{\mathrm{pol}}} : i < k\}$
            \State $D_{0,m_{\mathrm{pol}}} \leftarrow \mathcal{O}(\mathcal{Q}_{m_{\mathrm{pol}}}) = \bigoplus_{i \in \mathcal{Q}_{m_{\mathrm{pol}}}} x_{A,i}$
        \EndIf
        \State \Return
    \EndIf

    \State $h_{\mathrm{tr}} \leftarrow 2^{s-1}$

    \For{$i = 0, \dots, h_{\mathrm{tr}}-1$} \Comment{$f$-transform: left child}
        \State $\Gamma_{s-1,m_{\mathrm{pol}}+i} \leftarrow \Gamma_{s,m_{\mathrm{pol}}+i} \cdot \Gamma_{s,m_{\mathrm{pol}}+h_{\mathrm{tr}}+i}$
    \EndFor

    \State \Call{Node}{$m_{\mathrm{pol}},\; s-1$}

    \For{$i = 0, \dots, h_{\mathrm{tr}}-1$} \Comment{$g$-transform: right child}
        \State $r_u^{\mathrm{pol}} \leftarrow \Gamma_{s,m_{\mathrm{pol}}+i}$,\quad $r_v^{\mathrm{pol}} \leftarrow \Gamma_{s,m_{\mathrm{pol}}+h_{\mathrm{tr}}+i}$,\quad $b^{\mathrm{pol}} \leftarrow D_{s-1,m_{\mathrm{pol}}+i}$
        \State $\Gamma_{s-1,m_{\mathrm{pol}}+h_{\mathrm{tr}}+i} \leftarrow \begin{cases} \dfrac{r_u^{\mathrm{pol}} + r_v^{\mathrm{pol}}}{1 + r_u^{\mathrm{pol}} r_v^{\mathrm{pol}}} & b^{\mathrm{pol}} = 0 \\[6pt] \dfrac{r_v^{\mathrm{pol}} - r_u^{\mathrm{pol}}}{1 - r_u^{\mathrm{pol}} r_v^{\mathrm{pol}}} & b^{\mathrm{pol}} = 1 \end{cases}$
    \EndFor

    \State \Call{Node}{$m_{\mathrm{pol}}+h_{\mathrm{tr}},\; s-1$}

    \For{$i = 0, \dots, h_{\mathrm{tr}}-1$} \Comment{Combine decisions upward}
        \State $D_{s,m_{\mathrm{pol}}+i} \leftarrow D_{s-1,m_{\mathrm{pol}}+i} \oplus D_{s-1,m_{\mathrm{pol}}+h_{\mathrm{tr}}+i}$
        \State $D_{s,m_{\mathrm{pol}}+h_{\mathrm{tr}}+i} \leftarrow D_{s-1,m_{\mathrm{pol}}+h_{\mathrm{tr}}+i}$
    \EndFor

\EndFunction

\vspace{2pt}

\State $\hat{\mathbf{u}}_B \leftarrow (D_{0,0},\, D_{0,1},\, \dots,\, D_{0,N_p-1})$

\State $\hat{\mathbf{c}}_B \leftarrow \hat{\mathbf{u}}_B\, \mathbf{G}_{N_p} \bmod 2$,\quad where $\mathbf{G}_{N_p} = \mathbf{F}^{\otimes n_p}$,\quad $\mathbf{F} = \begin{pmatrix} 1 & 0 \\ 1 & 1 \end{pmatrix}$

\State $\hat{\mathbf{x}}_B \leftarrow \hat{\mathbf{c}}_{B,0:k}$

\State \Return $\hat{\mathbf{x}}_B$

\end{algorithmic}
\end{breakablealgorithm}

\section{Privacy Amplification}
\label{sec:Privacy_Amplification}
Privacy amplification is one of the most important steps in the post-processing phase of QKD. It allows reducing the information that Eve has about the shared key. After quantum transmission and error correction, Alice and Bob obtain a correlated key, but Eve may still possess partial information about the key due to noise, imperfections, or potential attacks on the quantum channel. Privacy amplification addresses this residual information by applying a randomly selected hash function, typically from a universal family of hash functions, to compress the reconciled key into a shorter final key. The amount of compression is determined by the estimated information available to Eve so that the resulting key is arbitrarily close to being uniformly random and independent of Eve's knowledge. In this way, even though Eve may have partial information about the reconciled key, the knowledge about the final secret key can be made negligibly small. The theoretical foundation of privacy amplification was established using universal hashing, and its security has subsequently been extended to scenarios where the eavesdropper possesses quantum information \cite{Bennett1995,Renner2004}. In this work, privacy amplification is performed using two methods, Toeplitz hashing and SHAKE256 hashing, along with their algorithms.  

\paragraph{Toeplitz Hashing}
After information reconciliation, Alice and Bob share a reconciled key \(K_{\mathrm{rec}} \in \{0,1\}^{k}\). A randomly chosen Toeplitz matrix is used as a universal hash function to compress this key into a shorter secret key. The public seed \(s_{\mathrm{pa}}\) initializes a pseudo-random generator \(\mathcal{G}_{\mathrm{prng}}\). This generator produces a random bit sequence \(\mathbf{w}\) of length \(k+L_{\mathrm{sec}}-1\). From this sequence, each of the \(L_{\mathrm{sec}}\) rows of the Toeplitz matrix is formed by taking a reversed window of \(k\) bits. The output bit for each row is the modulo-2 inner product of that row with the reconciled key. Repeating this for all \(L_{\mathrm{sec}}\) rows gives the final secret key \(K_{\mathrm{final}} \in \{0,1\}^{L_{\mathrm{sec}}}\). The Toeplitz matrix is specified by its first row and first column, so only a small amount of randomness is needed. The compression limits Eve's information, and the resulting key is statistically close to uniform if the output length is chosen correctly \cite{Renner2005}. The Toeplitz hashing is described in Algorithm~\ref{alg:pa-toeplitz}.
\begin{algorithm}
\caption{Privacy Amplification via Toeplitz Hashing}
\label{alg:pa-toeplitz}
\begin{algorithmic}[1]

\Require Reconciled key $K_{\mathrm{rec}} \in \{0,1\}^k$, QBER $Q \in (0, 0.5)$, security parameter $S_{\mathrm{sec}}$, single-photon probability $p_1$, EC leakage $\mathrm{leak}_{\mathrm{EC}}$, public seed $s_{\mathrm{pa}}$
\Ensure Final secret key $K_{\mathrm{final}} \in \{0,1\}^{L_{\mathrm{sec}}}$

\State $H_2(Q) \leftarrow -Q \log_2(Q) - (1-Q) \log_2(1-Q)$ \Comment{Binary entropy}
\State $t_{\mathrm{Eve}} \leftarrow \lceil k(1 - p_1) + k \cdot H_2(Q) + \mathrm{leak}_{\mathrm{EC}} \rceil$ \Comment{Eve's knowledge bound}
\State $L_{\mathrm{sec}} \leftarrow \max(0,\, k - t_{\mathrm{Eve}} - S_{\mathrm{sec}})$ \Comment{Extractable key length}

\If{$L_{\mathrm{sec}} \leq 0$}
    \State \Return $\emptyset$
\EndIf

\vspace{2pt}

\State $\mathcal{G}_{\mathrm{prng}} \leftarrow \Call{InitializePRNG}{s_{\mathrm{pa}}}$ \Comment{PRNG seeded with $s_{\mathrm{pa}}$}
\State $\mathbf{w} \leftarrow \Call{GenerateSequence}{\mathcal{G}_{\mathrm{prng}},\, k+L_{\mathrm{sec}}-1}$ \Comment{Draw $k+L_{\mathrm{sec}}-1$ random bits from PRNG}

\vspace{2pt}

\For{$i = 0, \dots, L_{\mathrm{sec}}-1$}
    \State $\mathbf{w}_i \leftarrow (w_{i+k-1},\, w_{i+k-2},\, \dots,\, w_{i})$ \Comment{Extract reversed slice}
    \State $K_{\mathrm{final},i} \leftarrow \left( \sum_{j=0}^{k-1} w_{i,j} \cdot K_{\mathrm{rec},j} \right) \bmod 2$ \Comment{GF(2) inner product}
\EndFor

\State \Return $K_{\mathrm{final}} = (K_{\mathrm{final},0}, \dots, K_{\mathrm{final},L_{\mathrm{sec}}-1})$

\end{algorithmic}
\end{algorithm}
\paragraph{SHAKE256 Hashing}
The SHAKE256 method uses a cryptographic extendable-output function (XOF) to extract a secret key directly from the reconciled key. The algorithm first absorbs the public seed \(s_{\mathrm{pa}}\) and the reconciled key \(K_{\mathrm{rec}}\) into the SHAKE256 state. It then squeezes \(B_{\mathrm{out}} = \lceil L_{\mathrm{sec}}/8 \rceil\) bytes of output. These bytes are converted into a bit string, and the first \(L_{\mathrm{sec}}\) bits are taken as the final secret key \(K_{\mathrm{final}}\). Because SHAKE256 is a cryptographic sponge function, its output is designed to be computationally indistinguishable from random, provided the seed is public but random and the input key is secret \cite{NISTFIPS202}. This method avoids explicitly constructing a Toeplitz matrix and provides a flexible output length. SHAKE256-based hashing is presented in Algorithm~\ref{alg:pa-shake}.
\begin{algorithm}
\caption{Privacy Amplification via SHAKE256}
\label{alg:pa-shake}
\begin{algorithmic}[1]

\Require Reconciled key $K_{\mathrm{rec}} \in \{0,1\}^k$, QBER $Q \in (0, 0.5)$, security parameter $S_{\mathrm{sec}}$, single-photon probability $p_1$, EC leakage $\mathrm{leak}_{\mathrm{EC}}$, public seed $s_{\mathrm{pa}}$
\Ensure Final secret key $K_{\mathrm{final}} \in \{0,1\}^{L_{\mathrm{sec}}}$

\State $H_2(Q) \leftarrow -Q \log_2(Q) - (1-Q) \log_2(1-Q)$ \Comment{Binary entropy}
\State $t_{\mathrm{Eve}} \leftarrow \lceil k(1 - p_1) + k \cdot H_2(Q) + \mathrm{leak}_{\mathrm{EC}} \rceil$ \Comment{Eve's knowledge bound}
\State $L_{\mathrm{sec}} \leftarrow \max(0,\, k - t_{\mathrm{Eve}} - S_{\mathrm{sec}})$ \Comment{Extractable key length}

\If{$L_{\mathrm{sec}} \leq 0$}
    \State \Return $\emptyset$
\EndIf

\vspace{2pt}

\State $B_{\mathrm{out}} \leftarrow \lceil L_{\mathrm{sec}} / 8 \rceil$

\vspace{2pt}

\State $\mathcal{H}_{\mathrm{shake}} \leftarrow \text{SHAKE256\_Init}()$ \Comment{Using python hashlib package}
\State $\Call{Update}{\mathcal{H}_{\mathrm{shake}},\, s_{\mathrm{pa}} \parallel K_{\mathrm{rec}}}$ \Comment{Absorb salt and key material}
\State $\mathbf{d}_{\mathrm{bytes}} \leftarrow \Call{Digest}{\mathcal{H}_{\mathrm{shake}},\, B_{\mathrm{out}}}$

\vspace{2pt}

\State $\mathbf{d}_{\mathrm{bits}} \leftarrow \emptyset$
\For{$b_{\mathrm{byte}} \in \mathbf{d}_{\mathrm{bytes}}$}
    \For{$j = 7 \text{ down to } 0$}
        \State $\mathbf{d}_{\mathrm{bits}} \leftarrow \mathbf{d}_{\mathrm{bits}} \parallel ((b_{\mathrm{byte}} \gg j) \mathbin{\&} 1)$
    \EndFor
\EndFor

\State \Return $\mathbf{d}_{\mathrm{bits}}[0 : L_{\mathrm{sec}}-1]$ \Comment{Truncate to exact target length $L_{\mathrm{sec}}$}

\end{algorithmic}
\end{algorithm}

\section{Simulation Results}
\label{Simulation_Results}

In this work, the QKD simulator is built around an asynchronous actor-based model combined with a client-server classical post-processing framework. The system components, including Alice, the quantum channel, Eve, the detector, and Bob, operate as independent actors that do not share memory directly and communicate by passing messages. Each actor runs asynchronously, so that the quantum transmission and classical post-processing can proceed without tight coupling. Alice and Bob are connected via a quantum channel (lossy fiber), where Eve can sit in between to intercept and resend photons, and a classical authenticated channel. Alice acts as the server and houses the quantum source: her role is to prepare and transmit weak coherent pulses, keep a record of what she has sent, and expose a small classical API (oracle interface) for the steps that require public discussion. Conversely, Bob acts as the client and houses the detector: once his detector measures the incoming pulses and records the detection events, Bob initiates basis sifting, estimates the QBER, runs the chosen reconciliation protocol, and performs privacy amplification. This architecture provides three major conveniences: (i) it separates the quantum and classical layers, which makes the simulator easier to debug and extend; (ii) it provides a common oracle interface, so different reconciliation protocols can be compared; and (iii) it closely follows the actual BB84 workflow, where Alice and Bob exchange only classical messages over an authenticated channel while the quantum channel carries the key material. 



All reconciliation protocols use the same oracle abstraction. Alice's key is never exposed directly. Instead, Bob can ask for the parity of any subset \(S\) of the key indices:
\begin{equation}
\mathcal{O}(S)=\bigoplus_{j\in S} x_{A,j},
\end{equation}
where \(x_{A,j}\) is Alice's bit at position \(j\). Alice returns only one bit, the XOR parity. This is exactly the information that would be sent over an authenticated classical channel. Cascade uses this oracle for block parities \(\mathcal{O}(\mathcal{B}_\ell)\) and binary search.  LDPC uses it for selected parity bits \(v_j\), syndrome contributions \(u_j\), and subblock confirmation, running a maximum of \(I_{\mathrm{BP}}=10\) BP iterations with an LLR saturation value \(\Lambda_{\mathrm{sat}}=20.0\) and a maximum segment length \(\kappa_{\mathrm{seg}}=6000\). Winnow uses it for block parities and Hamming syndrome checks \(\mathcal{O}(\mathcal{T}_{\ell,j})\). Polar codes use it when the successive-cancellation decoder has low confidence. Because every protocol talks to Alice only through the oracle, we can count leakage \(\Lambda_{\mathrm{leak}}\) uniformly and compare protocols fairly.


In the simulation, Alice prepares weak coherent pulses with a mean photon number \(\mu=0.1\), and the signals travel through a fiber spool with attenuation \(\alpha_f=0.2\)~dB/Km. Unless otherwise stated, the fiber length is $d=50$ Km. An intercept–resend eavesdropper (Eve) is positioned between the fiber channel and Bob’s detector, intercepting each pulse reaching her with probability 20\%; for intercepted pulses, she performs a measurement in a randomly chosen basis and resends a newly prepared single-photon pulse encoded with the measured bit and her chosen basis, while forwarding unintercepted pulses unchanged. Detection is performed by a single-photon detector with efficiency \(\eta_{\mathrm{det}}=0.8\) and dark-count probability \(P_{\mathrm{dark}}=0.01\). Each data point is averaged over 100 independent Monte Carlo runs, with each run transmitting \(N_{\mathrm{tx}}=100{,}000\) pulses. Bob performs sifting using public basis announcements, retains the \(n_{\mathrm{sift}}\) events where his basis matches Alice's, and reserves \(n_{\mathrm{test}} = 0.2 n_{\mathrm{sift}}\) bits as a sample to estimate the QBER \(Q\). The disclosed test bits are discarded from both sides before reconciliation, leaving a working key of length \(n_{\mathrm{key}} = n_{\mathrm{sift}} - n_{\mathrm{test}}\) (denoted as \(k = |\mathbf{x}_B|\) during reconciliation) and resulting in a reconciled key \(K_{\mathrm{rec}}\) of length \(|K_{\mathrm{rec}}| \le n_{\mathrm{key}}\). The measured QBER \(Q\) drives both the reconciliation engine (block sizing, code rate, and frame-error budget) and the privacy amplification compression. We also compute the single-photon probability \(p_1\) directly from the channel transmission efficiency \(\eta_{\mathrm{ch}}\), detector efficiency \(\eta_{\mathrm{det}}\), and the Poissonian photon-number distribution \(P(n_{\mathrm{ph}})\) of Alice's source so that multi-photon pulses vulnerable to photon-number-splitting (PNS) attacks are excluded from the extractable key length \(L_{\mathrm{sec}}\).

For a fair comparison of QKD post-processing algorithms, all error correction protocols share the same physical BB84 configuration described above, while using protocol-specific reconciliation parameters. Cascade and Winnow use \(N_{\mathrm{pass}}=4\) reconciliation passes. The LDPC protocol uses a maximum of \(I_{\mathrm{BP}}=30\) damped normalized min-sum BP iterations (\(\alpha_{\mathrm{BP}}=0.75\), \(\beta_{\mathrm{BP}}=0.45\)) followed by reliability-ordered GF(2) OSD-0 post-processing, a maximum segment length \(\kappa_{\mathrm{seg}}=4224\), a target frame error rate of \(0.05\), a coarse parity-budget sweep factor \(\gamma=1.07\) (with a fine sweep factor \(\gamma_{\mathrm{fine}}=1.035\) when the syndrome mismatch count \(\mathrm{syn}^* \le \tau_{\mathrm{syn}} = 12\)) across a maximum of \(P_{\mathrm{dec}}=12\) decoding passes, and a universal-hash parameter \(t_{\mathrm{hash}}=32\) bits. Polar codes use a frame-error budget \(\Delta_{\mathrm{FER}}=0.01\) with a confidence threshold \(\theta_c=0.5\). The privacy amplification stage defaults to the Toeplitz universal hashing algorithm \(H_{\mathrm{PA}}\) using a public seed \(s_{\mathrm{pa}}\). The key observations from these simulations are mentioned below. 

\begin{figure}
        \centering
        \includegraphics[width=\linewidth]{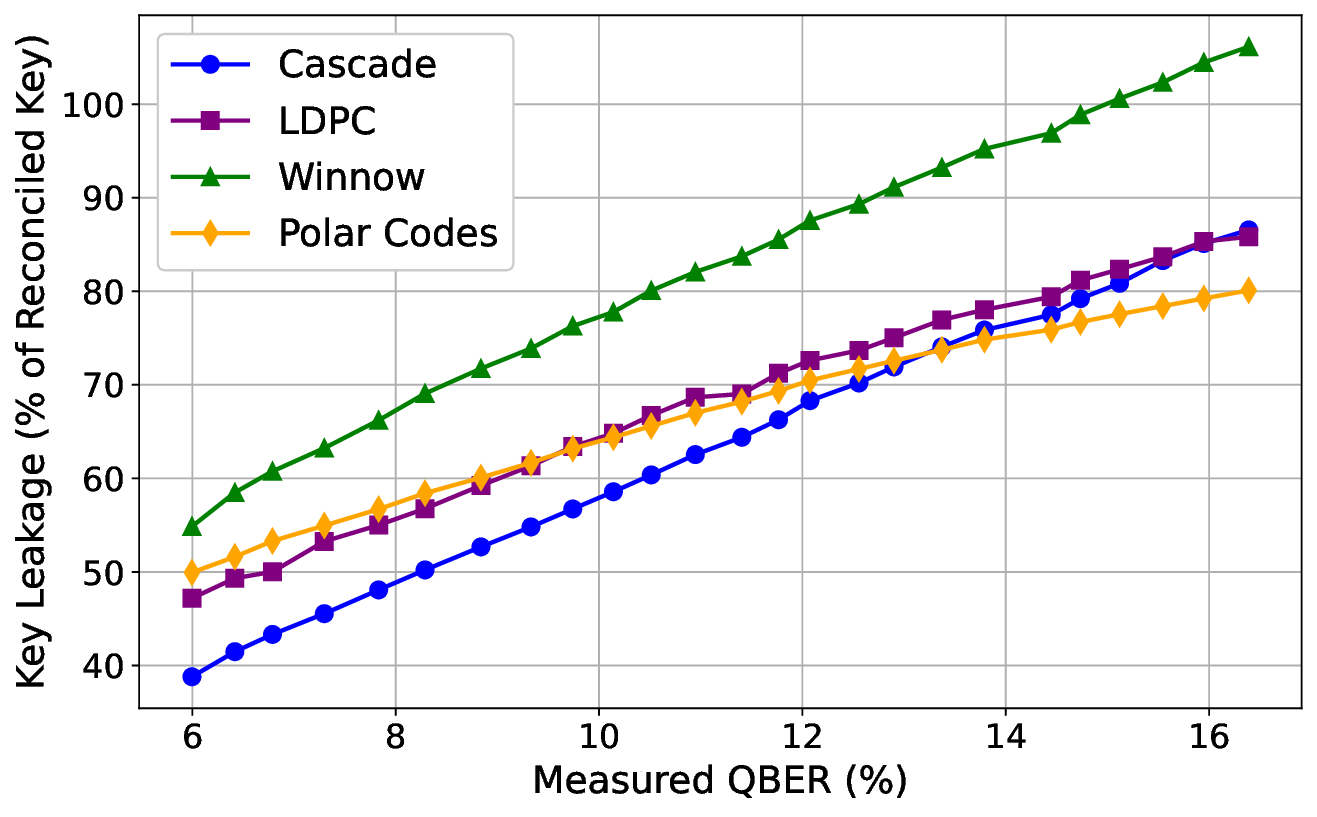}
        \caption{Key leakage as a percentage of the reconciled key i.e the key after error correction versus measured QBER \(Q\) for Cascade, LDPC, Winnow, and Polar codes over a \(d=50\)~km fiber link (\(\mu=0.1\), \(N_{\mathrm{tx}}=10^5\)).}\label{fig:leakage_qber}
 \end{figure}


To evaluate the trade-offs among the four error-correction protocols, we first examine the information they reveal during the reconciliation process. Fig.~\ref{fig:leakage_qber} compares the total key leakage against QBER. The key leakage is defined as the ratio of the total number of bits revealed over the classical channel \(\Lambda_{\mathrm{leak}}\) to the reconciled key length \(K_{\mathrm{rec}}\), expressed as a percentage: \((\Lambda_{\mathrm{leak}} / K_{\mathrm{rec}}) \times 100\). This quantifies the fraction of the key that an eavesdropper could potentially learn during the error correction phase. It is important to note that practical BB84 implementations typically abort key generation when the measured QBER \(Q\) exceeds the 11\% security threshold, so the analysis focuses primarily on the operational regime below this limit while also noting the extended behavior for completeness. The observed QBER range begins at approximately 6\% under the simulated channel, detector, and eavesdropping conditions. This non-zero baseline QBER partly reflects the intercept–resend attack, in which Eve intercepts a fraction of the transmitted pulses. This introduces additional bit errors whenever Eve's measurement basis differs from Alice's, providing a controlled, non-zero error baseline against which the reconciliation protocols can be evaluated under eavesdropping conditions.

As shown in Fig.~\ref{fig:leakage_qber}, the protocols exhibit distinct leakage profiles. Winnow consistently reveals the highest amount of information across all QBER values, starting near 50\% and exceeding 90\% at higher error rates. This occurs because Winnow transmits a full Hamming syndrome for every discordant block, which scales poorly as \(Q\) increases. In the operational regime below 11\% QBER, Cascade maintains the lowest leakage, growing smoothly. This is because Cascade relies on binary search on discordant blocks, querying only the parity information strictly necessary to locate errors. Polar codes and LDPC exhibit very similar but higher leakage profiles in this region, starting around 48\% and rising to roughly 70\%, both outperforming Winnow. While the protocol aborts above 11\% QBER, examining the extended range reveals important algorithmic scaling behaviors. Beyond 12\% QBER, LDPC and Polar catch upto Cascade performance. Cascade's leakage rises steeply and surpasses both LDPC and Polar at these higher error rates. This shift is attributed to the LDPC protocol's method of calculating parity budget using a finite-length starting point, and also by relying on sweeping up only when the hash check fails; LDPC strictly limits the revelation of excess parity bits even as the number of errors grows. Polar codes achieve a similarly competitive profile by only querying the oracle when the successive-cancellation decoder has low confidence. Cascade, despite its low leakage at low QBER, scales less efficiently at higher error rates due to the increased backtracking required by its binary search. 
\begin{figure}
        \centering
        \includegraphics[width=\linewidth]{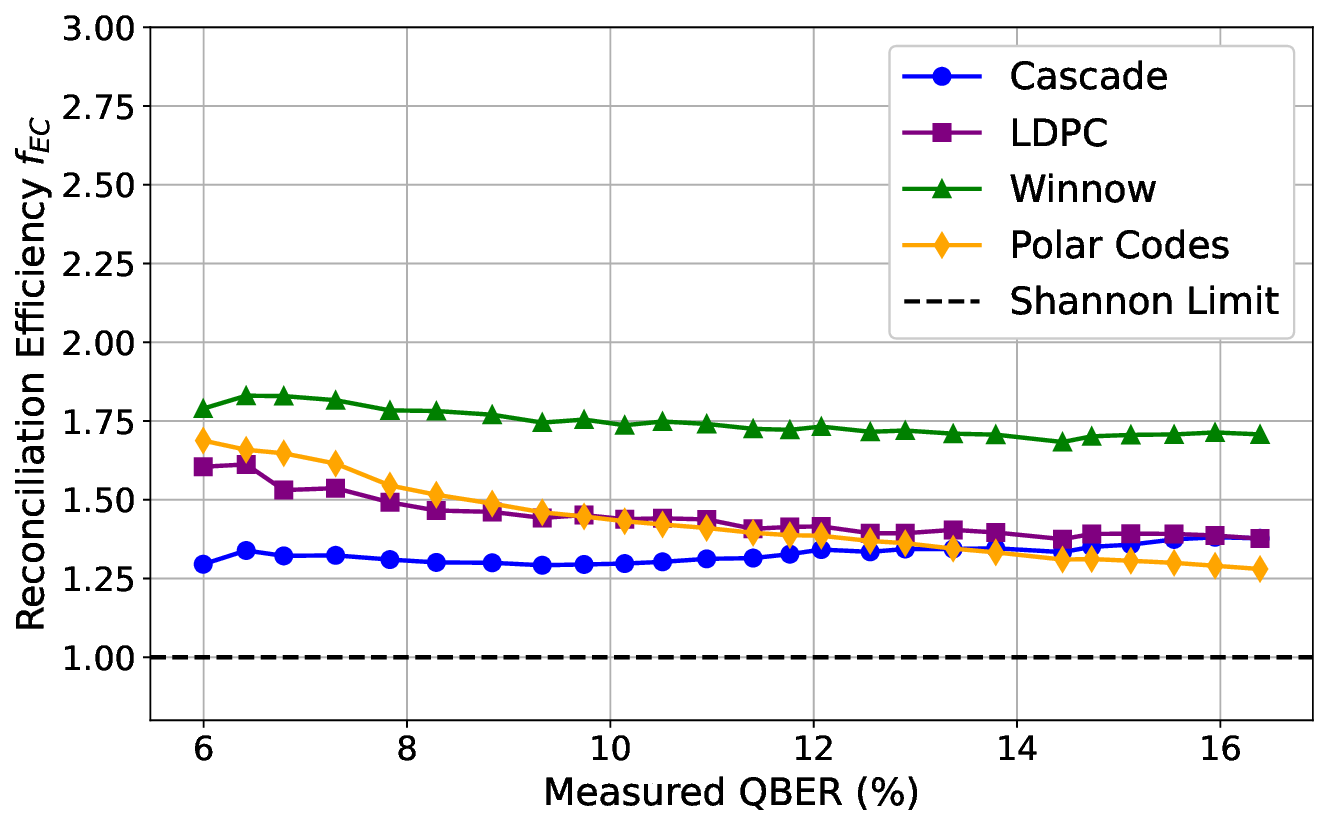}
        \caption{Information reconciliation efficiency  versus measured QBER \(Q\) compared against the theoretical Shannon limit (\(f_{\mathrm{EC}} = 1.0\)). Protocols closer to this limit reveal less redundant information.}\label{fig:IRE_qber}
    \end{figure}
    
Fig.~\ref{fig:IRE_qber} compares the performance metric reconciliation inefficiency \(f_{\mathrm{EC}}\), defined as $f_{\mathrm{EC}} = \Lambda_{\mathrm{leak}} / (K_{\mathrm{rec}} H_2(Q))$, where \(H_2(Q)\) is the binary entropy of the measured QBER \(Q\). Ideally, \(f_{\mathrm{EC}} \to 1\), meaning the protocol reveals only the minimum information required by the Shannon limit. In the operational regime below 11\% QBER \cite{shor2000simple}, Cascade maintains the lowest inefficiency, staying closest to the Shannon limit. This is because its binary-search mechanism reveals only the minimal parity required to locate discordant bits. Polar codes and LDPC exhibit highly similar efficiency profiles in this region, both starting higher than Cascade but catching up rapidly. Winnow consistently exhibits the highest inefficiency across all QBER values, as its block syndrome approach reveals highly redundant information. Beyond approximately 11\% QBER, LDPC and Polar show a very similar performance to Cascade. This trend is similar to the LDPC and Polar codes performance in information leakage, as discussed above. Thus, Cascade's efficiency degrades as the overhead of repeated backtracking accumulates with more errors.
   \begin{figure}
   \centering
\includegraphics[width=\linewidth]{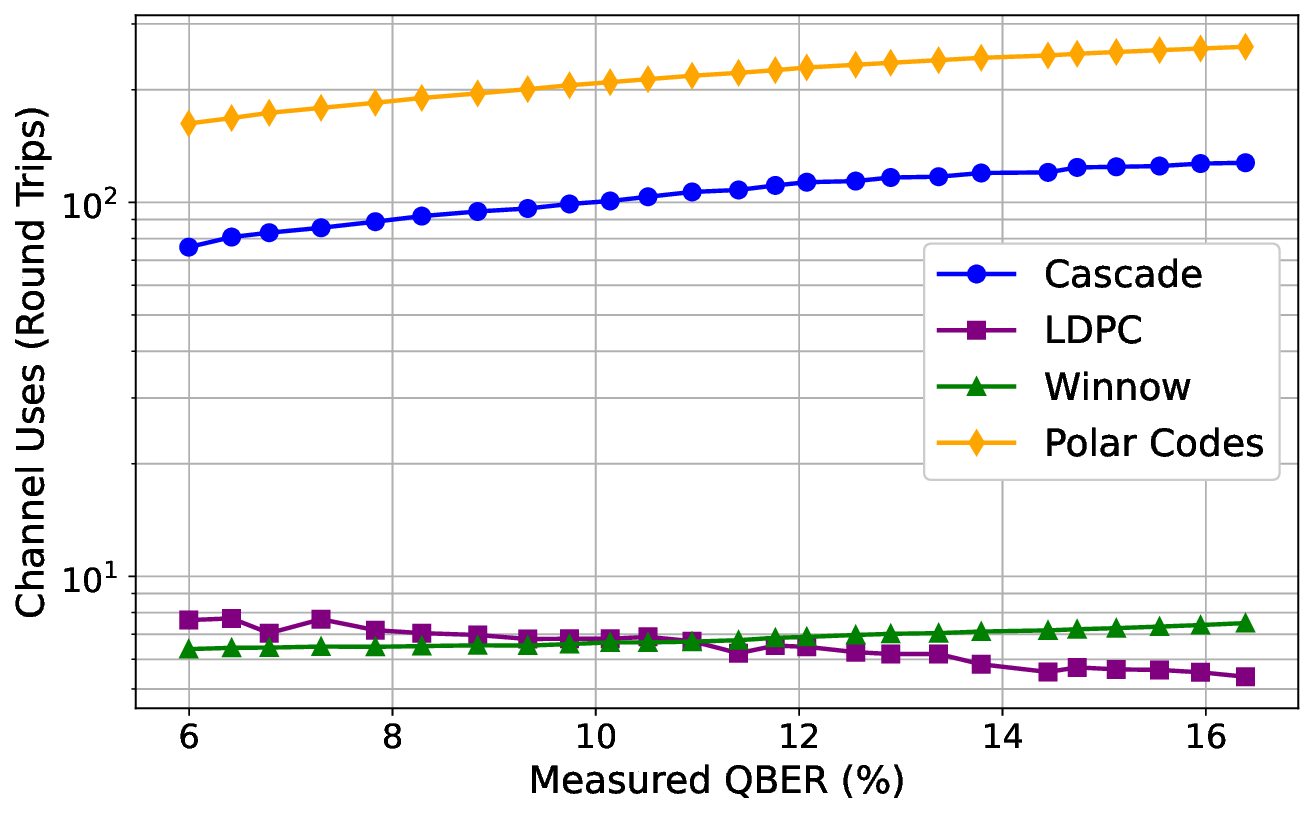}
        \captionof{figure}{Classical communication latency, measured in channel uses (round-trip oracle queries) on a logarithmic scale, versus measured QBER \(Q\).}\label{fig:latency_qber}
    \end{figure}
    
Fig.~\ref{fig:latency_qber} evaluates the classical communication latency on a logarithmic scale by plotting the number of channel uses (sequential round-trip oracle calls between Bob and Alice) as a function of the measured QBER \(Q\). Because network propagation delay over long-distance classical links often dominates post-processing throughput, minimizing round-trip interactions is critical for real-time QKD deployment. The curves in Fig.~\ref{fig:latency_qber} separate into two distinct latency classes separated by more than an order of magnitude: highly interactive protocols (Polar codes and Cascade) at the top and low-latency syndrome-based protocols (Winnow and LDPC) at the bottom. Polar codes require the highest number of channel uses. This high interaction count arises because the SC decoding tree traverses leaf nodes sequentially and must pause to query the oracle \(\mathcal{O}(\mathcal{Q}_{m_{\mathrm{pol}}})\) whenever a leaf reliability falls within the ambiguous confidence region defined by \(\theta_c\). Cascade is the second most interactive protocol, as each discordant block requires \(\lceil \log_2 \kappa \rceil\) sequential binary-search steps plus recursive backtracking queries across \(N_{\mathrm{pass}} = 4\) passes. In contrast, Winnow and LDPC require fewer than \(10\) round trips across the entire QBER range. Winnow remains nearly constant between round trips because it exchanges all block parities in batch and computes multi-bit Hamming syndromes in a single batched round trip per pass. While LDPC queries all \(m_c\) initial syndrome bits \(\mathbf{v} = (v_0, \dots, v_{m_c-1})\) in a single batched oracle call across the check-row supports \(S_j\), and likewise batches the \(t_{\mathrm{hash}}\) universal-hash verification queries \(V_j\) into a single round trip.
\begin{figure}
\centering
\includegraphics[width=\linewidth]{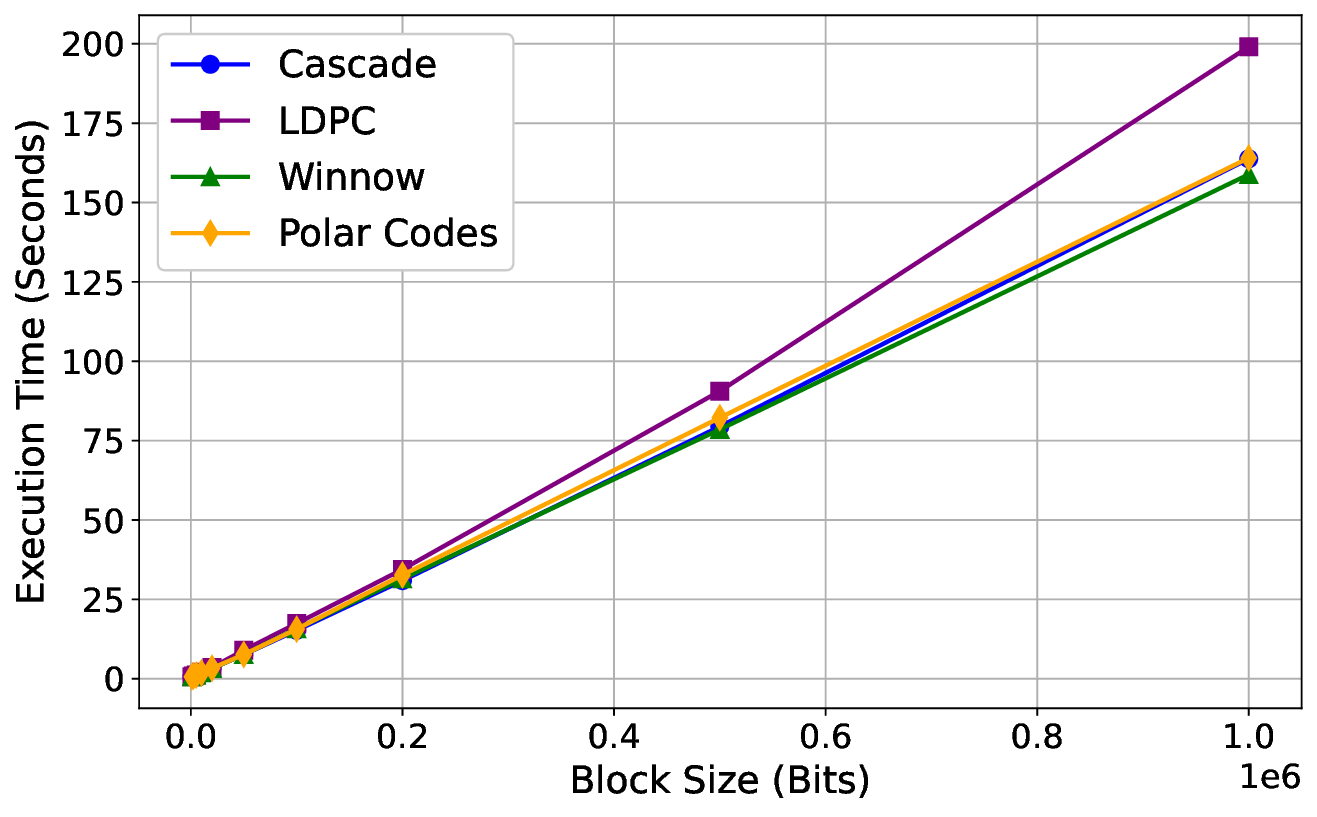}
        \caption{Total post-processing execution time (s) versus transmitted block size \(N_{\mathrm{tx}}\) (bits) from \(10^3\) to \(10^6\) bits.}\label{fig:E_time_block_size}
\end{figure}

Fig.~\ref{fig:E_time_block_size} presents the computational scalability of the four protocols by comparing the total end-to-end execution time (in seconds) against the transmitted block size \(N_{\mathrm{tx}}\) from \(10^3\) to \(10^6\) bits. All execution-time measurements were obtained on a machine equipped with an Intel i7-10750H processor, 16 GB RAM, and an NVIDIA GeForce RTX 2060 with Max-Q Design. It can be observed that the LDPC protocol requires the highest execution time because it uses expensive decoding steps. In each rate-adaptive pass, the decoder repeatedly passes damped normalized min-sum messages \(M_{i \to a}\) and \(M_{a \to i}\) along the edges of the Tanner graph defined by \(\mathbf{H}_{\mathrm{act}}\) until the residual syndrome \(\mathbf{s}_{\mathrm{res}}\) vanishes or the decoder reaches \(I_{\mathrm{BP}} = 30\) iterations. When BP alone leaves a non-zero syndrome mismatch (\(\mathrm{syn} > 0\)), the protocol invokes reliability-ordered GF(2) OSD-0 post-processing, which ranks variable nodes by their reliability score \(\psi_i = -\eta_{\mathrm{OSD}} \delta_i + \rho_{c,i}\) and performs Gaussian elimination over the \(m_c \times k_c\) active parity-check submatrix \(\mathbf{H}_{\mathrm{act}}\) to solve for the error pattern \(\mathbf{e}_{\mathrm{OSD}}\).  Cascade, Winnow, and Polar codes scale much more favorably. Cascade and Winnow rely on simple parity checks, and Polar's successive cancellation decoding has a complexity of \(O(N \log N)\). 

\begin{figure}
    \centering
    \includegraphics[width=\linewidth] {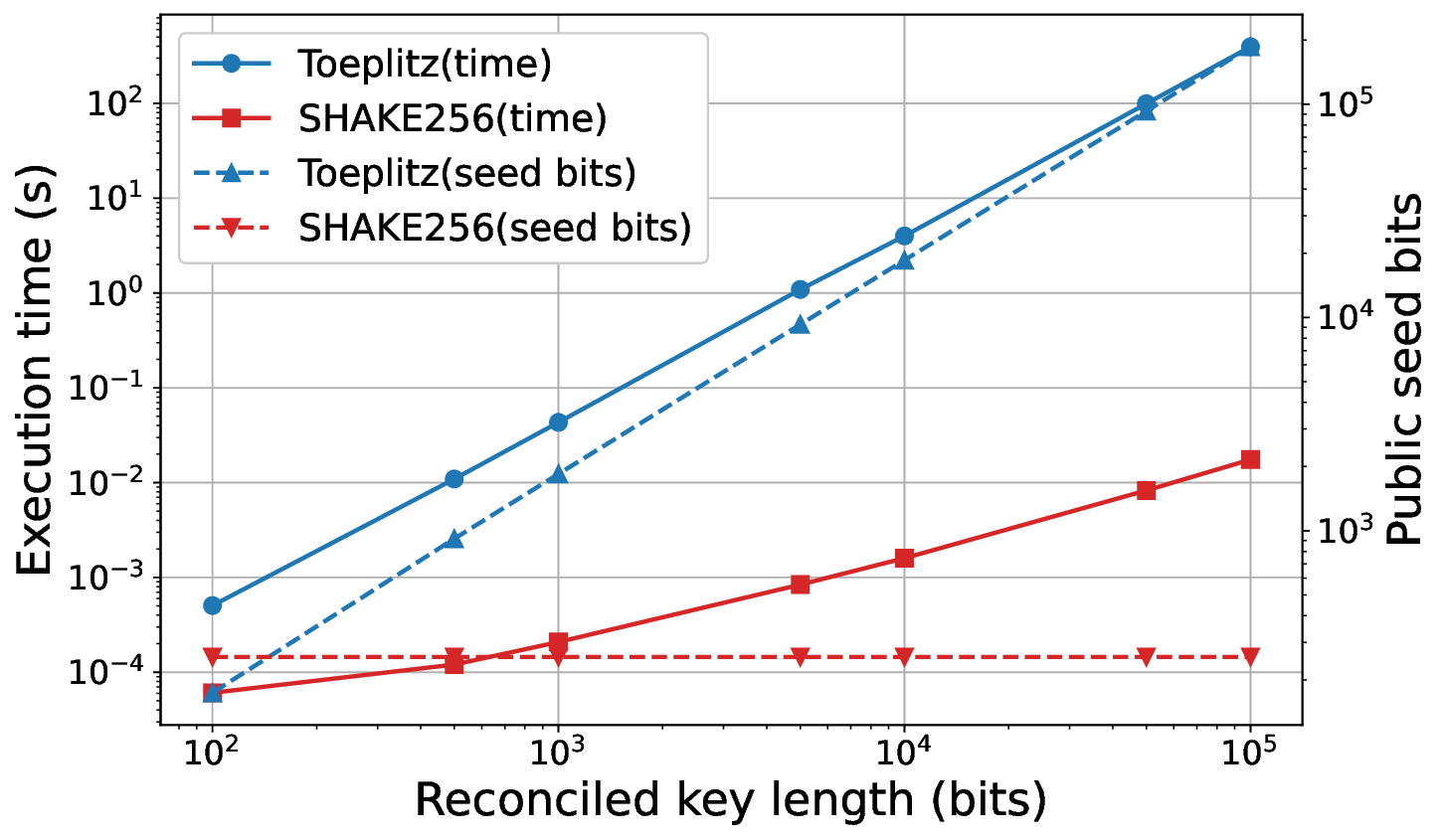}
    \caption{Privacy amplification benchmark comparing Toeplitz universal hashing and SHAKE256 across reconciled key lengths in \(k \in [10^2, 10^5]\) bits. Solid curves (left axis) depict execution time (s), and dashed curves (right axis) indicate required public seed bits.}\label{fig:pa_performance}
\end{figure}
Fig.~\ref{fig:pa_performance} compares Toeplitz universal hashing and SHAKE256 for the privacy amplification stage across reconciled key lengths in \(k \in [10^2, 10^5]\) bits. As the reconciled key length \(k\) grows, Toeplitz execution time increases much faster than SHAKE256, because Toeplitz uses the pseudo-random sequence \(\mathbf{w}\) from \(\mathcal{G}_{\mathrm{prng}}\) to form an \(L_{\mathrm{sec}} \times k\) matrix and computes \(L_{\mathrm{sec}}\) GF(2) inner products (\(\mathbf{w}_i \cdot K_{\mathrm{rec}}\)). SHAKE256 instead absorbs \(s_{\mathrm{pa}} \parallel K_{\mathrm{rec}}\) into the sponge state \(\mathcal{H}_{\mathrm{shake}}\) and squeezes \(B_{\mathrm{out}}\) output bytes, so its computational cost grows roughly linearly with \(k\). The trade-off also appears in the public seed size: Toeplitz requires \(k + L_{\mathrm{sec}} - 1\) random seed bits, which scale linearly with \(k\), whereas SHAKE256 requires only a small, fixed-length public seed \(s_{\mathrm{pa}}\) (256 bits). Thus, Toeplitz provides an information-theoretic universal-hashing security proof via the Leftover Hash Lemma but is slower and seed-intensive, whereas SHAKE256 is faster and seed-efficient but relies on the computational security assumptions of the cryptographic sponge construction. The optimal choice depends on whether unconditional information-theoretic security or practical throughput is the primary system requirement.

\begin{figure}
        \includegraphics[width=\linewidth]{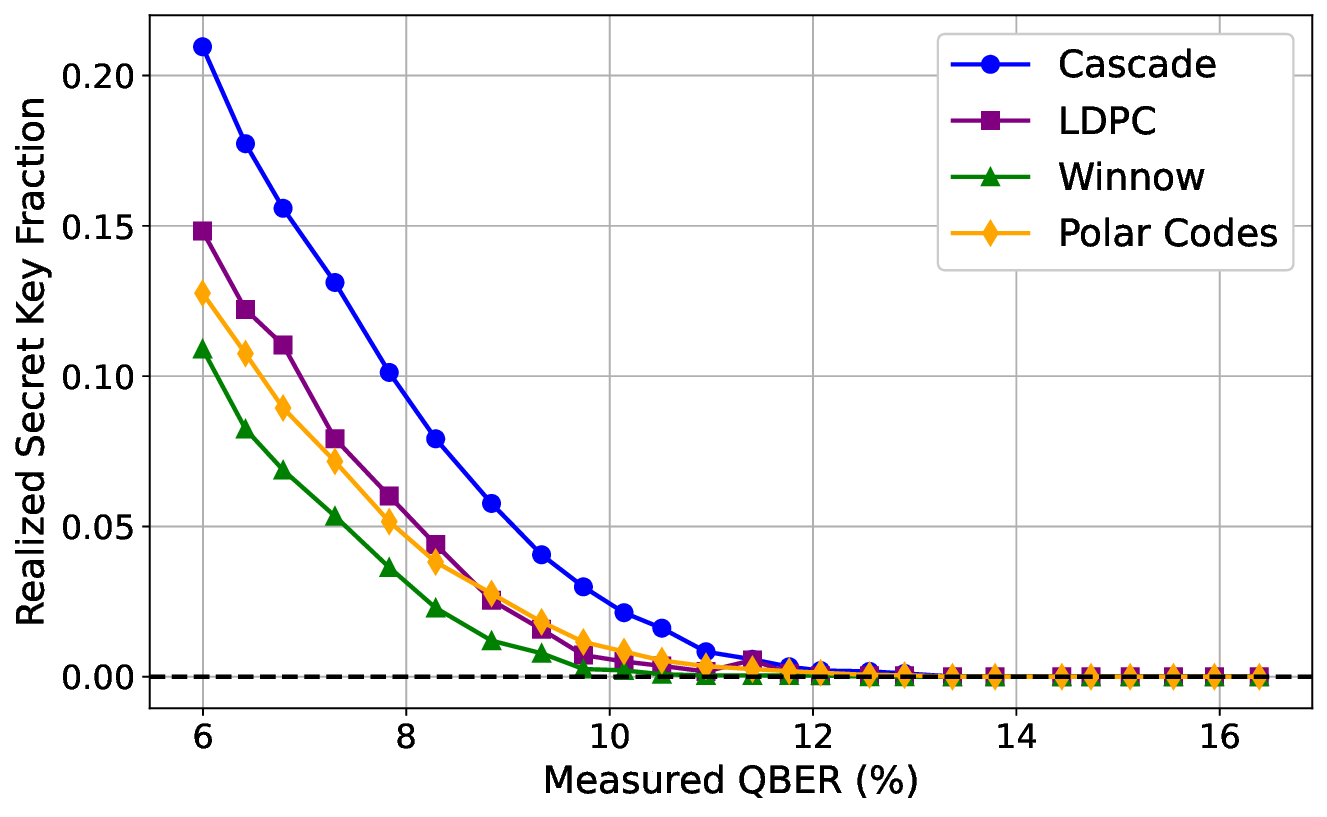}
        \caption{Realized Secret key fraction versus measured QBER \(Q\) after error reconciliation and Toeplitz privacy amplification over a \(50\)~km link.}\label{fig:skrf_qber}
\end{figure}
 
Fig.~\ref{fig:skrf_qber} compares the realized secret key fraction, which measures the actual secure key yield relative to the initial sifted key. Unlike theoretical estimates, this metric accounts for real-world operational losses, including parameter estimation, discarded segments, and the privacy amplification required to erase both the eavesdropper's potential knowledge and the data leaked during error correction. As the QBER rises, the key fraction drops for all protocols because correcting more errors inherently leaks more data. Cascade achieves the highest key yield across the viable operating range (up to roughly 11\% QBER) due to its minimal parity leakage during error correction. The 5G NR LDPC and Polar codes perform similarly and consistently outperform Winnow. LDPC holds a slight edge at very low error rates due to its highly optimized matrix design and efficient initial parity estimation. Conversely, polar codes perform slightly better at higher QBER, as LDPC's incremental parity sweeps introduce minor data leakage at higher error rates. Winnow yields the lowest key fraction overall, and around 11\% QBER, the combined security penalties and leakage overhead overwhelm the available key for all protocols, dropping the final yield to zero.

\begin{figure}
    \centering
    \includegraphics[width=\linewidth]{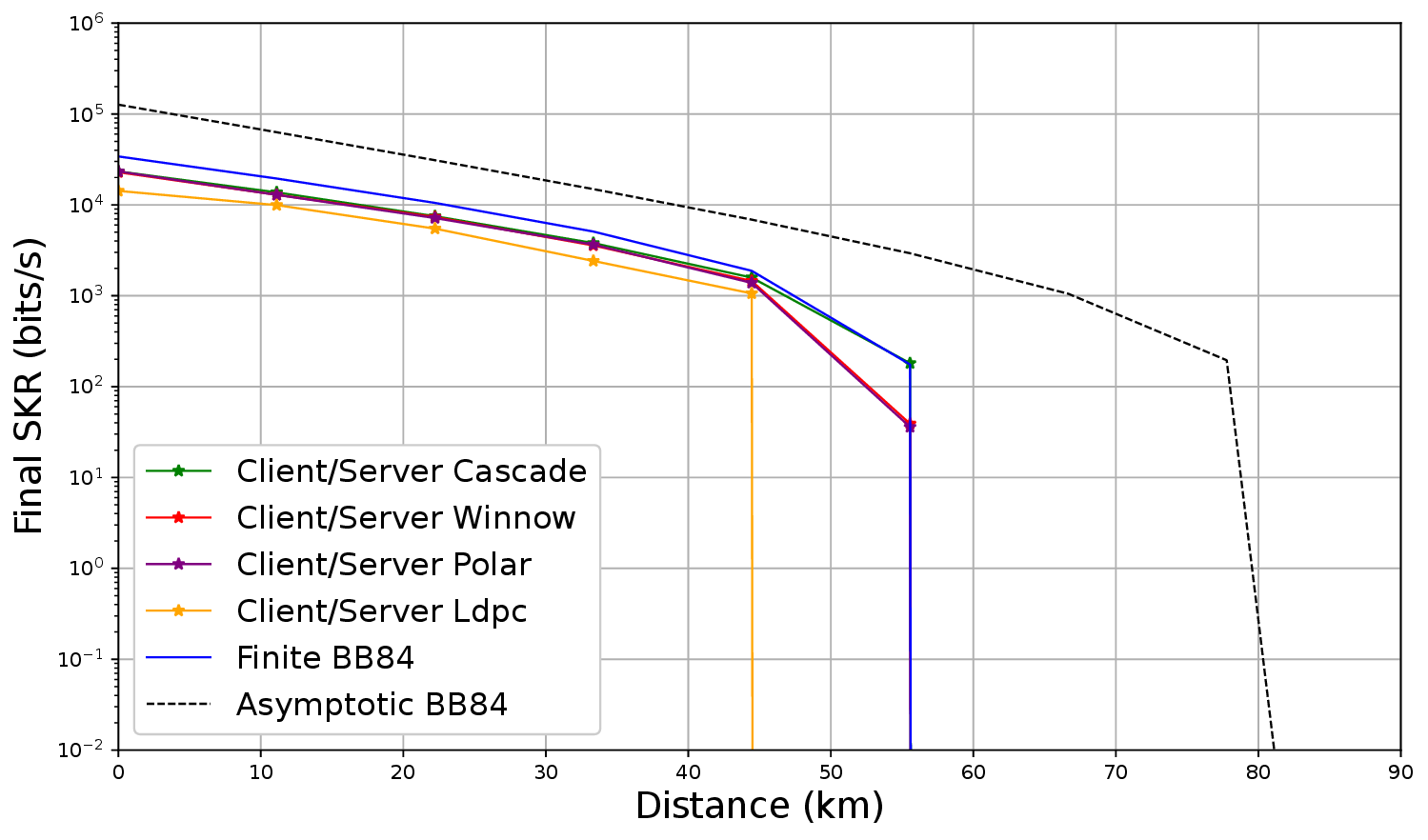}
    \captionof{figure}{Secret key rate \(R_{\mathrm{sk}}\) versus fiber transmission distance \(d\) (km), comparing the client-server BB84 post-processing protocols against theoretical asymptotic and finite-key BB84 bounds.}\label{fig:skr_distance}
\end{figure}
\begin{figure}
        \centering
        \includegraphics[width=\linewidth]{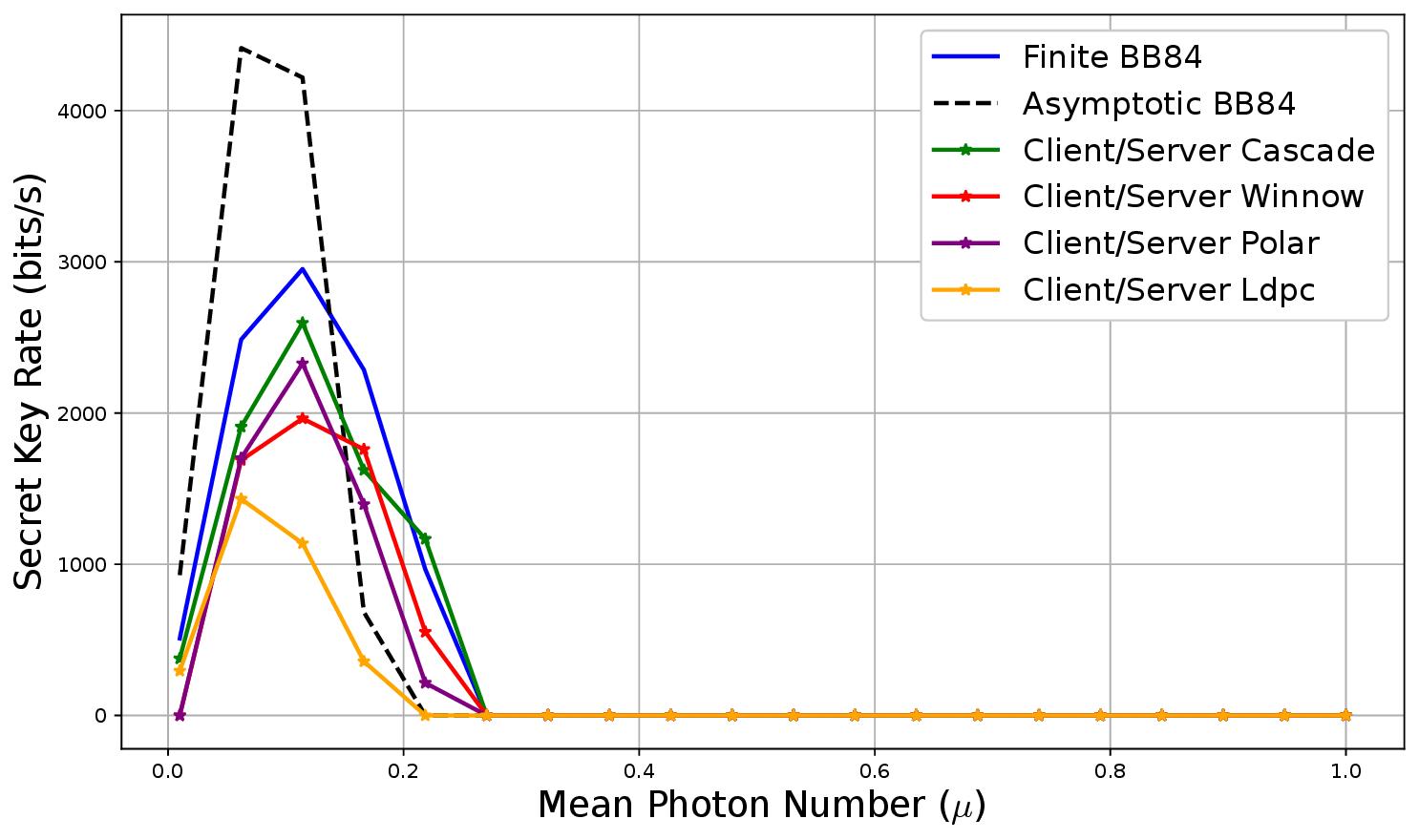}
        \captionof{figure}{Secret key rate \(R_{\mathrm{sk}}\) versus mean photon number \(\mu\) at \(d=40\)~km with a \(1\)~MHz source repetition rate, comparing practical client-server reconciliation protocols against finite-key and asymptotic BB84 theoretical bounds.}\label{fig:skr_mpn}
\end{figure}
Fig.~\ref{fig:skr_distance} and~\ref{fig:skr_mpn} compare the server-client protocols against finite-size and asymptotic BB84 theoretical SKR formulas. Fig.~\ref{fig:skr_distance} compares the SKR versus distance of the server-client-based practical post-processing protocols against the theoretical asymptotic and finite-size SKR expressions. It can be observed that the asymptotic SKR represents the upper bound, the finite-size SKR curve sits below it, and the server-client protocols sit slightly below the finite-size curve because they include practical losses. All protocols track each other closely up to about 40~km, then they drop sharply. LDPC drops out around 45~km, slightly earlier than Cascade and Polar (around 55~km), due to the residual overhead of BP decoding. In Fig.~\ref{fig:skr_mpn}, the SKR is plotted against the mean photon number \(\mu\) at 40~km. All curves peak at a low \(\mu\) (around 0.1--0.15) and then fall to zero. This peak occurs because increasing \(\mu\) raises the signal rate but also increases the multi-photon probability, which helps Eve's photon-number-splitting attack. LDPC, Cascade, and Polar codes peak at very similar values, while Winnow peaks slightly lower. Beyond \(\mu \approx 0.2\), all protocols go to zero as the secure key rate becomes negative after accounting for Eve's information.

\begin{figure}
    \centering
    \includegraphics[width=\linewidth]{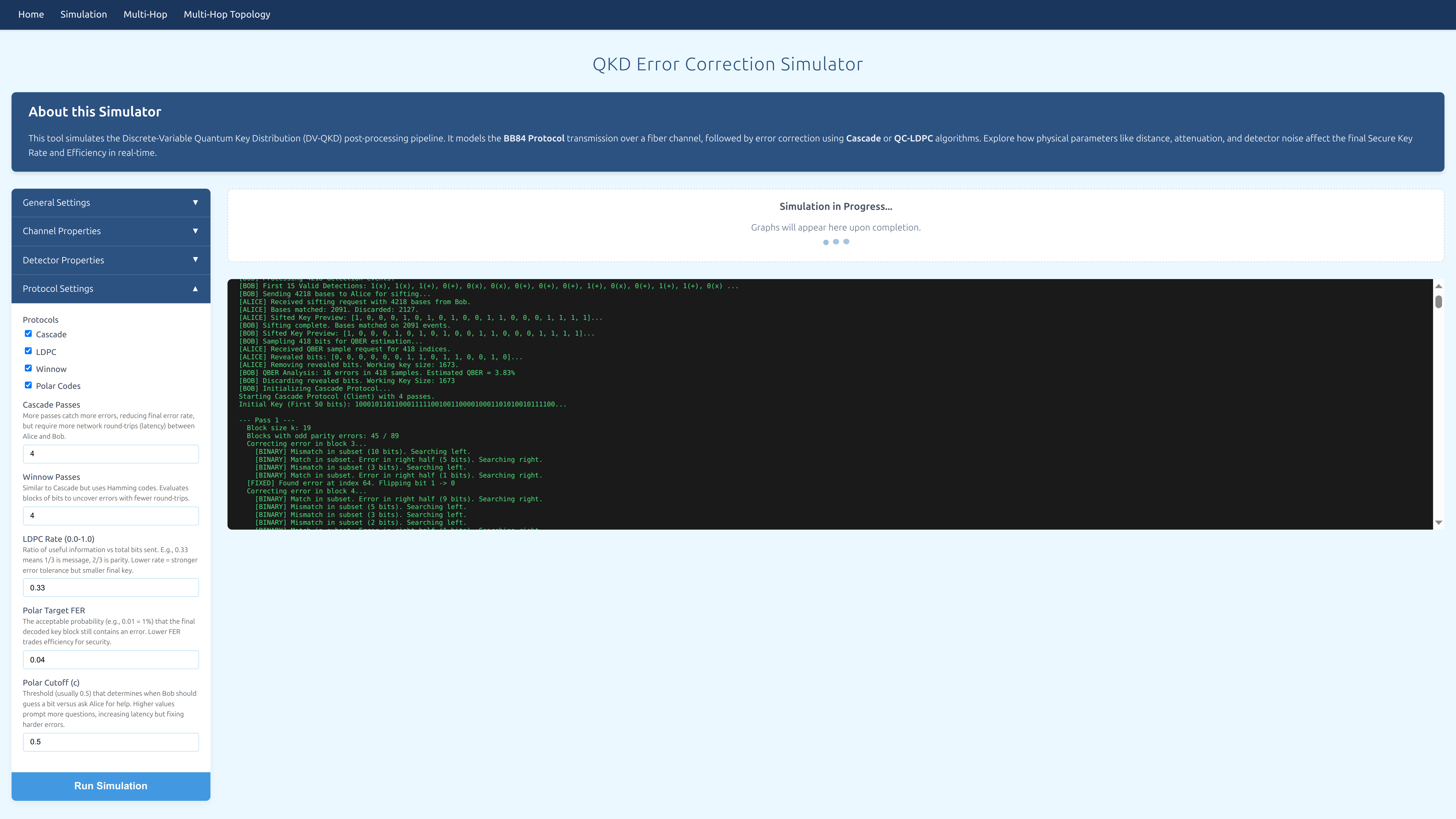}
        \captionof{figure}{Web-based GUI for the simulator using Django base}\label{fig:gui_home}
\end{figure}

\begin{figure}
        \centering
        \includegraphics[width=\linewidth]{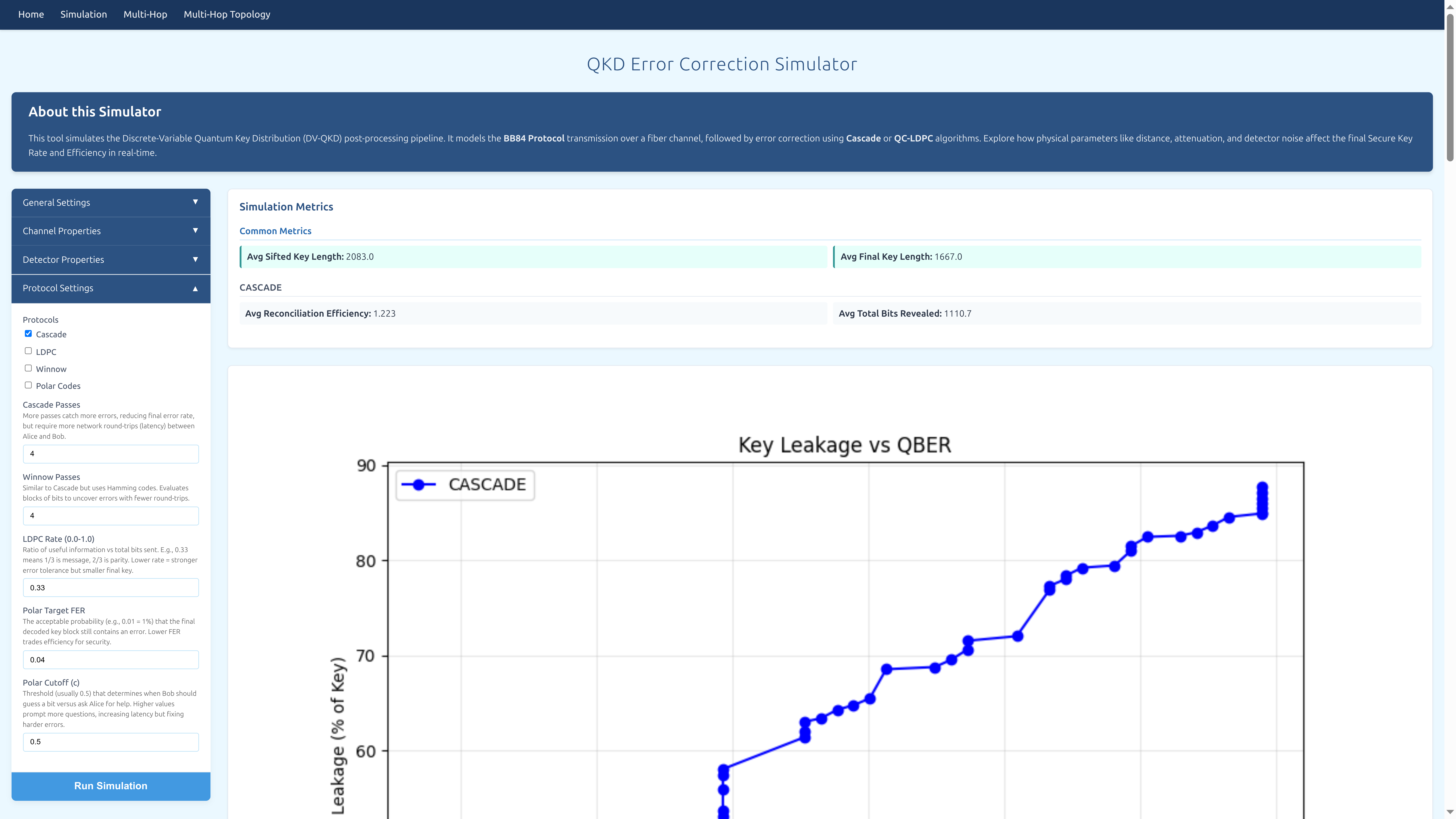}
        \captionof{figure}{Interactive Web-based GUI for the server client framework for students and researchers}\label{fig:gui_interactive}
\end{figure}

Overall, the results show a clear trade-off for different error correction protocols. Cascade provides the best balance of low leakage and low inefficiency, though it requires high interactivity. LDPC offers the lowest latency (fewest round trips) and competitive efficiency, making it highly suitable for high-speed networks, at the cost of higher computational time for decoding. Polar Codes offer a strong middle ground with good efficiency and lower computational complexity than LDPC, but higher latency. Winnow shows moderate performance but its high leakage and inefficiency limit its use to lower QBER regimes. The finite-size SKR analysis confirms that these practical protocols approach the theoretical finite-key bound closely at short distances and only diverge when the key becomes too short for efficient reconciliation.

In addition to these protocol-level simulations, the framework provides accessible graphical user interfaces, shown in Figs.~\ref{fig:gui_home} and~\ref{fig:gui_interactive}. These interfaces are designed for educational and exploratory use: beginners, students, and researchers can run simulations and inspect QBER, SKR, leakage, and reconciliation behaviour without writing code, while experienced researchers can use the server-client workflow to reproduce experiments, compare protocols, and prototype new configurations.

\section{Conclusion}
\label{sec:Conclusion}

In this paper, we have presented a comprehensive algorithmic-level tutorial on QKD post-processing algorithms, including cascade, winnow, LDPC, and polar code-based information reconciliation, along with Toeplitz and SHAKE256 hashing-based privacy amplification protocols. Furthermore, we have also presented a comprehensive performance comparison of these QKD post-processing algorithms using an actor-based server-client framework, developed and validated through an end-to-end simulator. We have also compared the SKR performance of the practical QKD post-processing algorithms against the theoretical asymptotic and finite-size SKR expressions. Finally, an interactive web-based GUI is built on a server-client actor architecture that can be used by researchers, students, and practitioners to configure the protocol parameters, run simulations, and visualize QBER, SKR, leakage, and reconciliation efficiency. This work presents a practical simulation framework for DV-QKD protocols that incorporates post-processing algorithms. For future work, the framework can be extended to CV-QKD protocols.


\bibliographystyle{IEEEtran}
\bibliography{qkd}

\end{document}